\documentclass[12pt]{article}
\usepackage{savesym}
\usepackage{cite}
\usepackage{wasysym}
\usepackage{amscd}
\usepackage{graphicx}
\usepackage{pifont}
\usepackage{float}
\usepackage{subfig}
\usepackage[all]{xy}
\usepackage{multicol}
\usepackage{amsfonts}
\usepackage{picture}
\usepackage{amssymb}
\savesymbol{iint}
\savesymbol{iiint}
\usepackage{amsmath}
\restoresymbol{TXF}{iint}
\restoresymbol{TXF}{iiint}
\usepackage{color}
\usepackage{heck}
\usepackage{hyperref}
\hypersetup{colorlinks=true}
\hypersetup{linkcolor=black}
\hypersetup{citecolor=black}
\hypersetup{urlcolor=black}
\usepackage{setspace}
\usepackage{wrapfig}
\usepackage{rotating}

\newcommand{\qed}{\nobreak \ifvmode \relax \else
      \ifdim\lastskip<1.5em \hskip-\lastskip
      \hskip1.5em plus0em minus0.5em \fi \nobreak
      \vrule height0.75em width0.5em depth0.25em\fi}
\usepackage{verbatim}
\numberwithin{equation}{section}
\begin{document}

\date{September, 2011}

\institution{HarvardU}{\centerline{${}^{1}$Jefferson Physical Laboratory, Harvard University, Cambridge, MA 02138, USA}}

\institution{SISSA}{\centerline{${}^{2}$Scuola Internazionale Superiore di Studi Avanzati, Via Bonomea 265 34100 Trieste, ITALY}}

\title{BPS Quivers and Spectra of Complete $\mathcal{N}=2$ Quantum Field Theories}

\authors{Murad Alim,\worksat{\HarvardU}\footnote{e-mail: {\tt alim@physics.harvard.edu}} Sergio Cecotti,\worksat{\SISSA}\footnote{e-mail: {\tt cecotti@sissa.it}} Clay C\'{o}rdova,\worksat{\HarvardU}\footnote{e-mail: {\tt clay.cordova@gmail.com}} Sam Espahbodi,\worksat{\HarvardU}\footnote{e-mail: {\tt espahbodi@physics.harvard.edu}} Ashwin Rastogi,\worksat{\HarvardU}\footnote{e-mail: {\tt rastogi@physics.harvard.edu}} and Cumrun Vafa\worksat{\HarvardU
}\footnote{e-mail: {\tt vafa@physics.harvard.edu}}}

\abstract{We study the BPS spectra of $\mathcal{N}=2$ complete quantum field theories in four dimensions.  For examples that can be described by a pair of M5 branes on a punctured Riemann surface we explain how triangulations of the surface fix a BPS quiver and superpotential for the theory.     The BPS spectrum can then be determined by solving the quantum mechanics problem encoded by the quiver.  By analyzing the structure of this quantum mechanics we show that all asymptotically free examples, Argyres-Douglas models, and theories defined by punctured spheres and tori have a chamber with finitely many BPS states.  In all such cases we determine the spectrum.

}

\maketitle

\enlargethispage{\baselineskip}

\setcounter{tocdepth}{3}
\begin{spacing}{1}
\tableofcontents
\end{spacing}
\section{Introduction}
It has long been understood that BPS spectra play a pivotal role in the study of
quantum field theories with extended supersymmetry. This was especially
underscored in the breakthrough Seiberg-Witten solution to $\mathcal{N}=2$ 4d
supersymmetric Yang-Mills theory \cite{SW1,SW2}. In that work, finding
singularities in moduli space where various BPS states became massless was the
fundamental insight that led to the full solution of the IR dynamics of the
theory. However, in spite of the general consensus about the importance of
BPS states and a wealth of recent progress in the subject \cite{KS,GMN08,GMN09,GMN10,GMN11,ADJM1,ADJM2,MPS,JS1,JS2,DG,DGS}, there are relatively few explicit examples, and hence little known about the general
structure of BPS spectra.

In this paper we take the first steps towards developing a theory of BPS
spectroscopy, and in particular we determine the BPS spectra for an infinite
set of new examples.  We investigate BPS spectra in a class of $\mathcal{N}=2$
quantum field theories known as \emph{complete theories} \cite{CV11}.  These
theories are defined by the property that as one varies all parameters
(including moduli, couplings and bare masses), the number of independent central
charges  is equal to the rank of the charge lattice.  Completeness is a strong
assumption about a field theory and is typically not satisfied.  However, a
rich class of examples of such theories includes all the four dimensional
$\mathcal{N}=2$ models that can be obtained by wrapping a pair of M5 branes on
a punctured Riemann surface.  These are the so-called rank two Gaiotto theories
\cite{Gaiotto, W, CD1, CD2}.  As determined in \cite{CV11}, such
examples have an additional
remarkable property: their BPS spectrum can be encoded by a \emph{BPS quiver}
\cite{Denef,DM,DFR1,DFR2,CNV,CV11,CDZ, Fiol, FM00}.   This dramatically simplifies the
problem of finding BPS states.  In place of some tedious
weak coupling physics or intractable strong coupling dynamics,
the BPS spectrum is governed by a quantum mechanics problem encoded in this quiver.

Because of their simplicity, the class of complete theories defined by pairs of
M5 branes on Riemann surfaces will be the focus of our investigation in this
work.  Broadly speaking, our aim is to determine and understand the BPS
quiver in such examples and, when possible, to solve the associated quantum
mechanics problem and determine the BPS spectrum.  With this in mind, section
\ref{review} provides a brief synopsis of some necessary background on BPS
quivers.

Next, to accomplish our first goal of determining the BPS quiver, in section
\ref{surfaces} we reconstruct these complete theories via geometric engineering
in type IIB string theory on a local Calabi-Yau threefold.\cite{KMV,KKV,KLMVW}.
Such an approach has the advantage that the BPS states can be explicitly
identified as D3-branes  wrapping special lagrangian cycles in the Calabi-Yau.
This makes the appearance of a quiver in the BPS state counting problem
manifest: the quiver simply encodes the world volume quantum mechanics of the
D3-branes.\cite{DM}  However, we can go further and pass from this implicit
description of the quantum mechanics of D3-branes to an explicit algorithm for
constructing the BPS quiver.  As we review there, the structure of the quiver
is completely encoded by a certain triangulation of the Gaiotto curve, the
Riemann surface where the pair of M5 branes lives.  Further, we explain how the
same triangulation allows one to compute the superpotential for the quiver, and
in this way makes the task of determining the full BPS quiver data for any
given example an algorithmic procedure.

Having accomplished our first task, in section \ref{finite} we focus on
determining what structure the resulting spectra possess and on computing some
explicit examples. The problem of determining the BPS spectrum is
computationally most tractable in a chamber where there are finitely many BPS
states; we restrict our attention to this case.  One of our most interesting
results is a determination of an infinite class of theories which have such a
finite chamber.  Indeed, as we prove in section \ref{finite}, theories with
finite chambers include all asymptotically free examples, Argyres-Douglas
models, and theories defined by punctured spheres and tori.  The latter
examples are particularly interesting: They are conformal field theories where
the only breaking of scale invariance is that introduced by adding bare mass
terms.  In all such cases the spectrum can be calculated explicitly and
algorithmically using the techniques developed herein.

Finally, in section \ref{exceptional} we undertake a brief investigation of complete theories with BPS quivers which do not come from Gaiotto type constructions.  In \cite{CV11} such theories were classified.  They consist of eleven exceptional theories which are not of the Riemann surface type.  For all these examples except one, we determine an associated superpotential and a finite chamber of BPS states.
\section{Review of BPS Quivers}
\label{review}
The role of BPS quivers in $\mathcal{N}=2$ 4d gauge theories was systematically developed in \cite{ACCERV}. Here we will review the key points, and refer the reader to that paper for the full dicussion.
\subsection{$\mathcal{N}=2$ Data}
To begin, we consider the data of an arbitrary $\mathcal{N}=2$ 4d theory that will enter into the discussion. The theory has a moduli space of vacua; we focus on the Coulomb branch, $\mathcal{U}$. At a generic point $u$ in this moduli space, we have a $U(1)^r$ gauge symmetry, with the following additional structure:
\begin{itemize}
\item A charge lattice $\Gamma$ with dimension $2r+f$, where $f$ is the rank of the flavor symmetry. The charges of BPS states form a set of occupied points in this lattice.
\item An antisymmetric electric-magentic inner product on the charge lattice $\circ :\Gamma\times\Gamma\rightarrow \mathbb{C}$.
\item A central charge function $\mathcal{Z}_u:\Gamma\rightarrow \mathbb{C}$.
\end{itemize}

For the theories we will study, the above data are conveniently encoded in a family of Riemann surfaces $\Sigma_{u}$, the Seiberg-Witten curve, which varies over the Coulomb branch.  This surface has the property that its homology lattice of one-cycles is naturally identified with the lattice of electric and magnetic charges in the theory.  Since the central charge function is linear on the charges, it can be presented as an integral of a one-form, the Seiberg-Witten differential, $\lambda_{u}$.  That is for $\gamma \in \Gamma$
\begin{equation}
\mathcal{Z}_{u}(\gamma)=\int_{\gamma}\lambda_{u}.
\end{equation}
Furthermore, the electric-magnetic inner product is naturally identified with the intersection product on one-cycles. Flavor symmetries are incorporated by allowing $\lambda_{u}$ to have a first order poles on $\Sigma_{u}$, whose residues encode the bare masses of the theory.  Together the family $(\Sigma_{u}, \lambda_{u})$ constitutes the complete solution to the low-energy physics.

\subsection{BPS Quiver}\label{quivs}
The Seiberg-Witten construction gives the extreme IR solution of $\mathcal{N}=2$ field theories; to study the massive BPS sector, we must introduce some new structure. The BPS quiver will allow computation of the full BPS spectrum of the theory at some fixed point of moduli space, $u$, given some data about a basis of elementary BPS states. Choose a half-plane $\mathcal{H}$ in the complex central charge $\mathcal{Z}$-plane. All states with central charge in $\mathcal{H}$ will be considered `particles', whereas the states in the opposite half-plane will be considered `anti-particles.' The choice of half-plane is clearly a convention, and the resulting total BPS spectrum of particles and anti-particles must be invariant under different choices. Suppose there exists a set of $2r+f$ hypermultiplet states, $\{\gamma_i\}$, in the chosen half-plane that form a positive integral basis for all particles.\footnote{Of course, finding such a basis would require us to know the BPS spectrum at a single point in parameter space. We will be able to get around even this by constructing the quiver from auxiliary means, such as the techniques of section \ref{surfaces}.} When such a basis exists, it is unique. There are theories for which such a basis fails to exist, but in this paper we will not consider such theories.

Given the basis, $\{\gamma_i\}$, we can construct a quiver as follows: For every charge, $\gamma_i$, draw a node associated to it.
For every pair of charges, $\gamma_i,\gamma_j$, with $\gamma_i\circ\gamma_j>0,$ draw $\gamma_i\circ\gamma_j$ arrows from $\gamma_j$ to $\gamma_i$.\footnote{Note that by construction, the resulting quiver is two-acyclic.  That is, the quiver has no oriented cycles of length two.} The problem of finding BPS bound states of the elementary BPS particles, $\{\gamma_i\}$ is solved by a quiver quantum mechanics problem on this quiver.

\subsection{Quiver Quantum Mechanics}
We now wish to check whether a particular site of the charge lattice, $\gamma=\sum_i n_i\gamma_i\in \Gamma$ is occupied by a BPS state, and if so, to determine the spin and degeneracy of the associated particles. If we do this for every $\gamma$, we will have solved the full BPS spectrum. The interactions of the elementary BPS states are encoded in the four supercharge quiver quantum mechanics of the quiver given above. In other words, we have a (0+1)d supersymmetric gauge theory with
\begin{equation}
\mathrm{Gauge \  Group}= \prod_{\mathrm{nodes}}U(n_{i}); \hspace{.5in} \mathrm{Matter}=\bigoplus_{\mathrm{arrows}}B^{a}_{ij}.
\end{equation}
where $i$ indexes the nodes, $a$ indexes the arrows, and $B^a_{ij}$ indicates a bifundamental field charged under the gauge groups at nodes $i,j$, with the arrow $a$ pointing from node $i$ to node $j$. To look for BPS states we study the moduli space of supersymmetric ground states, $\mathcal{M}_\gamma$, on the Higgs branch of this quantum mechanics theory.  This moduli space can be given a completely holomorphic description as follows:
\begin{itemize}
\item $\mathcal{M}_{\gamma}$ is the space of solutions to the F-term equations, subject to an additional stability condition, modulo the action of the complexified gauge group $\prod_{i}Gl(n_{i},\mathbb{C})$.  The relevant stability condition is $\mathit{\Pi}$\emph{-stability} \cite{DFR1} which we introduce in the next subsection.
\end{itemize}

In the above, the F-term equations are fixed by some choice of superpotential, which can be any arbitrary gauge-invariant holomorphic function of the matter fields. Note that non-trivial choices of superpotential only exist for quivers with closed oriented loops. There is nothing a priori which fixes this superpotential; it is an independent datum of our construction which must be computed by alternative means.   Later in subsection \ref{super} we discuss general constraints on $\mathcal{W},$ and construct the superpotential explicitly for the theories we study.  In the remainder of this section we simply assume that $\mathcal{W}$ is given.  This superpotential yields the F-term equations of motion
\begin{equation}
\frac{\partial \mathcal{W}}{\partial B^{a}_{ij}}=0.
\end{equation}

If $\mathcal{M}_{\gamma}$ is non-empty then there exists a BPS particle in the spectrum with charge $\gamma$.  To determine how many such particles and their spin we examine the structure of the cohomology of $\mathcal{M}_{\gamma}$.  Specifically, since $\mathcal{M}_{\gamma}$ is the moduli space of a theory with four supercharges it is a K\"{a}hler manifold, and as such its cohomology automatically forms representations of Lefschetz $SU(2)$.  Each Lefschetz $SU(2)$ representation gives a supersymmetric BPS multiplet, with spacetime spin determined up to an overall shift by the Lefschetz spin:
\begin{equation}
\mathrm{Spin}=\mathrm{Lefschetz}\otimes \left[\mathbf{\frac{1}{2}}\oplus 2(\mathbf{0})\right]. \label{lef}
\end{equation}
In our applications to complete theories we will only find BPS states which are either hypermutliplets or vector multiplets and hence we need only a basic version of the above.  If $\mathcal{M}
_{\gamma}$ is a point, we obtain a hypermultiplet, while if  $\mathcal{M}_{\gamma}\cong \mathbb{P}^{1}$ we find a vector multiplet.

\subsection{Quiver Representation Theory}\label{repthy}
$\Pi$-stability, the stability condition relevant to our description of the moduli space, is defined in the language of quiver representation theory.  A \emph{representation} $R$ of a quiver $Q$ consists of a complex vector space $\mathbb{C}^{n_{i}}$ for each node $i$ and a linear map $B^{a}_{ij}: \mathbb{C}^{n_{i}}\longrightarrow \mathbb{C}^{n_{j}}$ for each arrow $a$ from node $i$ to node $j$. This is simply a choice of expectation values of matter fields in the corresponding quiver quantum mechanics. A \emph{subrepresentation} $S$ of the representation $R$ is a choice of vector subspaces $\mathbb{C}^{m_{i}}\subset \mathbb{C}^{n_{i}}$ for each node, and maps, $b^{a}_{ij}: \mathbb{C}^{m_{i}}\longrightarrow \mathbb{C}^{m_{j}}$ for each arrow such that all diagrams of the following form commute.
\begin{equation}
\xymatrix{
\mathbb{C}^{n_{i}}  \ar[r]^{B^{a}_{ij}} & \mathbb{C}^{n_{j}} \\
\mathbb{C}^{m_{i}}\ar[r]^{b^{a}_{ij}} \ar[u]& \mathbb{C}^{m_{j}} \ar[u]}
\end{equation}

Let $\gamma_R=\sum_j \gamma_j n_j$ denote the charge of a rep $R$. The appropriate stability condition is as follows.
$R$ is called stable if for all subrepresentations $S$ other than $R$ and zero, one has
\begin{equation}
\arg\mathcal{Z}_{u}(S)<\arg\mathcal{Z}_{u}(R). \label{stable}
\end{equation}
This is the notion of $\Pi$-stability, introduced in \cite{DFR1}. Intuitively, we think of the subrep $S$ as a parton state of charge $\gamma_S$ which is bound to some other states to form the BPS particle of charge $\gamma_R$. $\Pi$-stability can be heuristically thought of as the condition that forbids the particle $\gamma_R$ from decaying to produce $\gamma_S$.

We can now explicitly define the moduli space $\mathcal{M}_{\gamma}$ as the set of stable quiver representations modulo the action of the complexified gauge group:
\begin{equation}
\mathcal{M}_{\gamma}=\left \{\phantom{\int}\hspace{-.2in}R =\{B^{a}_{ij}:\mathbb{C}^{n_{i}}\rightarrow \mathbb{C}^{n_{j}} \} \left |\phantom{\int} \hspace{-.15in} \frac{\partial \mathcal{W}}{\partial B^{a}_{ij}}=0, \  R \ \mathrm{is}\ \Pi-\mathrm{stable} \right. \right\}/\prod_{i}Gl(n_{i},\mathbb{C}).
\end{equation}
This is a completely holomorphic description of $\mathcal{M}_{\gamma}$, and in many examples it is explicitly computable.

As a very elementary application, we note that the nodes of a quiver are always $\Pi$-stable reps. That is, consider $\gamma_j$ as the representation given by choosing $n_i=\delta_{ij}$. This is always stable since it has no non-trivial subrepresentations, and thus in particular no destabilizing subreps. Furthermore, since there is only one non-zero vector space, all maps in the representation must be chosen zero; thus the moduli space $\mathcal{M}_{\gamma_j}$ is given by a single point. We find that each node of a quiver gives a multiplicity one hypermultiplet BPS state.

The phenomenon of BPS wall-crossing can be seen to arise from the stability condition above. As we move around in moduli space, the central charge function $\mathcal{Z}_u$ changes. On some real codimension 1 subspaces, central charges of various states will become aligned; consequently, we may have a situation in which $\arg\mathcal{Z}_{u}(S)=\arg\mathcal{Z}_{u}(R),$ where $S$ is some subrep of a rep $R.$ This is a wall of marginal stability for the state $\gamma_R$. On one side of the wall, the stability condition is satisfied, and $\gamma_R$ is a stable BPS bound state of the theory; on the other side, the stability condition is not satisfied, so no bound state with charge $\gamma_R$ exists in that region of moduli space.

\subsection{Quiver Mutation}\label{MUT}
We now return to the choice of half-plane that was fixed arbitrarily in subsection \ref{quivs}. We wish to consider how the analysis changes if we choose a different half-plane. Begin with some chosen half-plane $\mathcal{H}$, with its unique basis of BPS states $\{\gamma_i\}$ and resulting quiver $Q$. Let $\gamma_1$ be the left-most of the basis states. Consider rotating the half-plane clockwise by $\theta,$ $\mathcal{H} \rightarrow \mathcal{H}_\theta=e^{-i\theta}\mathcal{H}$.

As we tune $\theta$ up from zero, nothing happens as long as all the $\{\gamma_i\}$ remain in the half-plane $\mathcal{H}_\theta.$ At $\theta=\pi-\mathrm{arg}(\mathcal{Z}_u\gamma_1)$, the left-most state $\gamma_1$ exits the half-plane $H_\theta$ on the left, and simultaneously the anti-particle state $-\gamma_1$ enters the half-plane on the right. Thus we must find a new basis of elementary BPS states, $\{\widetilde{\gamma_i}\}$ for this choice of half-plane. The operation that gives this new basis is known in the study of quivers as \emph{mutation}. The new basis is given as follows:
\begin{eqnarray}
\widetilde{\gamma_{1}} & = & -\gamma_{1}\\
\widetilde{\gamma_{j}}  & = &
  \begin{cases}
   \gamma_{j}+( \gamma_{j} \circ   \gamma_{1})\gamma_{1} & \text{if }  \gamma_{j} \circ   \gamma_{1} >0 \\
   \gamma_{j}      & \text{if }   \gamma_{j} \circ \gamma_{1} \le 0
  \end{cases}
\end{eqnarray}
This result is obtained by using quiver representation theory to compute the possible stable reps of the original quiver involving $\gamma_1$ and ensuring that they can still be generated by the new basis \cite{ACCERV}.

The change in basis has a simple description at the level of the quiver and the superpotential. Denote the new quiver $\widetilde{Q}$ and superpotential $\widetilde{\mathcal{W}}.$
\begin{enumerate}
\item The nodes of $\widetilde{Q}$ are in one-to-one correspondence with the nodes in $Q$.
\item The arrows of $\widetilde{Q}$, denoted $\widetilde{B}^{a}_{ij}$, are constructed from those of $Q$, denoted $B^{a}_{ij}$ as follows:
\begin{enumerate}
\item For each arrow $B^{a}_{ij}$ in $Q$ draw an arrow $\widetilde{B}^{a}_{ij}$ in $\widetilde{Q}$.
\item For each length two path of arrows passing through node $1$ in $Q$, draw a new arrow in $\widetilde{Q}$ connecting the initial and final node of the length two path
\begin{equation}
B^{a}_{i1}B^{b}_{1j}\longrightarrow \widetilde{B}^{c}_{ij}.
\end{equation}
\item Reverse the direction of all arrows in $\widetilde{Q}$ which have node 1 as one of their endpoints.
\begin{equation}
\widetilde{B}^{a}_{i1}\longrightarrow \widetilde{B}^{a}_{1i}; \hspace{.5in}\widetilde{B}^{a}_{1j}\longrightarrow \widetilde{B}^{a}_{j1}.
\end{equation}
\end{enumerate}
\item The superpotential $\widetilde{\mathcal{W}}$ of $\widetilde{Q}$ is constructed from the superpotential $\mathcal{W}$ of $Q$ as follows:
\begin{enumerate}
\item Write the same superpotential $\mathcal{W}$.
\item For each length two path considered in step 2(b) replace in $\mathcal{W}$ all occurrences of the product $B^{a}_{i1}B^{a}_{1j}$ with the new arrow $\widetilde{B}^{a}_{ij}$.
\item For each length two path considered in step 2(b) $B^{a}_{i1}B^{b}_{1j}$ there is now a new length three cycle in the quiver $\widetilde{Q}$ formed by the new arrow created in step 2(b) and the reversed arrows in step 2(c)
\begin{equation}
\widetilde{B}^{a}_{1i}\widetilde{B}^{c}_{ij}\widetilde{B}^{b}_{j1}.
\end{equation}
Add to the superpotential all such three cycles.
\end{enumerate}
\item[4.]  In general, the mutated quiver $\widetilde{Q}$ now has some two-cycles. For each two-cycle in $\widetilde{Q}$ for which a quadratic term appears in $\widetilde{W}$ delete the two associated arrows.
\item[5.] For each deleted arrow $\widetilde{B}^{a}_{ij}$ in step 4, solve the equation of motion
\begin{equation}
\frac{\partial{\widetilde{W}}}{\partial \widetilde{B}^{a}_{ij}}=0.
\end{equation}
Use the solution to eliminate $\widetilde{B}^{a}_{ij}$ from the potential.
\end{enumerate}
It is known that the representation theory of this new quiver $\widetilde{Q}$ is appropriately equivalent to that of the original one, $Q$ \cite{FZ,DWZ,FST,BD}.\footnote{When we refer to the representation theory of the quiver $Q$, we implicitly include the choice of superpotential $\mathcal{W}$ and the charges $\{\gamma_i\}$ labeling nodes.} We refer to the set of all quivers equivalent to each other up to mutation as a \emph{mutation class}.

As a simple example of this procedure we study the mutation of the quiver shown below on the left at node 1. On the right is the quiver formed after step 3, prior to integrating out and canceling the two-cycle.

\noindent\begin{minipage}{.45\textwidth}
  \begin{equation}
\xy
(-20,0)*+{1}*\cir<8pt>{}="a" ; (20,0)*+{2}*\cir<8pt>{}="b" ; (0,28.2)*+{3}*\cir<8pt>{}="c" ;
\ar @{->} "a"; "b"
\ar @{->} "b"; "c"
\ar @{->} "c"; "a"
 \endxy
  \nonumber
\end{equation}

\begin{eqnarray}
\mathcal{W} & = &B_{12}B_{23}B_{31} \nonumber
\end{eqnarray}
\end{minipage}
\begin{minipage}{.55\textwidth}
\begin{equation}
\xy
(-20,0)*+{1}*\cir<8pt>{}="a" ; (20,0)*+{2}*\cir<8pt>{}="b" ; (0,28.2)*+{3}*\cir<8pt>{}="c" ;
\ar @{->} "b"; "a"
\ar @{->} "b"; "c" <3pt>
\ar @{->} "c"; "b" <3pt>
\ar @{->} "a"; "c"
 \endxy
  \nonumber
\end{equation}

\begin{eqnarray}
\widetilde{\mathcal{W}} & = & \widetilde{B}_{32}\widetilde{B}_{23}+ \widetilde{B}_{32} \widetilde{B}_{21} \widetilde{B}_{13} \nonumber
\end{eqnarray}
\end{minipage}
\vspace{.1in}

\noindent Since the two-cycle has an associated quadratic term in the superpotential, we integrate out, producing a vanishing superpotential and a quiver of the following form.
\begin{equation}
\xy
(-40,0)*+{2}*\cir<8pt>{}="a" ; (0,0)*+{1}*\cir<8pt>{}="b" ; (40,0)*+{3}*\cir<8pt>{}="c" ;
\ar @{->} "a"; "b"
\ar @{->} "b"; "c"
 \endxy
  \nonumber
\end{equation}

As a general rule, the study of quivers is greatly complicated by the existence of pairs of opposite arrows whose associated fields cannot be integrated out from the superpotential.  For one thing, a quiver with two-cycles could not possibly arise from our construction, in which the signed number of arrows is given by the electric-magnetic product. Furthermore, the mutation rule becomes more subtle for quivers with two-cycles \cite{DWZ}. Fortunately, we will see in section \ref{super} that the natural superpotential associated to the theories studied here always allows two-cycles to be integrated out as above.

As we have emphasized here, it is most straightforward to imagine mutation as occurring at a fixed point in moduli space, when the choice of half-plane sweeps past a BPS state in the $\mathcal{Z}$-plane. From that point of view, mutation is a duality that provides different descriptions of the same physics. However, we may also consider fixing the half-plane and moving around in moduli space. Since the central charge function $\mathcal{Z}_u$ varies over the moduli space, at some point the central charge of one of the states of the elementary basis $\{\gamma_i\}$ may exit the chosen half-plane. Then as we track our quiver around moduli space, we would be forced to do a mutation. We emphasize that this \emph{does not} signify a wall of marginal stability; there will be no change in the BPS spectrum, since no BPS phases have become aligned. In this picture, there are many mutation-equivalent quivers which each cover some patch of moduli space of the theory. So more carefully, we should assign a mutation class of quivers to a theory, and some subset of this class to each point of moduli space.

\subsection{Mutation Method}\label{mutmeth}
The role of mutation in this analysis in fact provides an extremely useful and elegant method for computing $\Pi$-stable quiver reps, without directly going through the quiver representation theory \cite{ACCERV}. The approach is as follows. Fix a point of moduli space $u,$ a half-plane $\mathcal{H}$, and its resulting quiver $Q$. Again consider rotating $\mathcal{H}\rightarrow \mathcal{H}_\theta,$ where we now run $\theta$ from 0 to $2\pi$. As we let $\theta$ sweep out this entire range, each BPS state will exit the half-plane $\mathcal{H}_\theta$ one-by-one, in phase order. Prior to exiting, every state becomes the left-most BPS state in the half-plane; note that the left-most BPS state must always appear as a node in the quiver, since positive linear combinations of nodes cannot be to the left of every node in the $\mathcal{Z}$-plane. As this state exits, we mutate on the associated node, so that we can track the relevant quiver over the full span of half-planes. We keep a list of the charges $\gamma$ associated to the nodes on which we mutate. At the end of this process, we are left with a choice of half-plane $\mathcal{H}_{2\pi}$ equivalent to the original one $\mathcal{H}$. Thus, the quiver returns to its original form, $Q$.

If the point of moduli space we are studying contains only finitely many states, then all stable BPS particles and anti-partices will appear in our list of mutated charges. In fact, since all the states found using this method appear at some stage as nodes of the quiver, they all correspond to multiplicity one hypermultiplets, as was seen in the first application in subsection \ref{repthy}. In the presence of higher spin or higher multiplicity states, the behavior will be more complicated. By our representation theory analysis, it is impossible for a higher spin or higher multiplicity state to become a node. The resolution is that there must be infinitely many hypermultiplets whose phases accumulate to the phase of this exotic state. The tower of hypermultiplets protect us from seeing it appear as a node. In many cases it is still possible to deduce the full BPS spectrum using this approach, though there are some situations (for example, when infinitely many vectors are present) where this method becomes highly intractable.

The result of the mutation method is determined by the relative phase orderings of the central charges. If we restrict to complete theories, we are allowed to pick any ordering, since all central charges can be varied independently over parameter space. Thus, there is no need to explicitly find a corresponding point in parameter space.\footnote{Moreover, it turns out that any path through mutation space is realizable by some choice of central charges, as long as we mutate on all positive charges before mutating on any negative charges \cite{ACCERV}. Thus in practice we can simply look for sequences of mutations which compose to the identity, and it is guaranteed that there will be a corresponding chamber in parameter space.}

To illustrate the mutation method, we use it to compute a finite chamber of the conformal Argyres-Douglas theory $A_3$. Theories of this type will be further discussed in section \ref{finite}. The quiver for this theory is given by the associated Dynkin diagram,

\vspace{4mm}
  \centerline{\begin{xy}
(0,0) *+{\gamma_1}*\cir<8pt>{} ="1",
(30,0) *+{\gamma_2}*\cir<8pt>{} ="2",
(60,0) *+{\gamma_3}*\cir<8pt>{} ="3",
"3", {\ar"2"},
"2", {\ar"1"}
\end{xy}}
\vspace{4mm}

\noindent To apply the mutation method, we can simply choose an ordering of central charges, since we are working with a complete theory. Then we do mutations in decreasing phase order of the central charges. Suppose we choose an ordering with $\arg\mathcal{Z}_u(\gamma_2)>\arg\mathcal{Z}_u(\gamma_1+\gamma_2)>\arg\mathcal{Z}_u(\gamma_1)>\arg\mathcal{Z}_u(\gamma_3).$ As described above, we begin rotating the half-plane, and $\gamma_2$ exits the half-plane first, so we mutate on the corresponding node and record $\gamma_2$ as a BPS state. The sequence of mutations is shown below:

    \begin{minipage}[b]{0.5\linewidth}
    \centering
    \[
\begin{xy}
(0,0) *+{\gamma_1+\gamma_2} ="1",
(15,20) *+{-\gamma_2}="2",
(30,-0) *+{\gamma_3}="3",
"2", {\ar"3"},
"1", {\ar"2"},
"3", {\ar"1"}
\end{xy}
    \]
    \small (i) Mutated on $\gamma_2$
    \end{minipage}
    \hspace{0.25cm}
    \begin{minipage}[b]{0.5\linewidth}
    \centering
    \[
\begin{xy}
(0,0) *+{-\gamma_1-\gamma_2} ="1",
(15,20) *+{\gamma_1}="2",
(30,-0) *+{\gamma_3}="3",
"2", {\ar"1"},
"1", {\ar"3"}
\end{xy}
    \]
    \small (ii) Mutated on $\gamma_1+\gamma_2$
    \end{minipage}

    \vspace{4mm}

    \begin{minipage}[b]{0.5\linewidth}
    \centering
    \[
\begin{xy}
(0,0) *+{-\gamma_2} ="1",
(15,20) *+{-\gamma_1}="2",
(30,-0) *+{\gamma_3}="3",
"1", {\ar"2"},
"1", {\ar"3"}
\end{xy}
    \]
    \small (iii) Mutated on $\gamma_1$
    \end{minipage}
    \hspace{0.25cm}
    \begin{minipage}[b]{0.5\linewidth}
    \centering
    \[
\begin{xy}
(0,0) *+{-\gamma_2} ="1",
(15,20) *+{-\gamma_1}="2",
(30,-0) *+{-\gamma_3}="3",
"1", {\ar"2"},
"3", {\ar"1"}
\end{xy}
    \]
    \small (iv) Mutated on $\gamma_3$
    \end{minipage}
    \vspace{4mm}

\noindent At this point all particles have exited the half-plane, and the quiver consists entirely of anti-particles. If we continue rotating the half-plane, we will simply generate all the anti-particles to the particles that were found above. We find that the BPS spectrum is given by multiplicity one hypermultiplets $\{\gamma_2,\gamma_1+\gamma_2,\gamma_1,\gamma_3\},$ along with the associated anti-particles.

\section{BPS Quivers of Complete Theories}
\label{surfaces}
Having finished our review of the basics of quiver representations, we now turn to our primary interest of determining the BPS quivers, superpotentials, and spectra for complete theories.  In this section we focus on determining the BPS quiver for those complete theories that coincide with the rank two Gaiotto theories.\footnote{In fact, among such theories, BPS quivers exist only for theories given by a Riemann surface \emph{with} some punctures.  The case with no punctures describes an exactly conformal theory and its BPS states do not admit a simple description.}   By construction, all such theories are intrinsically determined by a Riemann surface $\mathcal{C}$ decorated by a number of marked points defined by the punctures.  By the conclusion of this analysis, we will see that the BPS quiver, together with its superpotential, is encoded combinatorially in a triangulation of this decorated surface.

We will construct these models using geometric engineering \cite{KLMVW, KKV, KMV, KaMV} in type IIB string theory on a non-compact Calaibi-Yau threefold.  The threefolds in question can be built up starting from a Riemann surface $\mathcal{C}$.  We start with a four complex-dimensional space described by a rank three complex vector bundle over $\mathcal{C}$.  Explicitly
\begin{equation}
K_{\mathcal{C}}\oplus K_{\mathcal{C}} \oplus K_{\mathcal{C}} \rightarrow \mathcal{C}, \label{bundle}
\end{equation}
where in the above $K_{\mathcal{C}}$ denotes the canonical line bundle of holomorphic one-forms on the Riemann surface $\mathcal{C}$.  In general the surface $\mathcal{C}$ is punctured at a finite number of points $p_{i} \in \mathcal{C}$ and thus is non-compact.

Next we select a particular holomorphic quadratic differential $\phi$ on $\mathcal{C}$.  As a quadratic differential, $\phi$ transforms under holomorphic changes of coordinates on $\mathcal{C}$ as follows
\begin{equation}
\phi'(x')=\phi(x)\left(\frac{dx}{dx'}\right)^{2}.
\end{equation}
To completely specify the problem, we must also fix the limiting behavior of $\phi$ at the ideal boundaries of $\mathcal{C}$, namely the punctures $p_{i}$.  Near each such puncture the quadratic differential is permitted to have a pole of finite order.  We fix the non-normalizable behavior of $\phi$ as a boundary condition and therefore impose that near $p_{i}$
\begin{equation}
\phi(x)\sim \frac{1}{x^{k_{i}+2}}dx^{2}+\mathrm{less \ singular \ terms}. \label{bc}
\end{equation}
The integer $k_{i} \geq0$ associated to each puncture is invariant under changes of coordinates. It is an important aspect of the construction, which we return to in section \ref{triangulation}.\footnote{The reason for the exclusion of the case $k_{i}=-1$ is that such fluctuations in $\phi$ are normalizable, and hence are not fixed as part of the boundary conditions.}

Given this data our Calabi-Yau threefold is then defined by introducing local coordinates $(u,v,y)$ on the fiber of the vector bundle \eqref{bundle} and solving the following equation
\begin{equation}
uv=y^{2}-\phi(x). \label{cydef}
\end{equation}
The associated holomorphic three-from $\Omega$ is given by
\begin{equation}
\Omega = \frac{du}{u}\wedge dy \wedge dx.
\end{equation}
It is then known that finite mass strings probing the singularity of this geometry engineer a 4d field theory with $\mathcal{N}=2$ supersymmetry.  The Seiberg-Witten curve $\Sigma$ of such a theory is given by a double cover of $\mathcal{C}$, and we obtain the Seiberg-Witten differential by integrating $\Omega$ over a non-trivial 2-cycle in the fiber.
\begin{equation}
\Sigma=\{(x,y)|y^{2}=\phi(x)\}; \hspace{.5in}\lambda=\int_{S^2(x)} \Omega = ydx=\sqrt{\phi}.
\end{equation}
By varying the quadratic differential we obtain a family of Seiberg-Witten curves, and in this way the Coulomb branch $\mathcal{U}$ of the theory is naturally identified with the space of quadratic differentials obeying the boundary conditions \eqref{bc}.

It is also known that many of the simplest interesting gauge theories can be geometrically engineered in this fashion.  For example taking $\mathcal{C}$ to be a sphere with two punctures $p_{i}$ both with $k_{i}=1$ constructs the pure $SU(2)$ theory.  In general the class of field theories constructed in this way yields asymptotically free or conformal theories with gauge groups given by a product of $SU(2)$'s, together with various scaling and decoupling limits of such field theories.  They are exactly the type IIB version of the rank two Gaiotto theories constructed using M-theory in \cite{Gaiotto}, and, as we have mentioned above, in that context $\mathcal{C}$ is referred to as the Gaiotto curve.

For our present purposes, the primary advantage of building an $\mathcal{N}=2$ quantum field theory in string theory is that the set of supersymmetric objects in string theory, the BPS branes, is known.  In our case we seek a brane whose physical interpretation in four-dimensions is a charged supersymmetric particle of finite mass.  Thus the worldvolume of the brane should be an extended timelike worldline in Minkowski space times a volume minimizing compact cycle in the Calabi-Yau \eqref{cydef}.  Since type IIB has only odd dimensional branes, the only possibility is that BPS states are described geometrically by Dirichlet three-branes wrapping special lagrangian three-cycles.

Thus we are reduced to a classical, if difficult, geometric problem of counting special lagrangians \cite{Joyce, SV}.  These are compact lagrangian three-manifolds $N$ on which the holomorphic three-form has a constant phase
\begin{equation}
\Omega|_{N}=e^{i\theta}|\Omega|. \label{slag}
\end{equation}
The central charge of such a brane is given by
\begin{equation}
\mathcal{Z}_u(N)=\int_N \Omega,
\end{equation}
and the phase $\theta$ in the above is identified with the argument of the central charge of the 4d particle defined by $N$
\begin{equation}
\theta=\arg\mathcal{Z}(N).
\end{equation}

\begin{figure}[here!]
  \centering

  \subfloat[$S^{3}$]{\label{fig:slag1}\includegraphics[width=0.4\textwidth, height=.2\textheight]{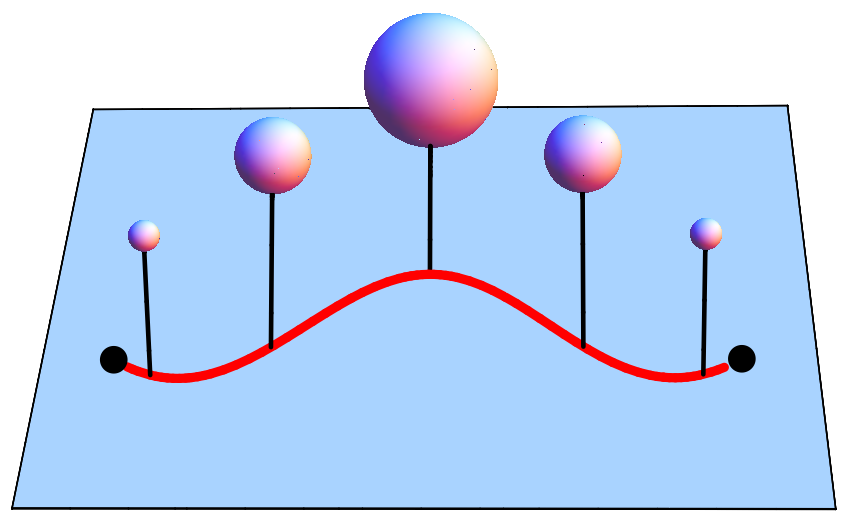}}
\hspace{.25in}
  \subfloat[$S^{1}\times S^{2}$]{\label{fig:slag2}\includegraphics[width=0.4\textwidth, height=.2\textheight]{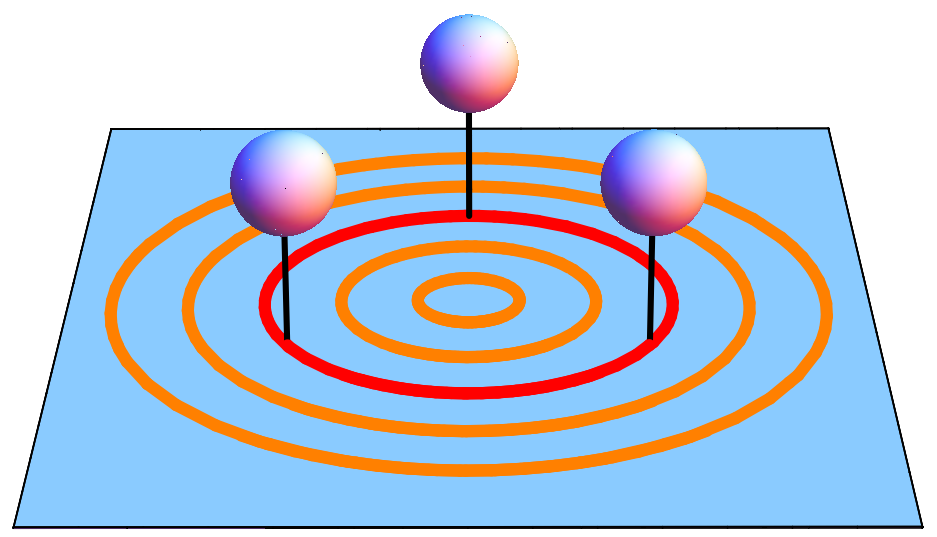}}
  \caption{Special-Lagrangian geometry in the Calabi-Yau.  The blue denotes a patch of the surface $\mathcal{C}$.  The red trajectory denotes the cycle $\eta$ and the $S^{2}$ fibers are indicated schematically above $\mathcal{C}$.  In (a) the topology of the cycle $\eta$ is an interval which terminates at two zeros of $\phi$.  The $S^{2}$ fibers shrink at these end points yielding a total space of an $S^{3}$.  In (b), the cycle $\eta$ has the topology of a circle, and the total space is $S^{1}\times S^{2}$.  Such special-lagrangians always come in one parameter families indicated in orange. }
  \label{fig:slag}
\end{figure}

Now one of the key observations of \cite{KLMVW} is that, in the geometries described by \eqref{cydef}, the counting of special lagragians can in fact be phrased entirely as a problem in $\mathcal{C}$.  To exhibit this feature we use the fact that all of our special lagrangians are embedded inside the vector bundle \eqref{bundle} and hence admit a natural projection to $\mathcal{C}$.  The image of this projection is a certain one cycle $\eta$ in $\mathcal{C}$ whose topology depends on the topology of $N$.  Each special lagrangian also wraps a non-trivial $S^2$ in the fiber, which shrinks to zero at the zeros of $\phi$. The possibilities in our examples are as follows, and are illustrated in Figure \ref{fig:slag}:
\begin{itemize}
\item $N\cong S^{3}$.  Such special-lagrangians are discrete.  Their quantization yields hypermultiplets in 4d.  When this three-sphere is projected to $\mathcal{C}$ we obtain an interval $\eta$ stretching between two zeros of the quadratic differential $\phi$.
\item $N\cong S^{1}\times S^{2}$.  This class of special-lagrangians always come in one-parameter families.  Their quantization yields a vector multiplet in 4d.  The projection of any such $S^{1}\times S^{2}$ to $\mathcal{C}$ is a closed loop $\eta$.
\end{itemize}

The shape of $\eta$ in $\mathcal{C}$ is constrained by the special Lagrangian condition \eqref{slag} on $N$.  Explicitly if we let $t \in \mathbb{R}$ parametrize $\eta$ then the condition of constant phase $\Omega$ reduces to
\begin{equation}
\sqrt{\phi}|_{\eta}=e^{i\theta}dt. \label{flow}
\end{equation}
The ambiguity in choosing the square root appearing in the above reflects the physical fact that for every BPS particle there is also an associated BPS antiparticle of opposite charge.  Choosing the opposite sign for the square root then sends $\theta\rightarrow \theta+\pi$, i.e. it replaces a BPS particle by its antiparticle.

We have now arrived at an elegant statement of the problem of calculating BPS states in this class of quantum field theories.  Our goal, however, is not directly to use this structure to compute the BPS states, but rather to extract the BPS quiver of this theory. In the following we will explain a natural way to extract such a quiver from a global analysis of the flow equations \eqref{flow}.

\subsection{Triangulations from Special-Lagrangian Flows}
\label{triangulation}
Our goal in this section will be to encode certain topological and combinatorial data about the special lagrangian flow in terms of a triangulation of the surface $\mathcal{C}$. Our basic strategy  will be to analyze the local and asymptotic properties of the flow on $\mathcal{C}$ defined by \eqref{flow}.  This is a problem which is well-studied in mathematics \cite{Strebel} and has recieved much attention in the present physical context \cite{SV,GMN08,GMN09,GMN10,GMN11}. We will confine ourselves to a brief self-contained review.  Since a quiver is constructed from hypermultiplets, our focus will be on the trajectories of this flow which interpolate between the zeros of $\phi$. Thus a special role will be played by these trajectories.

\begin{figure}[here!]
  \centering

  \includegraphics[width=0.5\textwidth,height=.25\textheight]{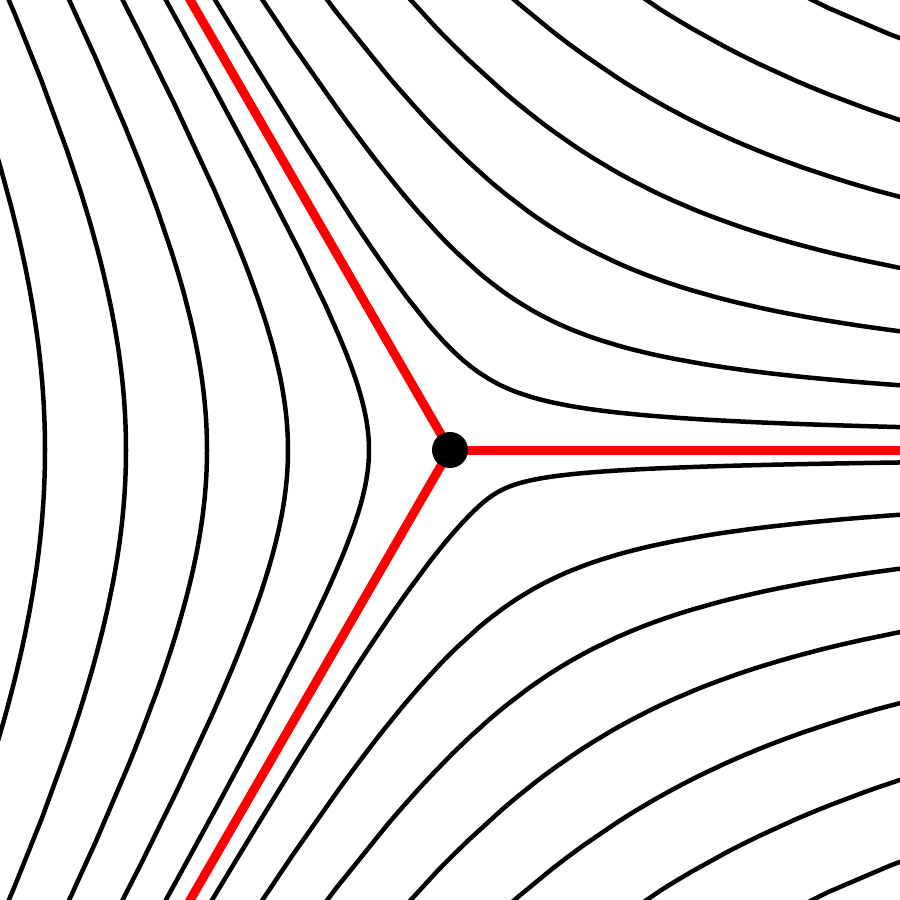}
  \caption{The local structure of the flow near a zero of $\phi$ shown as a black dot at the center of the diagram.  The red trajectories are the three flow lines which pass through the zero.  The black trajectories denote other generic flow lines.}
    \label{fig:trifurcate}
\end{figure}

To begin, we investigate the local nature of the flow near each zero.  We assume that this is a simple zero so that, in some holomorphic coordinate $w(x)$ centered at the zero of $\phi$, the flow equation \eqref{flow} takes the local form
\begin{equation}
\sqrt{w}dw=e^{i\theta}dt \Longrightarrow w(t)=\left(\frac 3 2 e^{i\theta} t+w_{0}^{3/2}\right)^{2/3}.
\end{equation}
Because of the three roots of the right-hand-side of the above, each zero has three trajectories emanating from it.  These trajectories make angles of $2\pi/3$ with each other and separate a local neighborhood centered on them into three distinct families of flow lines, as illustrated in Figure \ref{fig:trifurcate}.

Aside from the zeros, which can serve as endpoints for BPS trajectories, the other distinguished points for the flow are the punctures of $\mathcal{C}$.  Since the punctures form ideal boundaries of $\mathcal{C},$ they should be thought of as lying at strictly infinite distance.  Thus the behavior of the flow equation near these points governs the asymptotic properties of trajectories at very late and early times.  In a local neighborhood centered on the puncture $p_{i}\in \mathcal{C},$ the flow equation is asymptotically given by
\begin{equation}
\frac{dw}{w^{1+k_{i}/2}}=e^{i\theta}dt. \label{asymlate}
\end{equation}
We split our analysis of the solutions into two cases depending on the order $k_{i}+2$ of the pole in $\phi$ at the puncture:
\begin{itemize}
\item \emph{Regular Punctures}: $k_{i}=0$

The regular punctures in $\mathcal{C}$ are naturally associated to flavor symmetries and hence mass parameters of the engineered field theory \cite{Gaiotto}.  In our analysis this manifests itself in the following way: the residue of the pole in the flow equation is a coordinate invariant complex parameter that is part of the boundary data of the geometry.  Restoring this parameter to the asymptotic flow equation we then have.
\begin{equation}
m\frac{dw}{w}=e^{i\theta}dt. \label{logflow}
\end{equation}
The parameter $m$ is the residue of a first order pole in the Seiberg-Witten differential and can be interpreted as a bare mass parameter.

We deduce the behavior of the late time trajectories by integrating \eqref{logflow}.  The solution with initial condition $w_{o}$ takes the form
\begin{equation}
w(t)=w_{o}\exp\left(m^{-1}e^{i\theta}t\right). \label{logsol}
\end{equation}
Assume that the BPS angle $\theta$ has been chosen so that $m^{-1}e^{i\theta}$ is not purely imaginary.  Then the solution \eqref{logsol} is a logarithmic spiral.  Asymptotically all trajectories spiral in towards the puncture as illustrated in Figure \ref{fig:spiral}.
\begin{figure}[here!]
  \centering
  \includegraphics[width=0.5\textwidth,height=.25\textheight]{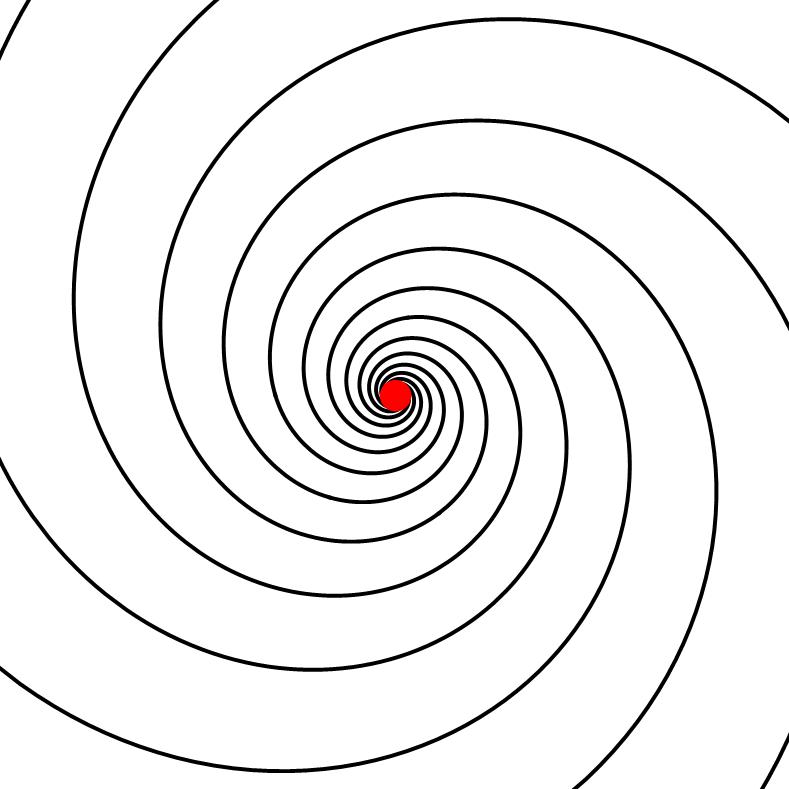}
  \caption{The local flow near a regular puncture indicated in red.  The flow lines are spirals terminating at the puncture.}
    \label{fig:spiral}
\end{figure}
\item \emph{Irregular Punctures}: $k_{i}>0$

In the case of irregular punctures, we find power law behavior for the asymptotic trajectories upon integrating (\ref{asymlate}):
\begin{equation}
w(t)=\left(\frac{-2e^{i\theta}}{k_i}t+\frac{1}{w_{o}^{k_{i}/2}}\right)^{-2/k_{i}}.
\end{equation}
A key feature of this solution is that it exhibits Stokes phenomena.  For large $|t|$ the trajectories converge to the origin $w=0$ along $k_{i}$ distinct trajectories.  We account for this behavior of the flows by cutting out a small disk in the surface $\mathcal{C}$ centered on the origin in the $w$ plane.  In terms of the metric structure of $\mathcal{C}$ this hole is to be considered of strictly infinitesimal size.  The modified surface now has a new ideal boundary $S^{1}$, and the $k_{i}$ limiting rays of the flows are replaced by $k_{i}$ marked points on this boundary.  This procedure is illustrated in Figure \ref{fig:bound}.
\begin{figure}[here!]
  \centering

  \subfloat[]{\label{fig:bound1}\includegraphics[width=0.4\textwidth]{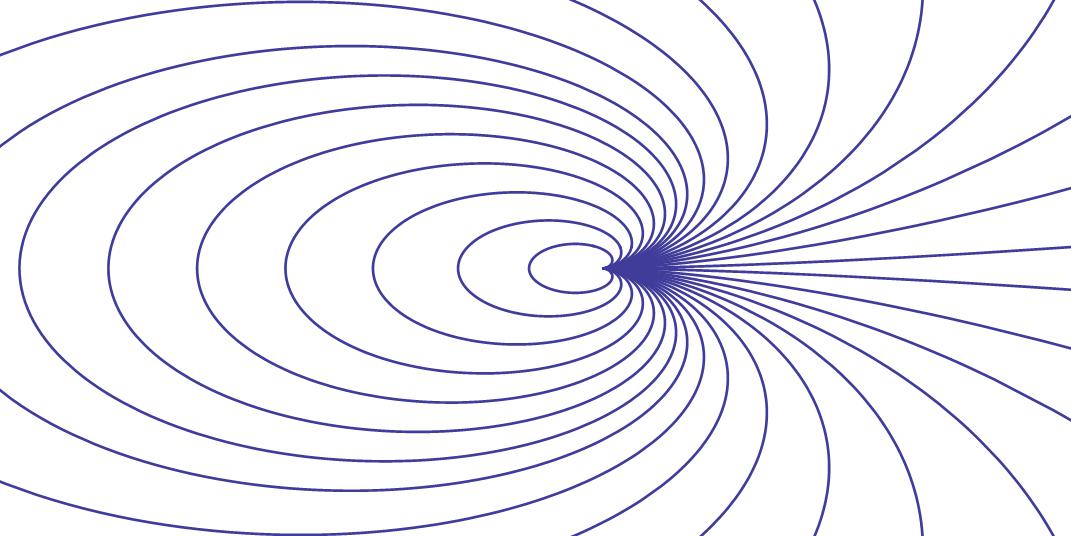}}
\hspace{.25in}
  \subfloat[]{\label{fig:bound2}\includegraphics[width=0.4\textwidth]{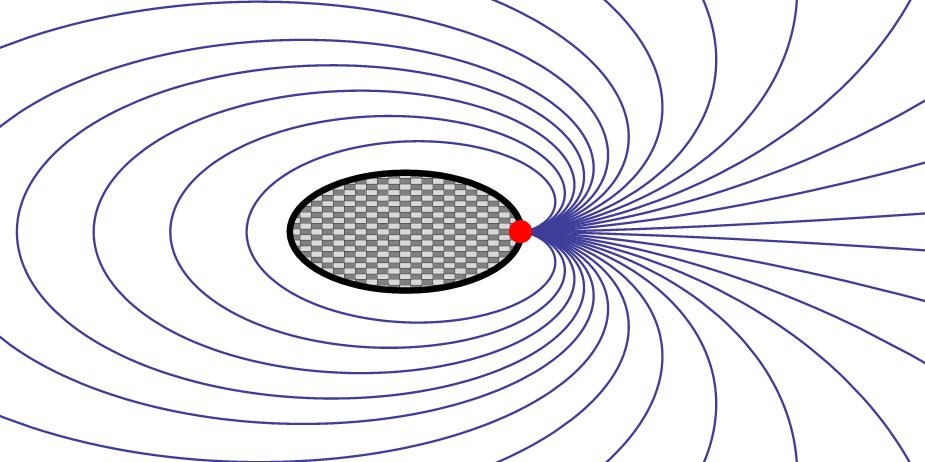}}
  \caption{Asymptotic flows near an irregular puncture with $k=1$.  In (a) the flow lines converge along a single ray, the rightward horizontal direction.  In (b), the surface $\mathcal{C}$ is modified by cutting out the small gray checkered region.  This surface now has a boundary, depicted by the black curve.  On the modified surface with boundary, generic flows terminate at a point, indicated in red, on the boundary. }
  \label{fig:bound}
\end{figure}

For each puncture $p_{i}$ with $k_{i}>0$ we perform the operation described above.  At the conclusion of this procedure our modified surface $\mathcal{C}$ now has an ideal boundary component $S^{1}_{i}$ for each irregular puncture $p_{i}$ and further each $S^{1}_{i}$ is decorated with $k_{i}$ marked points.  From now on, when discussing flows with irregular punctures, the symbol $\mathcal{C}$ shall mean this modified surface, equipped with  boundary components containing marked points for each irregular puncture.
\end{itemize}
Armed with the above, it is easy to deduce the global structure of the flow diagram on $\mathcal{C}$, that is, the global picture of the solutions to
\begin{equation}
\sqrt{\phi}=e^{i\theta}dt.
\end{equation}
We first choose the BPS angle $\theta$ \emph{generically}.  This means that there are no BPS trajectories in the flow, and hence no finite length trajectories connecting zeros of $\phi$ as well as no closed circular trajectories.  There are then two types of flow lines:
\begin{itemize}
\item \emph{Separating Trajectories}

These are flow lines which have one endpoint at a zero of $\phi$ and one endpoint at a regular puncture or marked point on the boundary of $\mathcal{C}$.  Separating trajectories are discrete and finite in number.

\item \emph{Generic Trajectories}

These are flow lines which have both endpoints at either regular punctures or marked points on the boundary.  Generic trajectories always come in one parameter families.
\end{itemize}

\begin{figure}[here!]
  \centering

  \subfloat[Flow Diagram]{\label{fig:flowcart1}\includegraphics[width=0.4\textwidth]{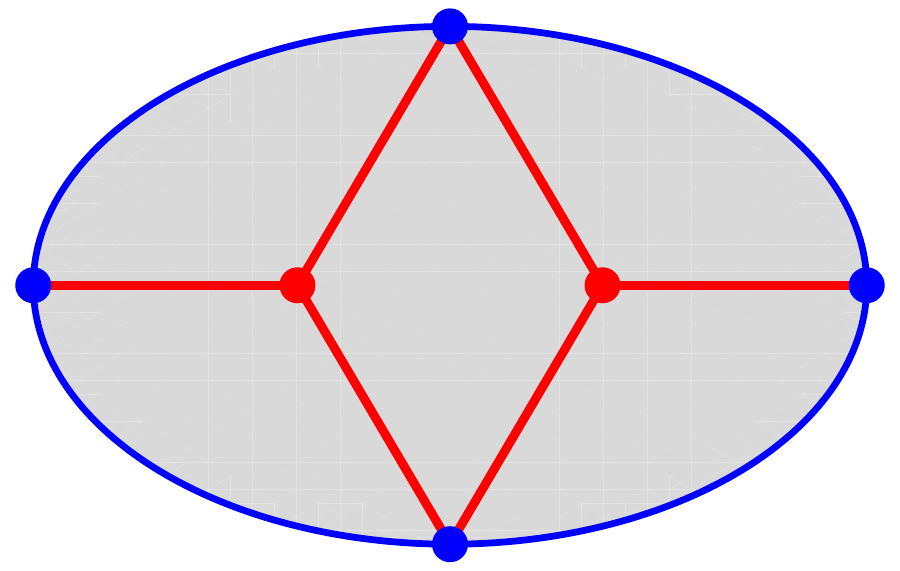}}
\hspace{.25in}
  \subfloat[Triangulation]{\label{fig:flowcart2}\includegraphics[width=0.4\textwidth,height=.19\textheight]{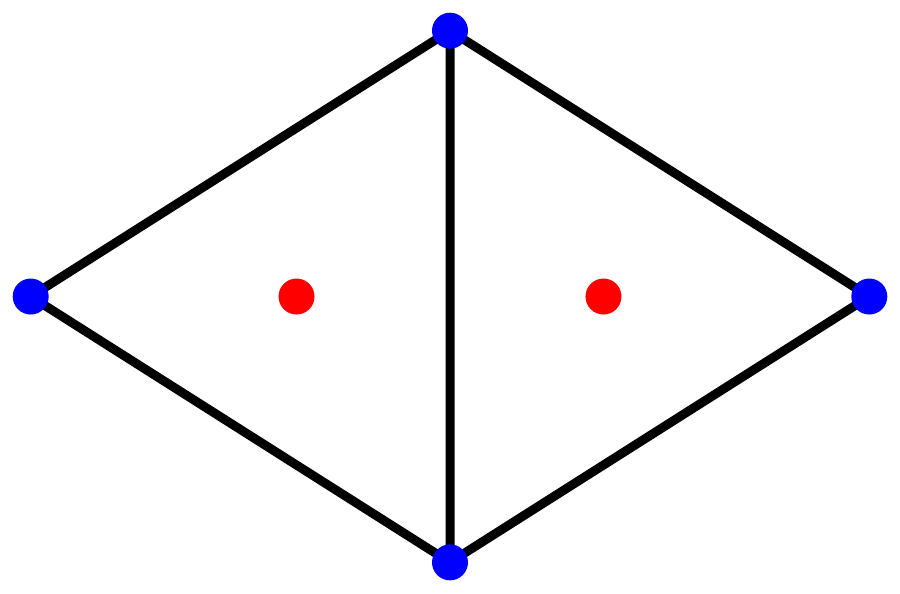}}
  \caption{An example flow diagram and its associated triangulation.  In (a) we have a global flow diagram on a disc with four marked points on the boundary.  The red dots are the zeros of $\phi$ and the associated separating trajectories are the red lines.  The gray cells denote one parameter families of generic flows.  All flow lines end on the four marked blue dots on the boundary.  In (b) we have extracted the associated triangulation.  Each black line is a generic flow line selected from each one parameter family.  The resulting triangles each contain one zero of $\phi$ by construction.}
  \label{fig:flowcart}
\end{figure}

A useful way to encode the topological structure of these flow diagrams is the following.  We consider our surface $\mathcal{C}$ with boundary.  It has marked points in the interior for each regular puncture, and marked points on the boundary given by the order of the pole of $\phi$ at the associated irregular puncture.  Then, for each one parameter family of generic trajectories, we choose exactly one representative trajectory and draw an arc on $\mathcal{C}$ connecting the indicated marked points.  An example is indicated in Figure \ref{fig:flowcart2}.  This procedure produces an \emph{ideal triangulation} of $\mathcal{C}$ where each diagonal of the triangulation terminates at two marked points.  Further, by construction, each triangle contains exactly one zero of $\phi$. Generally it is possible for the flow to produce an ideal triangulation with \emph{self-folded} triangles; these result in some technical complications which we address in appendix \ref{selffold}. 

In summary, for a fixed quadratic differential $\phi$ and generic angle $\theta$, we have produced an ideal triangulation of $\mathcal{C}$ by studying trajectories of
\begin{equation}
\sqrt{\phi}=e^{i\theta}dt.
\end{equation}
The combinatorial structure of this triangulation encodes properties of the flow, and we will see in the remainder of this section how to directly extract a BPS quiver and superpotential from this triangulation. Throughout the discussion it will be important to inquire how the triangulation varies as the data $(\phi,\theta)$ varies.  The quadratic differential $\phi$ labels a point in the Coulomb branch of the gauge theories in question, and thus it is natural to fix this data and study the BPS spectrum at fixed point in moduli space.  By contrast, the angle $\theta$ is completely arbitrary.  Any generic angle $\theta$ can be used, and different angles will produce distinct triangulations.  Demanding that ultimately our results are independent of $\theta$ will give a powerful constraint in the upcoming analysis.

\subsection{BPS Quivers from Ideal Triangulations}
\label{triangles}
We have now arrived at the structure of an ideal triangulation on the surface $\mathcal{C}$.  From this data there is a simple algorithmic way to extract a quiver \cite{FST}.  As a preliminary definition, we refer to an edge in the triangulation as a diagonal, $\delta$, if the edge does not lie on a boundary of $\mathcal{C}$.   Then proceed as follows:
\begin{itemize}
\item  For each diagonal $\delta$ in the triangulation, draw exactly one node of the quiver.
\item For each pair of diagonals $\delta_{1}, \delta_{2}$ find all triangles for which the specified diagonals are both edges.  For each such triangle, draw one arrow connecting the nodes defined by $\delta_{1}$ and $\delta_{2}$.  Determine the direction of the arrow by looking at the triangle shared by $\delta_{1}$ and $\delta_{2}$.  If $\delta_{1}$ immediately precedes $\delta_{2}$ going counter-clockwise around the triangle, the arrow points from $\delta_{1}$ to $\delta_{2}$.
\end{itemize}
In \cite{CV11} many aspects of these quivers were explored and it was argued that these are exactly the BPS quivers of the associated quantum field theories.  We now provide a full explanation of this proposal.

\begin{figure}[here!]
  \centering

  \subfloat[$\theta<\theta_{critical}$]{\label{fig:precat}\includegraphics[width=0.3\textwidth]{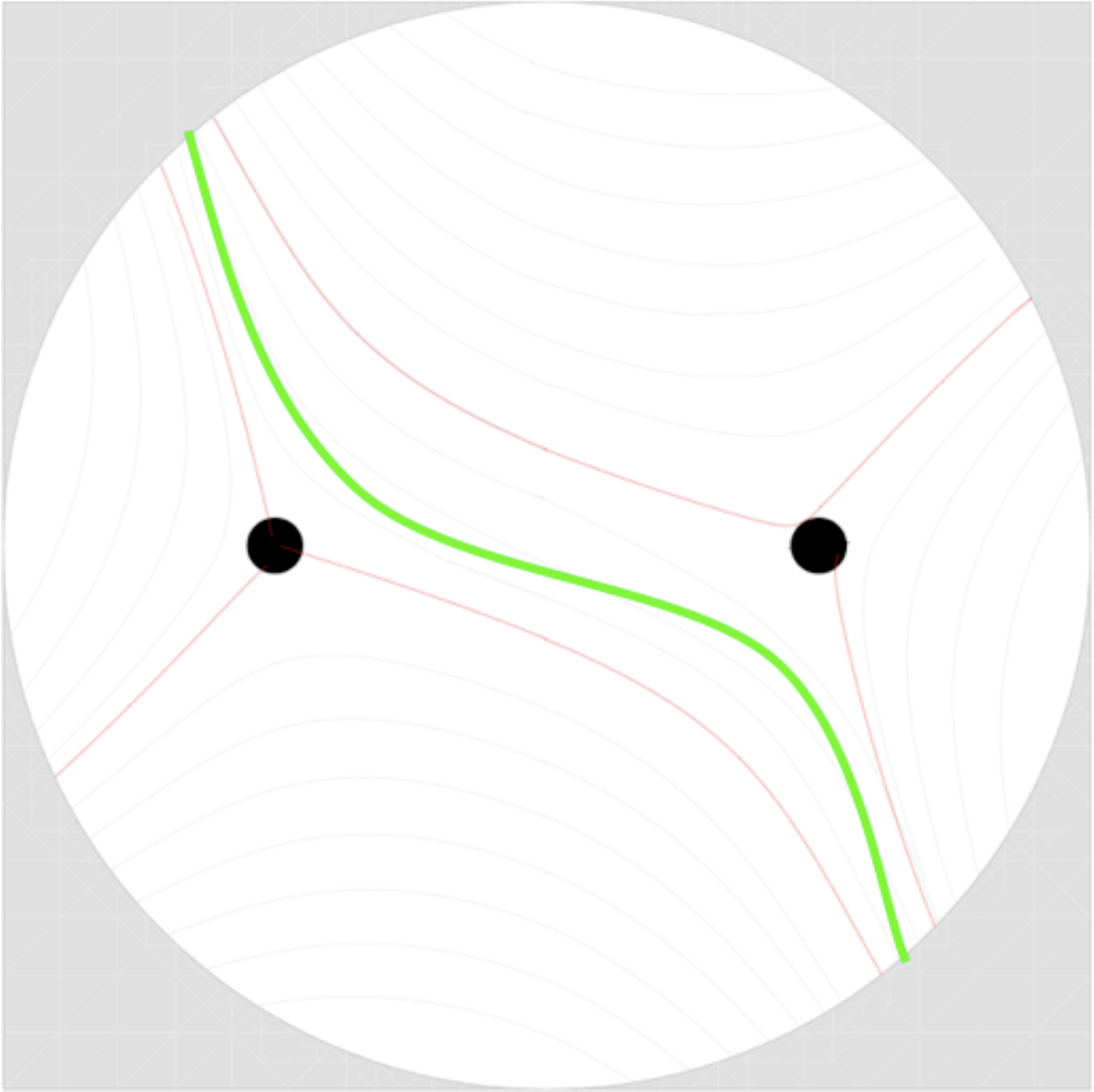}}
 \hspace{.2in}
  \subfloat[BPS State $\theta=\theta_{critical}$]{\label{fig:a1catas}\includegraphics[width=0.3\textwidth]{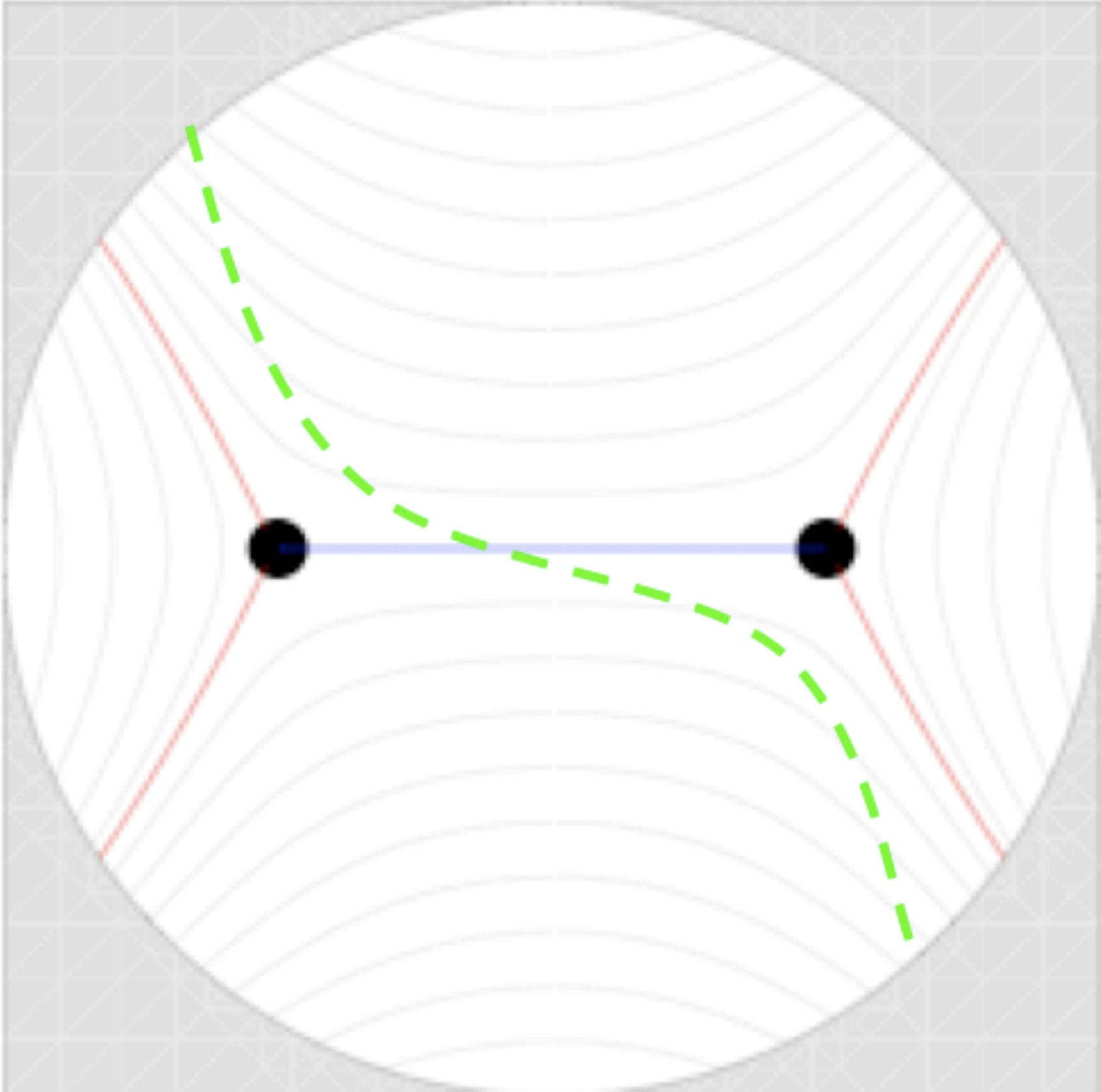}}
  \hspace{.2in}
  \subfloat[$\theta>\theta_{critical}$]{\label{fig:postcat}\reflectbox{\includegraphics[width=0.3\textwidth]{Figures/precata1.pdf}}}
  \caption{Evolution of the special lagrangian flows with the BPS angle $\theta$.  In each picture the black dots indicate the branch points of the cover where flows emerge.  Red trajectories are flows that emerge from the branch points and terminate on the boundary at $|x|=\infty$, while gray trajectories indicate generic flow lines.  The green trajectory denotes a representative of a generic flow line which can serve as an edge in the triangulation.  In (b) the BPS angle of the flow aligns with the phase of the central charge and a new kind of trajectory, shown in blue, traverses between branch points.  Afterwards in (c) the green line has flipped.}
\label{catas}
\end{figure}

We first address the identification of the diagonals of the triangulation with the nodes of the quiver.  As we have previously explained, our triangulation is constructed at a fixed value of the central charge angle $\theta$ appearing in \eqref{flow}.  This angle has been chosen such that no BPS states have a central charge occupying this angle.  Now let us imagine rotating $\theta$.  Eventually we will reach a critical value $\theta_{c}$ where a BPS hypermultiplet occurs and the structure of the flow lines will jump discontinuously.  The key observation is that each triangle in the triangulation contains exactly one zero of $\phi$.  Then, since BPS hypermultiplets are trajectories which connect zeros of $\phi$, a BPS hypermultiplet trajectory must cross some number of diagonals in the triangulation to traverse from one zero to another.  A simple example of this is illustrated in Figure \ref{catas}(b).

What the above example illustrates is that each diagonal $\delta$ labels an obvious candidate BPS hypermultiplet trajectory, connecting the two zeros in the two triangles which have $\delta$ as a common boundary.  Further any hypermultiplet trajectory which crosses multiple diagonals can be viewed homologically as a sum of the elementary BPS trajectories which cross only one diagonal.  Therefore, diagonals should be nodes of the BPS quiver.

Next let us justify why arrows in the quiver should be described by triangles in the triangulation.  Each elementary hypermultiplet, corresponding to a diagonal in the triangulation, lifts to a three-sphere in the Calabi-Yau.  Since these three spheres form nodes of the quiver, the lattice generated by their homology classes is naturally identified with the charge lattice $\Gamma$ of the theory.  Further the symplectic pairing given by the electric magnetic inner-product is precisely the intersection pairing on these homology classes.  Thus for each intersection point of the three-spheres, we should put an arrow connecting the associated nodes.  On the other hand it is clear that this intersection number can be calculated by projecting the three-spheres to $\mathcal{C}$ and then simply counting the signed number of endpoints that the associated trajectories share.  Each shared endpoint is naturally associated to the triangle containing it; so the triangles correspond to arrows between nodes.

The result of this section is that, given a Riemann surface $\mathcal{C}$ defining a 4d, $\mathcal{N}=2$ quantum field theory, we have produced a natural candidate BPS quiver.  It is quite interesting to note that as a result of recent mathematical work \cite{FST}, these quivers are all of \emph{finite mutation type}.  In other words, repeated mutations of vertices produce only a finite number of distinct quiver topologies.  In fact this property is equivalent to the more physically understandable property of completeness\cite{CV11}. The set of finite mutation type quivers (or equivalently, the set of complete theories) consists precisely of the quivers associated to triangulated surfaces, as described above, along with a finite number of exceptional cases, discussed in section \ref{exceptional}\cite{FST08}.

We can give one strong consistency check on our proposal for the BPS quivers as follows.  Observe that, to a given Riemann surface theory $\mathcal{C}$ we have in fact produced not one quiver but many.  Indeed our quivers are constructed from the triangulation produced from a fixed value $\theta$ of the BPS angle where there are no BPS states.  So in fact our assignment is
\begin{equation}
(\mathcal{C},\theta)\longrightarrow Q_{\theta}=\mathrm{BPS \ Quiver}. \label{thetaindep}
\end{equation}
As the central charge phase $\theta$ varies over a small region, the flow evolves continuously and the incidence data of the triangulation encoded in $Q_{\theta}$ remains fixed.  However,
as $\theta$ varies past a BPS state, the flow lines and triangulation will jump discontinuously, as illustrated in the basic example of Figure \ref{catas}.  This results in a new quiver $Q_{\theta'},$ distinct from $Q_{\theta}$.  Both of these quivers $Q_{\theta}$ and $Q_{\theta'}$ are natural candidates for the BPS quiver of theory defined by $\mathcal{C}$, and hence we should expect that the quantum mechanics theories they define are equivalent. In other words consistency of our proposal demands that all quivers of the from $Q_{\theta}$ for any given $\theta$ are mutation equivalent.  Happily, a simple theorem \cite{FST} shows that this is indeed the case: the set of quivers obtained from triangulations of a given surface precisely forms a mutation class of quivers.

Actually, we can say more. If we tune $\theta$ from $0$ to $2\pi$, we will see that every BPS hypermultiplet corresponds to a jump of the triangulation, and gives a new choice of quiver. This approach to computing BPS spectra was studied in \cite{GMN09}.  As was described there, the discontinuous jump of triangulation, or \emph{flip}, at each BPS state $\gamma$ is given by simply removing the diagonal crossed by $\gamma$, and replacing it with the unique \emph{other} diagonal that gives an ideal triangulation.\footnote{To clarify, once we remove the diagonal of the appropriate BPS state, we are left with some quadrilateral in our `triangulation.' To produce a true triangulation, we may add one of the two possible diagonals that would cut the quadrilateral into a triangle. A flip is simply given by taking the choice that differs from the original triangulation.} As argued in \cite{FST}, at the level of the quiver, this flip corresponds precisely to a mutation at the associated node. Thus, if we forget about the surface $\mathcal{C}$ and triangulation, and instead focus on the quiver itself, we see that we are simply applying the mutation method to compute $\Pi$-stable representations! This seems to be a deep insight into how the naively unrelated problems of finding special lagrangians and computing $\Pi$-stable quiver representations are in fact equivalent. Recall, however, that the mutation method made no reference to completeness of the theory. While the triangulations and flips exist for some set of complete theories, the mutation method is more general, and can be applied any BPS quiver. In \cite{ACCERV} we explored applications of the mutation method to non-complete theories.

\begin{figure}[here!]
  \center

  \includegraphics[width=0.4\textwidth]{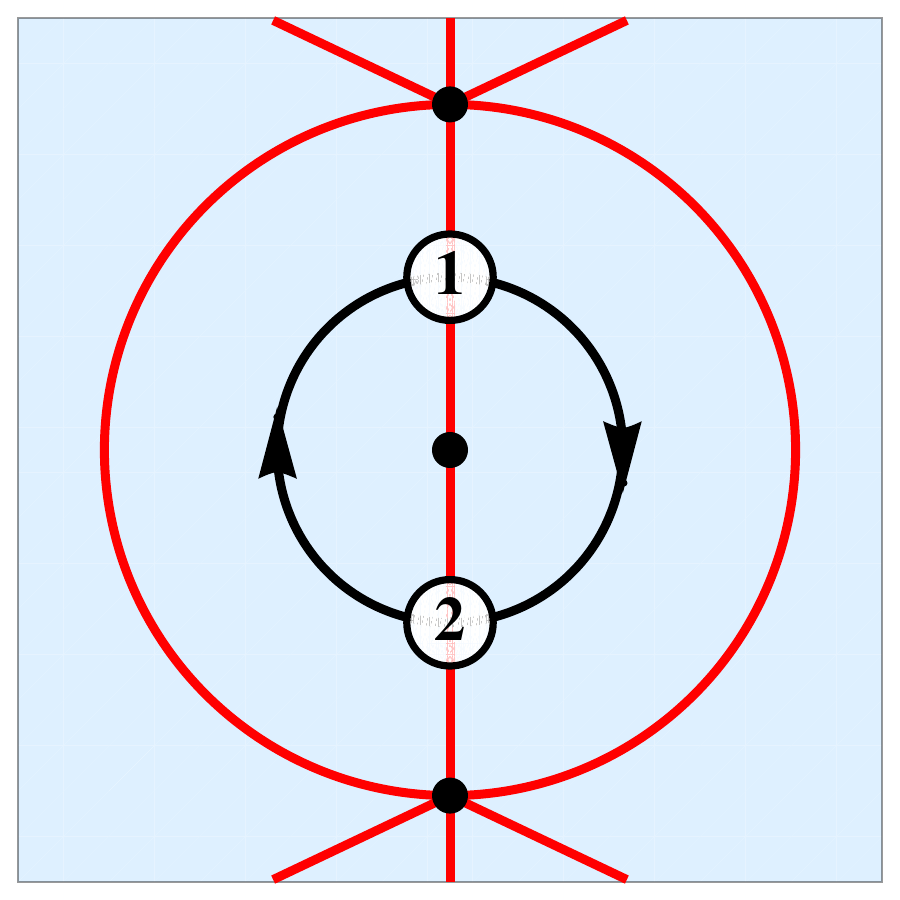}
    \caption{A bivalent puncture in the triangulation gives rise to a two-cycle in $Q$.  The blue denotes a patch of $\mathcal{C}$.  Red lines indicate diagonals and marked points are punctures.  The nodes of the quiver for the two indicated diagonals are drawn.  The bivalent puncture implies that there is a two cycle in the quiver indicated by the black arrows. }
  \label{fig:bivalent}
\end{figure}

In later sections of this paper we will see further evidence for this proposal by recovering the BPS quivers of well-known quantum field theories.  However, before reaching this point let us illustrate one important subtlety which we have glossed over in the above.  Consider the possible structure in an ideal triangulation of some Riemann surface $\mathcal{C}$, as illustrated in Figure \ref{fig:bivalent}.
According to the rules of this section, for each bivalent puncture in the triangulation we will obtain, as indicated, a cycle of length two in the quiver.  These are fields in the quiver theory which could, in principle, admit a gauge invariant mass term in the superpotential.  As mentioned in subsection \ref{repthy}, the quantum mechanics described by the quiver will be rather complicated, if no such mass term is generated.  In the next section we will argue that the natural potential for these theories does indeed generate all possible gauge invariant mass terms and therefore simplifies the resulting quivers considerably.
\subsection{The Superpotential}
\label{super}
The previous subsection identified a quiver associated to any ideal triangulation, and further suggested that this quiver is naturally the BPS quiver of the associated gauge theory.  In this subsection we will complete this picture by describing a natural superpotential for such a quiver, recently developed in the mathematics literature \cite{LF1,LF2,LF3}.  We will then argue on general grounds, essentially as a consequence of completeness, that this superpotential yields the necessary F-flatness conditions for the quiver quantum mechanics theory.

We will build up the superpotential starting from the elementary case of an acyclic quiver.  Since such a quiver has no cycles, there are simply no gauge invariant terms to be written and $\mathcal{W}=0$.

Next we consider an arbitrary quiver $Q$ which, by a sequence of mutations, is connected to an acyclic quiver.  Since $Q$ is the quiver of a complete theory, all of its central charges are free parameters that can be varied arbitrarily as one scans over parameter space.  It follows that the sequence of mutations connecting $Q$ to its dual acyclic form is in fact realizable by physical variation of parameters. Hence, following the mutation rules of section \ref{MUT}, the superpotential for the quiver $Q$ is completely fixed by the acyclic quiver with trivial potential.

\begin{figure}[here!]
  \centering
  \subfloat[Internal Triangle]{\label{fig:inttri}\includegraphics[width=0.4\textwidth, height=.4\textwidth]{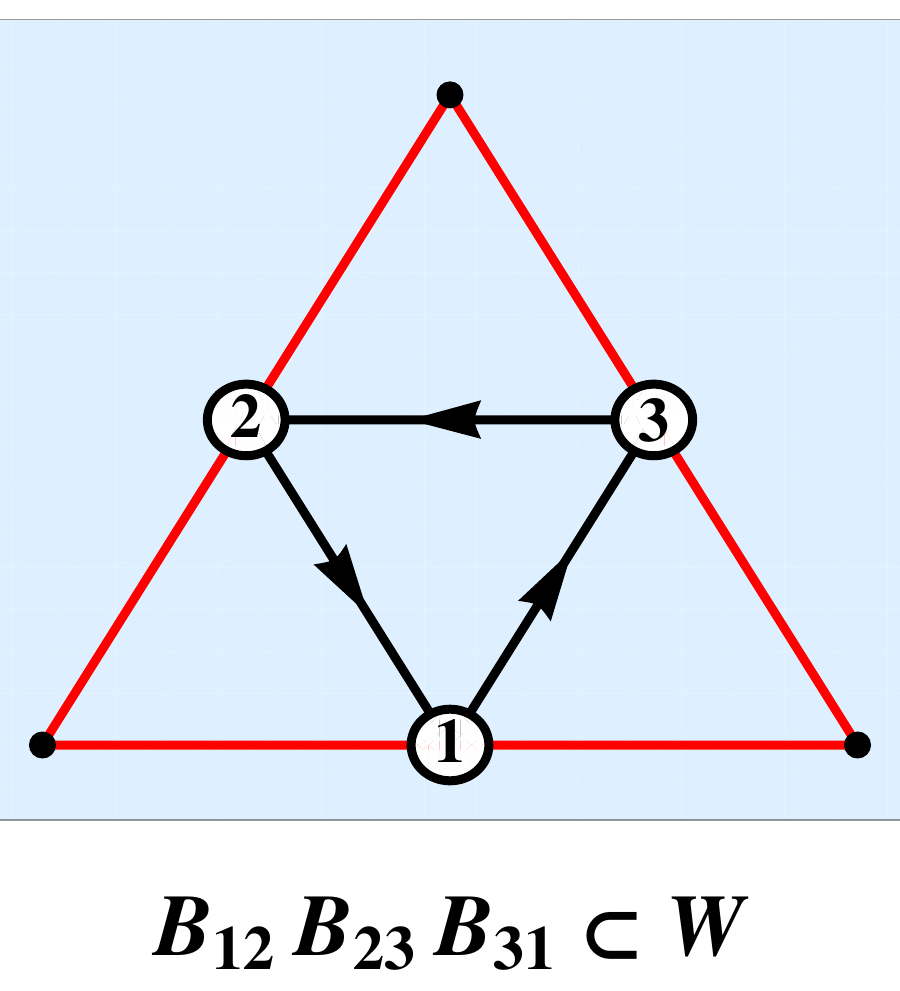}}
  \hspace{.2in}
  \subfloat[Internal Puncture]{\label{fig:intpunc}\includegraphics[width=0.4\textwidth, height=.4\textwidth]{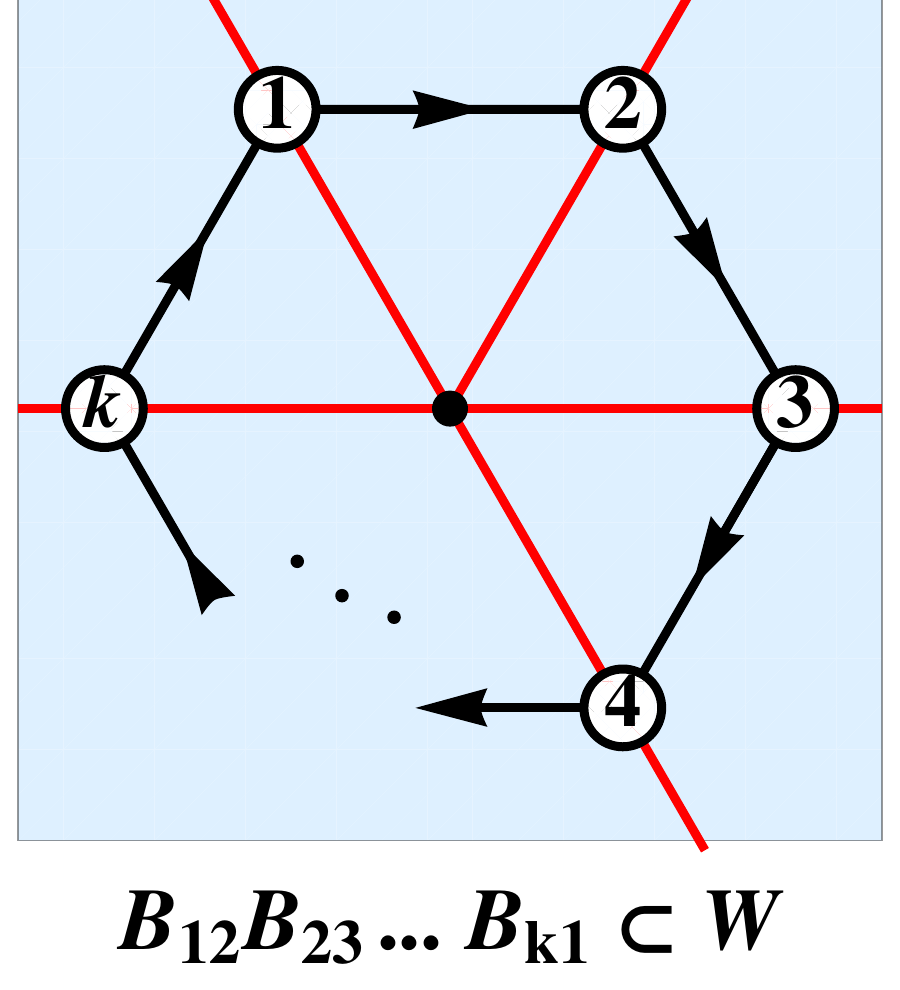}}
  \caption{The two distinct structures in the triangulation which contribute to the potential.  The blue region denotes a patch of $\mathcal{C}$, the red edges are diagonals in the triangulation.  These correspond to nodes of the quiver which we have indicated on the triangulation.  The black arrows connecting the nodes are the arrows in the quiver induced by the shared triangles shown in the diagram.  In (a) an internal triangle gives rise to a three-cycle in $\mathcal{W}$ in (b) an internal puncture of valence $k$ gives rise to a $k$-cycle in $\mathcal{W}$.}
\end{figure}

The argument of the previous paragraph shows that the $\mathcal{W}$ assigned to any such quiver $Q$ is completely fixed, however complicated the sequence of mutations leading from the acyclic form to $Q$ may be.  Surprisingly, there exists an elementary description of this superpotential in terms of the local incidence data of the triangulation of $\mathcal{C}$ which gives rise to $Q$.  This description has been developed in \cite{LF1}. For any quiver $Q$ mutation equivalent to an acyclic quiver, the superpotential $\mathcal{W}$ is computed as follows:
\begin{itemize}

\item Let $T$ denote a triangle in $\mathcal{C}$.  We say $T$ is \emph{internal} if all of its edges are formed by diagonals, that is none of the sides of $T$ are boundary edges in $\mathcal{C}$.  Then each edge of $T$ represents a node of the quiver and the presence of the internal triangle $T$ implies that these nodes are connected in the quiver in the shape of a three-cycle.  For each such triangle $T$ we add the associated three-cycle to $\mathcal{W}$. This situation is illustrated in Figure \ref{fig:inttri}.

\item Next let $p$ be an internal, regular puncture in $\mathcal{C}$.  Then some number $n$ of edges in the triangulation end at $p$.  Further since $p$ is an internal puncture which does not lie on the boundary of $\mathcal{C}$ it follows that each such edge terminating at $p$ is in fact a diagonal and hence a node of the quiver.  The $n$ distinct nodes are connected in an $n$-cycle in the quiver and we add this cycle to $\mathcal{W}.$  This situation is illustrated in Figure \ref{fig:intpunc}.
\end{itemize}

For quivers with multiple arrows between two given nodes, it is important to keep track of which triangle the arrow arises from when writing down the superpotential. The superpotential must be written with a fixed, consistent assignment of arrows to triangles; inconsistent choices are \emph{not} equivalent, and will generally give the wrong answer.

The observation that the superpotential can be determined in such an elementary way from the incidence data of the triangulation is striking.  It strongly suggests that $\mathcal{W}$ is a local object that can be determined patch by patch on $\mathcal{C}$.  Granting for the moment that this is so allows us to immediately generalize to any theory determined by an arbitrary Riemann surface $\mathcal{C}$.  We can simply extend the simple rules given above to all quivers.

One important consequence of this extension is that the it automatically ensures that all of our superpotentials will be compatible with mutation.  That is, just as in equation \eqref{thetaindep}, we have now constructed a map from a Riemann surface $\mathcal{C}$ and an angle $\theta$ to a quiver $Q$ and superpotential $\mathcal{W}$.  However the angle $\theta$ is arbitrary.  As $\theta$ rotates, in general the triangulation $\mathcal{T}$ of $\mathcal{C}$ will undergo a series of flips and arrive at a new triangulation $\widetilde{\mathcal{T}}$.  From this new triangulation we can determine the quiver $(\widetilde{Q},\widetilde{\mathcal{W}})$.  On the other hand we have previously noted that flips in the triangulation are the geometric manifestation of quiver mutation.  Thus we have two independent ways of determining the dual quiver and superpotential:
\begin{itemize}
\item Compute $(\widetilde{Q},\widetilde{\mathcal{W}})$ from $(Q,\mathcal{W})$ by performing a sequence of mutations.
\item Compute $(\widetilde{Q},\widetilde{W})$ from the new triangulation $\widetilde{\mathcal{T}}$
\end{itemize}
A necessary condition for a consistent superpotential is that the two computations yield the same answer.  In \cite{LF1} it was proved that this is the case.

The above argument shows that our proposal for the superpotential is consistent with the quiver dualities described by mutation.  However, it depends fundamentally on our locality hypothesis for the superpotential.
As we will now argue, using the completeness property of the field theories in question, we can give a strong consistency check on this assumption.

All of our arguments thus far involve constraints on $\mathcal{W}$ that arise from mutation.  As we mentioned in section \ref{MUT} mutations may be forced when, as we move around in moduli space, the central charges rotate out of the chosen half-plane.  Most importantly, all these rotations are physically realized, since in a complete theory all central charges are free parameters.

Of course the central charges of the theory come not just with phases but also with magnitudes.  In a complete theory we are also free to adjust these magnitudes arbitrarily.  Let us then consider the limit in parameter space where the magnitude of the central charge associated to a node $\delta$ becomes parametrically large compared to all other central charges
\begin{equation}
|\mathcal{Z}(\delta)|\longrightarrow \infty.
\end{equation}
In this limit, the BPS inequality implies that all particles carrying the charge $\delta$ become enormously massive and decouple from the rest of the spectrum.  At the level of the quiver $Q$ this decoupling operation is described as follows: simply delete from the quiver the node $\delta$ and all arrows which start or end at $\delta$.  This produces a new quiver $\widetilde{Q}$ with one node fewer than $Q$.  The superpotential for the resulting quiver theory $\widetilde{Q}$ is then determined simply by setting to zero all fields transforming under the gauge group indicated by $\delta$.

\begin{figure}[here!]
  \centering
  \subfloat[$\mathcal{C}$ pre-surgery]{\label{fig:surg1}\includegraphics[width=0.4\textwidth]{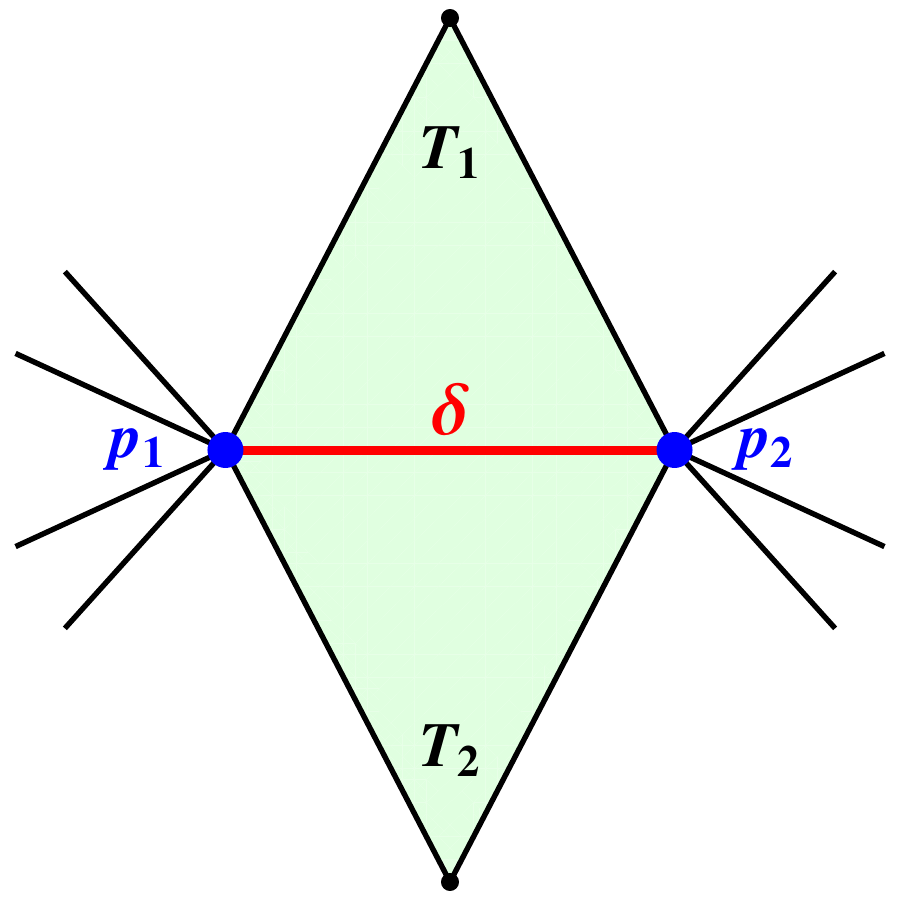}}
  \hspace{.2in}
  \subfloat[$\delta$ Decoupled]{\label{fig:surg2}\includegraphics[width=0.4\textwidth]{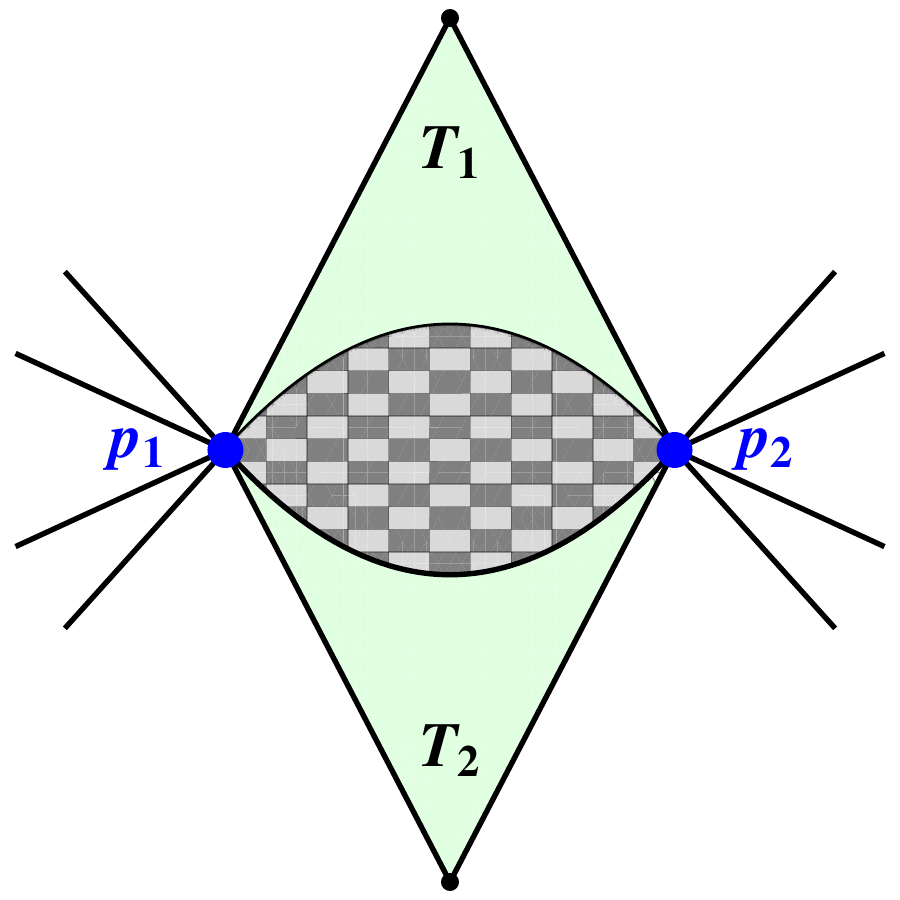}}
  \caption{The node decoupling surgery for a typical diagonal $\delta$.  In (a) we see a patch of $\mathcal{C}$ focused on the region involving a typical diagonal $\delta$.  In (b) $\delta$ has decoupled leaving a new a new Riemann surface $\tilde{\mathcal{C}}$ which differs from $\mathcal{C}$ by the addition of a new boundary component which encloses the checkered region and has two marked points $p_{i}$.}
  \label{fig:surg}
\end{figure}

 Following our interpretation of nodes of the quiver as diagonals in a triangulation, it is possible to describe this decoupling operation at the level of the Riemann surface $\mathcal{C}$ itself.  Consider the  diagram of Figure \ref{fig:surg1} which depicts the local region in $\mathcal{C}$ containing a diagonal $\delta$ traversing between two punctures or marked points $p_{i}$.  The decoupling operation to destroy the node $\delta$ is then realized by excising a small disc containing $\delta$ as a diameter and no other diagonals.  The result of this procedure is shown in Figure \ref{fig:surg2}.
It is clear from our construction of BPS quivers from triangulations that this decoupling operation produces a new surface $\widetilde{\mathcal{C}}$, whose BPS quiver is exactly $\widetilde{Q}$, the quiver with the node $\delta$ decoupled.  We may therefore determine the superpotential $\mathcal{W}$ for $\widetilde{Q}$ by applying the incidence rules described in this section to the new surface $\widetilde{\mathcal{C}}$.

In summary, we see that there are two distinct ways for computing the superpotential for the quiver $\widetilde{Q}$:
\begin{itemize}
\item Determine from $\mathcal{C}$ the superpotential for the quiver $Q$.  Then reduce to $\widetilde{Q}$ by deleting the node $\delta$.
\item Determine directly from the surface $\widetilde{\mathcal{C}}$ the superpotential for the quiver $\widetilde{Q}$.
\end{itemize}
Consistency of our proposal demands that the two methods give rise to the same superpotential.  It is easy to see directly that this is the case.  Indeed the effect of the surgery operation illustrated in Figure \ref{fig:surg} is to change the two triangles $T_{i}$ to external ones, and to change the points $p_{i}$ to marked points on the boundary.  Clearly this eliminates from the superpotential exactly those terms in which fields charged under the node $\delta$ appear.

By completeness, the decoupling limit argument can be applied to an arbitrary node in a BPS quiver and yields a strong consistency check on the locality hypothesis and thus our proposal for the superpotential.

Let us remark that the superpotential we have constructed naturally resolves the headache proposed at the end of section \ref{triangles}.  By construction, every two-cycle in a quiver arises from a bivalent puncture of the corresponding triangulation. For each bivalent puncture there is now a quadratic term in the superpotential that lifts the fields involved in the associated two-cycle. Thus we may integrate out and cancel all possible two-cycles to produce a two-acyclic quiver.

Finally, before turning to examples, we point out that it would be interesting to calculate this superpotential directly from a string theory construction. While several plausibility and consistency arguments have been given, a direct calculation may certainly lead to further insight.
\subsection{Examples from $SU(2)$ Gauge Theory}
\label{SQCD}
In this section we illustrate the rules developed above by cataloguing the BPS quivers, with their required superpotential, for simple theories given by a single $SU(2)$ gauge group with matter and asymptotically free or conformal coupling.  Of course each theory comes with a number of quivers related by mutations and we need only derive one.  Consistent with our previous discussion, for those examples involving irregular punctures, we will present triangulations of surfaces with boundary. In \cite{ACCERV}, the representation theory of these quivers was studied, and found to agree with the well known BPS spectra of the associated theories.

Before enumerating the examples, we take a moment to fix conventions.  Throughout, in all triangulations, red labeled lines denote diagonals, which appear as nodes of the quiver, while black lines denote boundary components.  Both regular punctures and marked points on the boundary are indicated by black dots.  Bifundamental fields corresponding to arrows in the quiver will be denoted by $X_{ij}$ and $Y_{ij}$ where $i$ and $j$ label the initial and final vertex of the arrow respectively.
\subsubsection{Asymptotically Free Theories}\label{asympex}
We first study quivers for $SU(2)$ theories with asymptotically free gauge coupling.

\begin{itemize}
\item$SU(2)$

This theory is constructed on an annulus with one marked point at each boundary.

\noindent\begin{minipage}{.45\textwidth}
  \includegraphics[width=\textwidth]{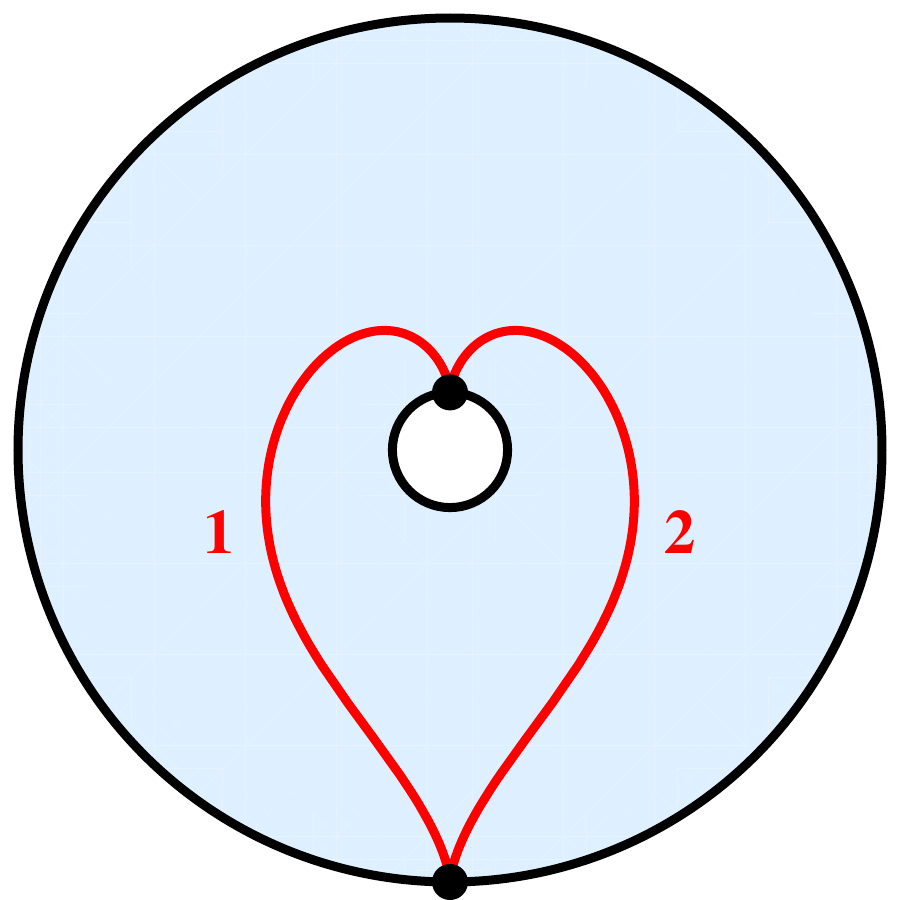}
\end{minipage}
\begin{minipage}{.55\textwidth}
\begin{equation}
\xy
(-20,0)*+{1}*\cir<8pt>{}="a" ; (20,0)*+{2}*\cir<8pt>{}="b"
\ar @{->} "a"; "b" <3pt>
\ar @{->} "a"; "b" <-3pt>
 \endxy
 \nonumber
\end{equation}

\begin{eqnarray}
\mathcal{W}& = &0.
\nonumber
\end{eqnarray}
\end{minipage}

Of course this is exactly the quiver for $SU(2)$ Yang-Mills.

\item$SU(2) \  N_{f}=1$

This theory is constructed on an annulus with one marked point on one boundary component, and two marked points on the remaining boundary component.

\noindent\begin{minipage}{.45\textwidth}
  \includegraphics[width=\textwidth]{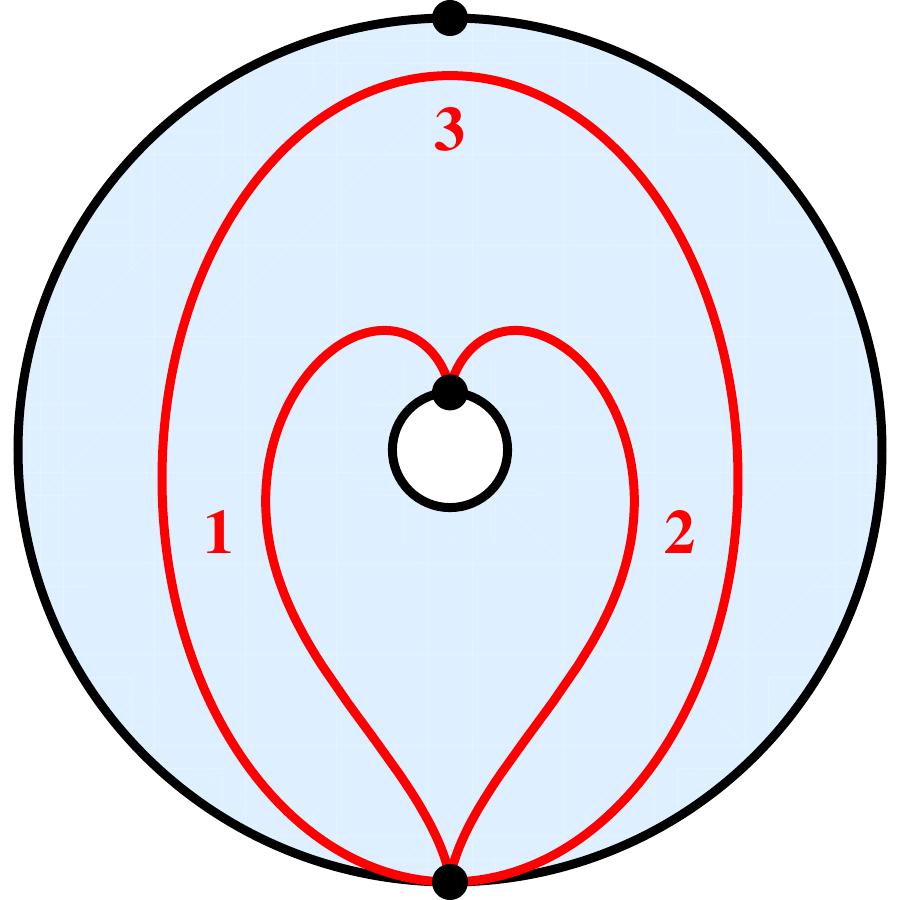}
\end{minipage}
\begin{minipage}{.55\textwidth}
\begin{center}
\begin{equation}
\begin{xy}
 (-20,0)*+{1}*\cir<8pt>{}="a" ;
(20,0)*+{2}*\cir<8pt>{}="b" ;
(0,28.2)*+{3}*\cir<8pt>{}="c" ;
\ar @{->} "a"; "b" <3pt>
\ar @{->} "a"; "b" <-3pt>
\ar @{->} "b"; "c"
\ar @{->} "c"; "a"
\end{xy}
 \nonumber
\end{equation}

\begin{eqnarray}
\mathcal{W} &= &X_{12}X_{23}X_{31}.
\nonumber
\end{eqnarray}
\end{center}
\end{minipage}

\item$SU(2)  \ N_{f}=2$

This theory is constructed on an annulus with two marked points on each boundary component.

\noindent\begin{minipage}{.45\textwidth}
  \includegraphics[width=\textwidth]{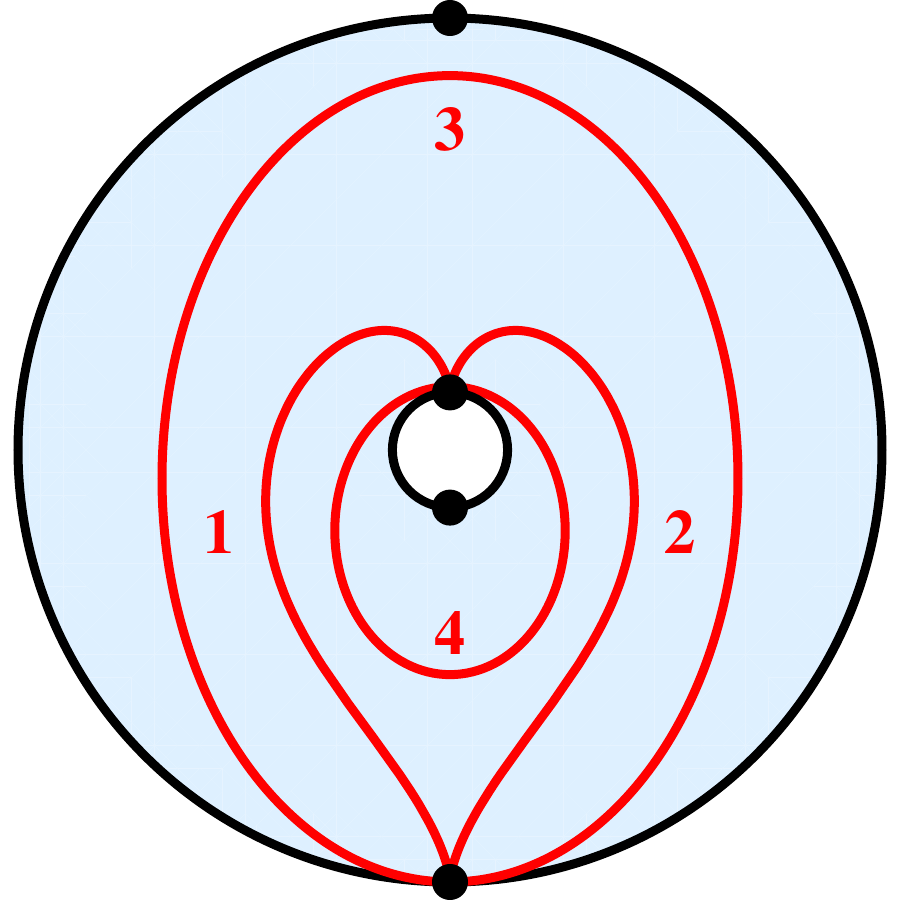}
\end{minipage}
\begin{minipage}{.55\textwidth}
\begin{equation}
\xy
(-20,0)*+{1}*\cir<8pt>{}="a" ; (20,0)*+{2}*\cir<8pt>{}="b" ;(0,28.2)*+{3}*\cir<8pt>{}="c";(0,-28.2)*+{4}*\cir<8pt>{}="d"
\ar @{->} "a"; "b" <3pt>
\ar @{->} "a"; "b" <-3pt>
\ar @{->} "b"; "c"
\ar @{->} "b"; "d"
\ar @{->} "c"; "a"
\ar @{->} "d"; "a"
 \endxy
 \nonumber
 \end{equation}

\begin{eqnarray}
\mathcal{W}& = &X_{12}X_{23}X_{31}+Y_{12}X_{24}X_{41}.
\nonumber
\end{eqnarray}
\end{minipage}

\item$SU(2)  \ N_{f}=3$

This theory is constructed on a disc with two marked points on the boundary and two punctures.

\noindent\begin{minipage}{.45\textwidth}
  \includegraphics[width=\textwidth]{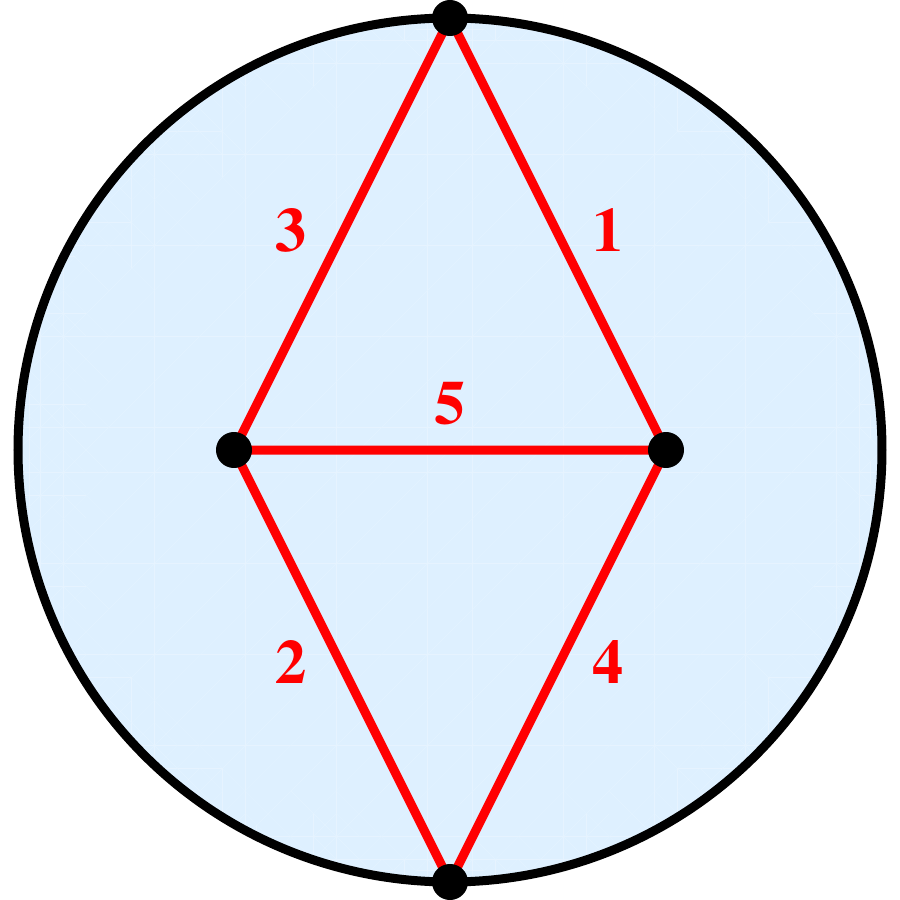}
\end{minipage}
\begin{minipage}{.55\textwidth}
\begin{equation}
\xy
(-20,0)*+{1}*\cir<8pt>{}="a" ; (20,0)*+{2}*\cir<8pt>{}="b" ;(0,28.2)*+{3}*\cir<8pt>{}="c";(0,-28.2)*+{4}*\cir<8pt>{}="d"; (0,0)*+{5}*\cir<8pt>{}="e"
\ar @{->} "e"; "a"
\ar @{->} "a"; "c"
\ar @{->} "c"; "e"
\ar @{->} "e"; "b"
\ar @{->} "b"; "c"
\ar @{->} "a"; "d"
\ar @{->} "d"; "e"
\ar @{->} "b"; "d"
\endxy
\nonumber
\end{equation}

\begin{eqnarray}
\mathcal{W} &= &X_{13}X_{35}X_{51}+X_{23}X_{35}X_{52} \nonumber \\
 & + & X_{14}X_{45}X_{51}+X_{24}X_{45}X_{52}. \nonumber
\end{eqnarray}
\end{minipage}
\end{itemize}
\subsubsection{Conformal Theories}
While the previous examples illustrate many general features, all the quivers given there are mutation equivalent to quivers without oriented cycles.  Thus for those cases the potential is completely fixed by the mutation rules of section \ref{MUT}.  Now we will consider the case of $SU(2)$ Yang-Mills theories with vanishing beta functions where the conformal invariance is broken only by mass terms.  Such quivers arise from triangulations of closed Riemann surfaces and never have acyclic quivers.  As such, our proposal for the superpotential is the only known way of constructing $\mathcal{W}$.
\begin{itemize}
\item $SU(2) \ N_{f}=4$

This theory is constructed on a sphere with four punctures.  We draw the associated triangulation on a plane omitting the point at infinity.

\noindent\begin{minipage}{.45\textwidth}
  \includegraphics[width=\textwidth]{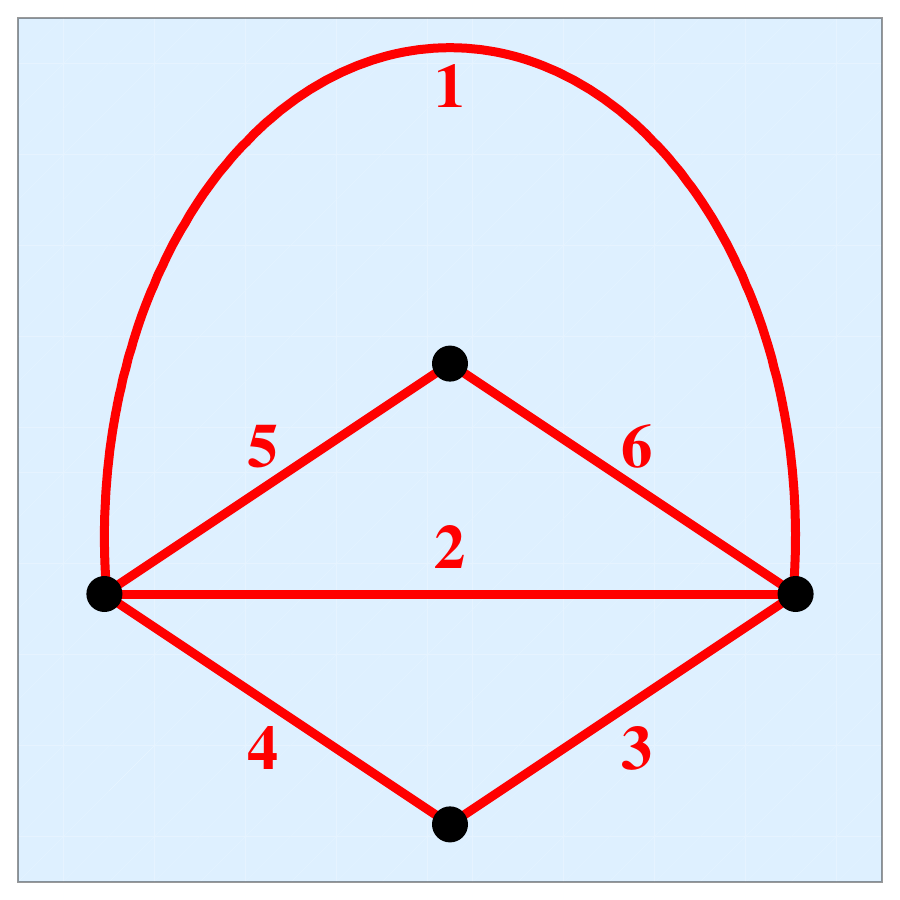}
\end{minipage}
\begin{minipage}{.55\textwidth}
\begin{equation}
\xy
(-20,0)*+{1}*\cir<8pt>{}="a" ; (20,0)*+{2}*\cir<8pt>{}="b" ;(0,15)*+{4}*\cir<8pt>{}="c";(0,30)*+{3}*\cir<8pt>{}="d"; (0,-15)*+{5}*\cir<8pt>{}="e"; (0,-30)*+{6}*\cir<8pt>{}="f";
\ar @{->} "a"; "e"
\ar @{->} "e"; "b"
\ar @{->} "b"; "c"
\ar @{->} "c"; "a"
\ar @{->} "f"; "a"
\ar @{->} "b"; "f"
\ar @{->} "d"; "b"
\ar @{->} "a"; "d"
\endxy
\nonumber
\end{equation}

\begin{eqnarray}
\mathcal{W}&=&X_{15}X_{52}X_{24}X_{41}+X_{13}X_{32}X_{26}X_{61} \nonumber\\
 & + & X_{15}X_{52}X_{26}X_{61}+X_{13}X_{32}X_{24}X_{41}. \nonumber
\end{eqnarray}
\end{minipage}

Notice that this triangulation contains two bivalent punctures; the quiver and superpotential above are obtained after integrating out the corresponding two-cycles.

\item $SU(2) \ \mathcal{N}=2^{*}$.

This theory is constructed on a torus with one puncture.  We draw the triangulation on a quadrilateral where opposite sides are identified.

\noindent\begin{minipage}{.45\textwidth}
  \includegraphics[width=\textwidth]{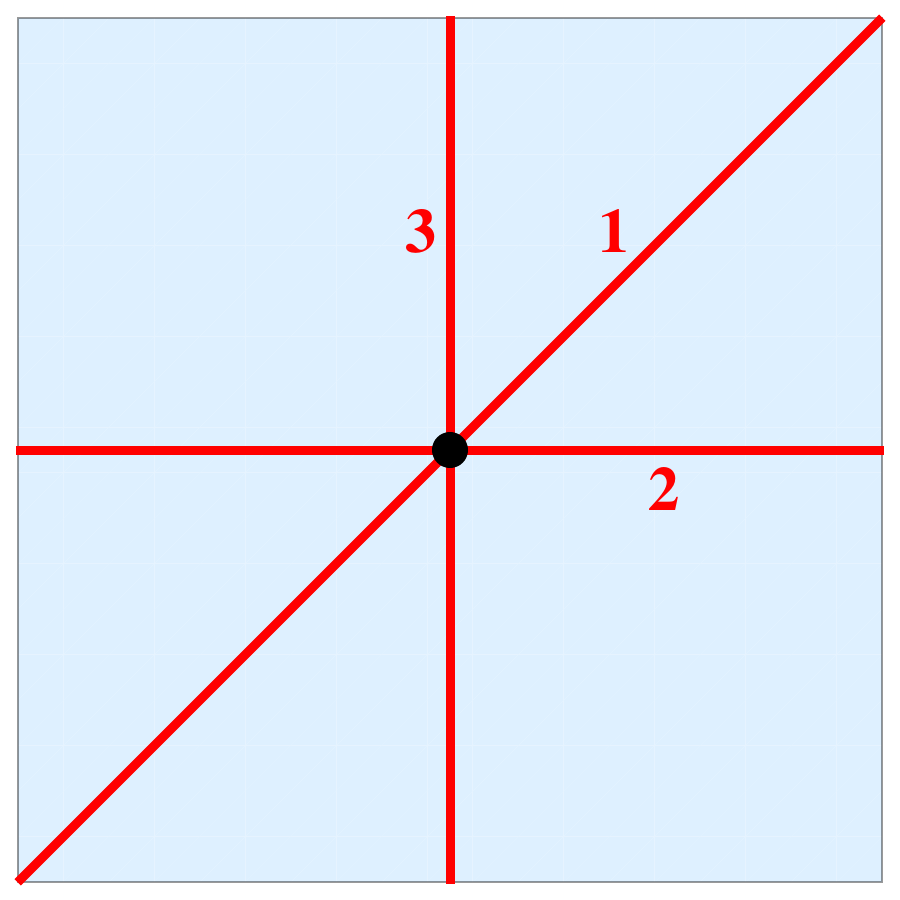}
\end{minipage}
\begin{minipage}{.55\textwidth}
\begin{equation}
\xy
(-20,0)*+{1}*\cir<8pt>{}="a" ; (20,0)*+{2}*\cir<8pt>{}="b" ; (0,28.2)*+{3}*\cir<8pt>{}="c" ;
\ar @{->} "a"; "b" <3pt>
\ar @{->} "a"; "b" <-3pt>
\ar @{->} "b"; "c" <3pt>
\ar @{->} "b"; "c" <-3pt>
\ar @{->} "c"; "a" <3pt>
\ar @{->} "c"; "a" <-3pt>
 \endxy
  \nonumber
\end{equation}

\begin{eqnarray}
\mathcal{W} & = & X_{12}X_{23}X_{31}+Y_{12}Y_{23}Y_{31}  \nonumber \\
 & + & X_{12}Y_{23}X_{31}Y_{12}X_{23}Y_{31}. \nonumber
\end{eqnarray}
\end{minipage}

It is amusing to note that the this quiver for the $\mathcal{N}=2^{*}$ theory is in fact invariant under mutation and, consistent with our general discussion, our potential is also mutation invariant.
\end{itemize}
Building from the examples in this section the reader can easily construct the BPS quiver for a complete theory associated to any arbitrary Riemann surface.

\section{Theories with Finite Chambers}\label{finite}
In this section we will identify a subset of complete $\mathcal{N}=2$ theories for which there exists some chamber containing only finitely many BPS states. In particular, we will show that all asymptotically free $SU(2)^n$ gauge theories, Argyres-Douglas models, and conformal theories with genus zero and genus one surfaces and sufficiently many punctures, meet this criterion. Our main motivation for studying theories with finite chambers is that they are especially well-adapted to the mutation method. As described in \cite{ACCERV}, the mutation method is most straightforward for computing BPS spectra which consist of only finitely many states. Additionally, as was mentioned in subsection \ref{mutmeth}, finite chambers have BPS spectra which consist exclusively of multiplicity one hypermultiplets.

\begin{figure}
\centering
\includegraphics[scale=1]{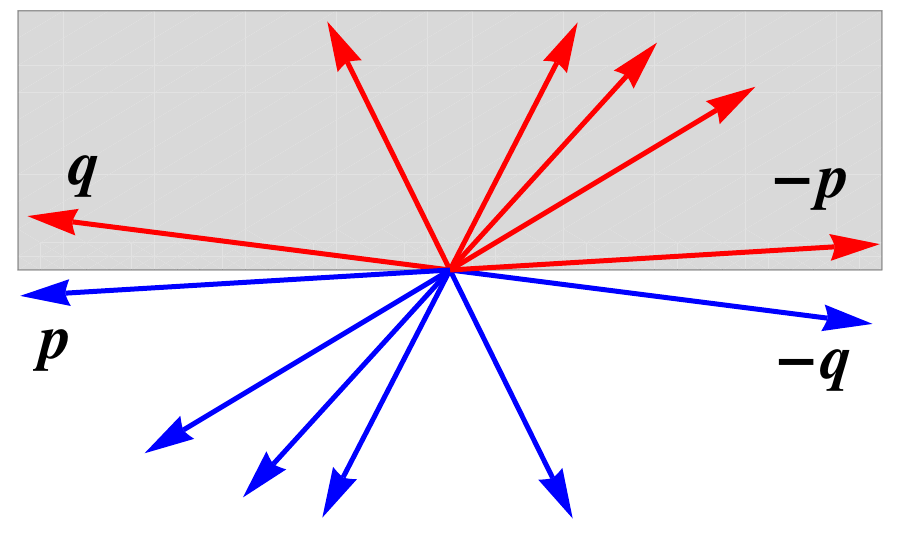}
\caption{Here we illustrate a choice of half-plane that forces $q,-p$ to be nodes of the quiver, for any arbitrary adjacent hypermultiplet BPS states $p,q.$ The grey region indicates the choice of particle half-plane, $\mathcal{H},$ while red vectors are BPS charges of particles, and blue vectors are BPS charges of anti-particles}
\label{adjacent}
\end{figure}

Complete theories also have especially well-behaved wall-crossing phenomena. It is a fact that the quiver of any complete theory has at most two arrows between any two nodes.\footnote{This can be understood via the triangulation construction. Two diagonals can share at most two triangles between them, and therefore the resulting quiver can have at most two arrows between any two nodes.} Consider some wall crossing of two adjacent hypermultiplet states $p,q,$ and choose the half-plane for the quiver such that $p$ is just outside of the half-plane on the left and $q$ is just inside the half-plane. This situation is illustrated in Figure \ref{adjacent}. The quiver must contain both $q$ and $-p$ as nodes since they form the boundary of the cone of positive states. Since we are studying a complete theory, we must have $|p\circ q |\le 2.$ The hypermultiplet wall-crossing is completely straightforward and explicit for any of the three possibilities.
\begin{itemize}
\item $|p\circ q|=0$: there is no change in the spectrum across the wall,\\
\item $|p\circ q|=1$: pentagon identity, which gives two states $p,q$ on one side of the wall and three states $p,p+q,q$ on the other side of the wall,\\
\item $|p\circ q|=2$: $SU(2)$ identity, which gives two states $p,q$ on one side of the wall, and the vector $p+q$ with infinite tower of dyons $(n+1)p+nq,np+(n+1)q$ for $n\ge 0$ on the other.
\end{itemize}
While the hypermultiplet wall-crossings are highly simplified, we should point out that it is still possible to have wall crossing of vector multiplets in a complete theory. This may produce some wild behavior involving infinitely many vectors, which is not so explicitly understood.

Of course, for complete theories the central charges for a basis of states can all be varied independently by tuning parameters; thus in principle, all chambers found via the wall-crossings described above should be physically realized in parameter space. Combining the mutation method and the wall crossing formulae above, explicit computation of BPS spectra for any complete theory with a finite chamber is now reduced to a completely algorithmic procedure for a large region of parameter space.

We devote the rest of this section to finding finite chambers of complete theories. The result of this study will produce finite chambers for the following theories:
\begin{itemize}
\item Conformal Argyres-Douglas type theories,
\item Asymptotically free $SU(2)^n$ gauge theories,
\item Conformal $SU(2)^n$ gauge theories with bifundamentals charged under the $i$th and $i+1$th $SU(2)$s for $i=1,\dots,k$, and 2 additional fundamentals each for the first and last $SU(2),$
\item Conformal $SU(2)^n$ gauge theories with bifundamentals charged under the $i$th and $i+1$th $SU(2)$s for $i=1,\dots,k$, and a bifundamental charged under the first and last $SU(2).$
\end{itemize}
The first two classes of theories arise from surfaces with boundary, which will be the main focus of the abstract arguments to follow. For the third and fourth class, which correspond to boundaryless spheres and tori with arbitrary punctures, some ad hoc techniques are applied to find finite chambers. Of the complete theories associated to Riemann surfaces, we have failed to find finite chambers for boundaryless $g\ge2$ surfaces.\footnote{Among the exceptional theories, dicussed in \ref{except}, we will find finite chambers for all except one, $X_7$} We note that there is another distinguishing feature of these boundaryless higher genus theories, namely, that they contain some matter fields in half-hypermultiplets, which cannot be given masses. As a result, it is impossible to take various decoupling limits with large masses. It would interesting to understand if this fact somehow precludes the existence of finite chambers for such theories.

\subsection{Examples}\label{exfinite}
Before we study the abstract arguments to prove existence of various finite chambers, we will present some explicit examples in this subsection to illustrate the objective of this program. The examples will also illustrate the three classes of theories which we will explore in this section. Our main tool here is the mutation method. We recall that, when applying the mutation method to complete theories, we are free to simply choose any ordering of central charges we wish. In the examples below, we demonstrate the existence of the finite chamber by providing an ordering of central charges that yields finitely many mutations in the mutation method; completeness guarantees that a corresponding region of parameter space exists.
\begin{itemize}
\item \emph{Argyres-Douglas $D_4$ theory.}\\
The BPS structure of Argyres-Douglas $A_n$ theories was studied systematically in \cite{SV}. There exist analogues of Argyres-Douglas theory associated to $ADE$ Dynkin diagrams, which were studied in \cite{CNV,CD2}. The quivers of these theories are precisely their associated Dynkin diagrams.\footnote{The underlying graph of the quiver, where we ignore orientation of arrows, exactly agrees with the associated Dynkin diagram. It can be checked that all orientations of arrows for such quivers are mutation equivalent.}  Here we study the Argyres-Douglas theory associated to $D_4.$ The Gaiotto curve of this theory is given by a sphere with one regular puncture, and one puncture with $k=4.$ The resulting surface with boundary and quiver are given below.

\noindent
  \begin{minipage}{0.5\textwidth}
  \includegraphics[scale=0.6]{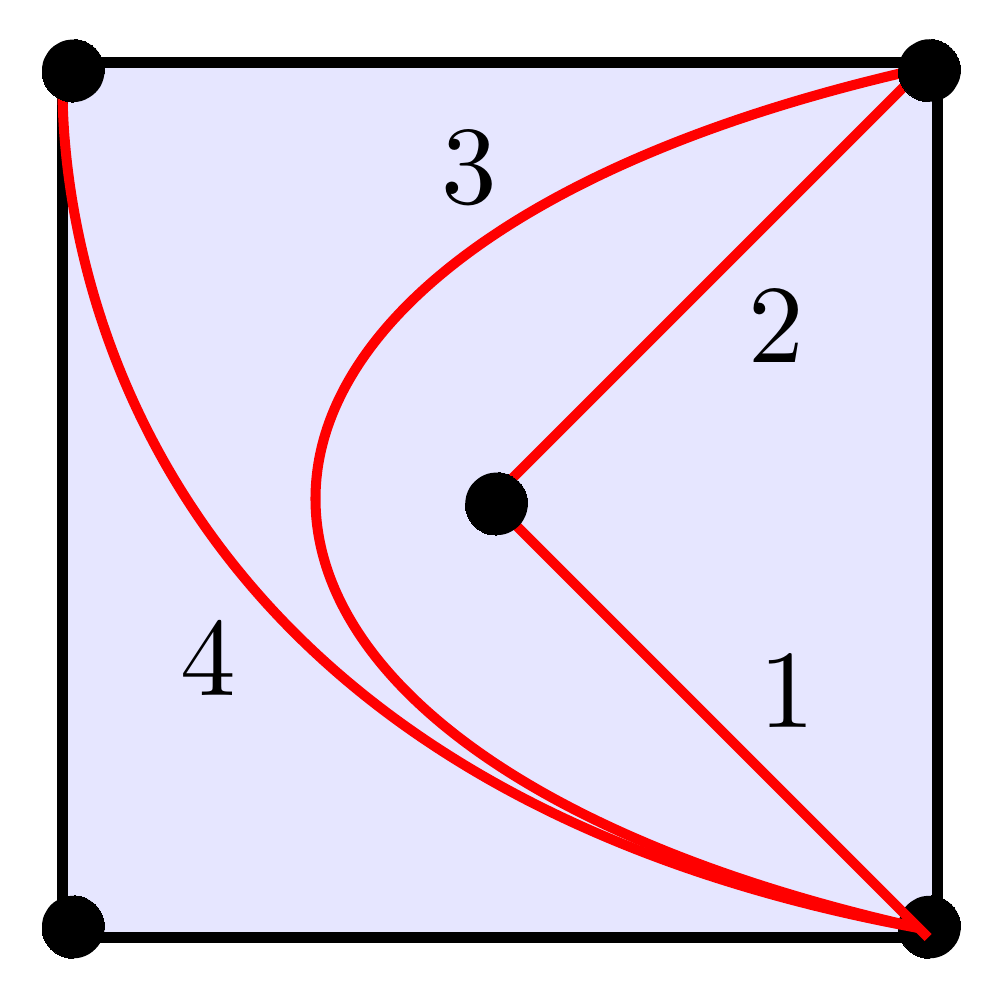}
  \end{minipage}
  \begin{minipage}{0.4\textwidth}
  \centerline{\begin{xy} 0;<1pt,0pt>:<0pt,-1pt>::
(0,0) *+{3}*\cir<8pt>{} ="3",
(-45,45) *+{1}*\cir<8pt>{} ="1",
(-45,-45) *+{2}*\cir<8pt>{} ="2",
(64,0) *+{4}*\cir<8pt>{}="4",
"2", {\ar"3"},
"3", {\ar"1"},
"4", {\ar"3"}
\end{xy}}
\end{minipage}

\noindent
It is quite easy to identify a finite chamber for this theory via the mutation method. For example, take $\arg\mathcal{Z}(\gamma_1)>\arg\mathcal{Z}(\gamma_3)>\arg\mathcal{Z}(\gamma_2)>\arg\mathcal{Z}(\gamma_4);$ then we mutate on $1,3,2,4$ in that order. This gives a chamber whose BPS stable states are precisely those associated to nodes of this quiver, without any additional bound states. In fact, a chamber with just the nodes themselves always exists for any acyclic quiver: choose an ordering on the nodes so that $\mathrm{arg}\mathcal{Z}(\gamma_i)>\mathrm{arg}\mathcal{Z} (\gamma_j)$ if and only if $\gamma_j\circ\gamma_i\ge0$. That such a choice is possible is due to the fact that the quiver has no oriented cycles. Then we can see that the resulting chamber will have only its nodes as $\Pi$-stable representations, via either the mutation method or directly from quiver representation theory.

\item \emph{$SU(2)^2,$ one bifundamental hypermultiplet.}\\
This theory corresponds to the surface and quiver shown below.

\noindent
  \begin{minipage}{0.5\textwidth}
  \includegraphics[scale=0.6]{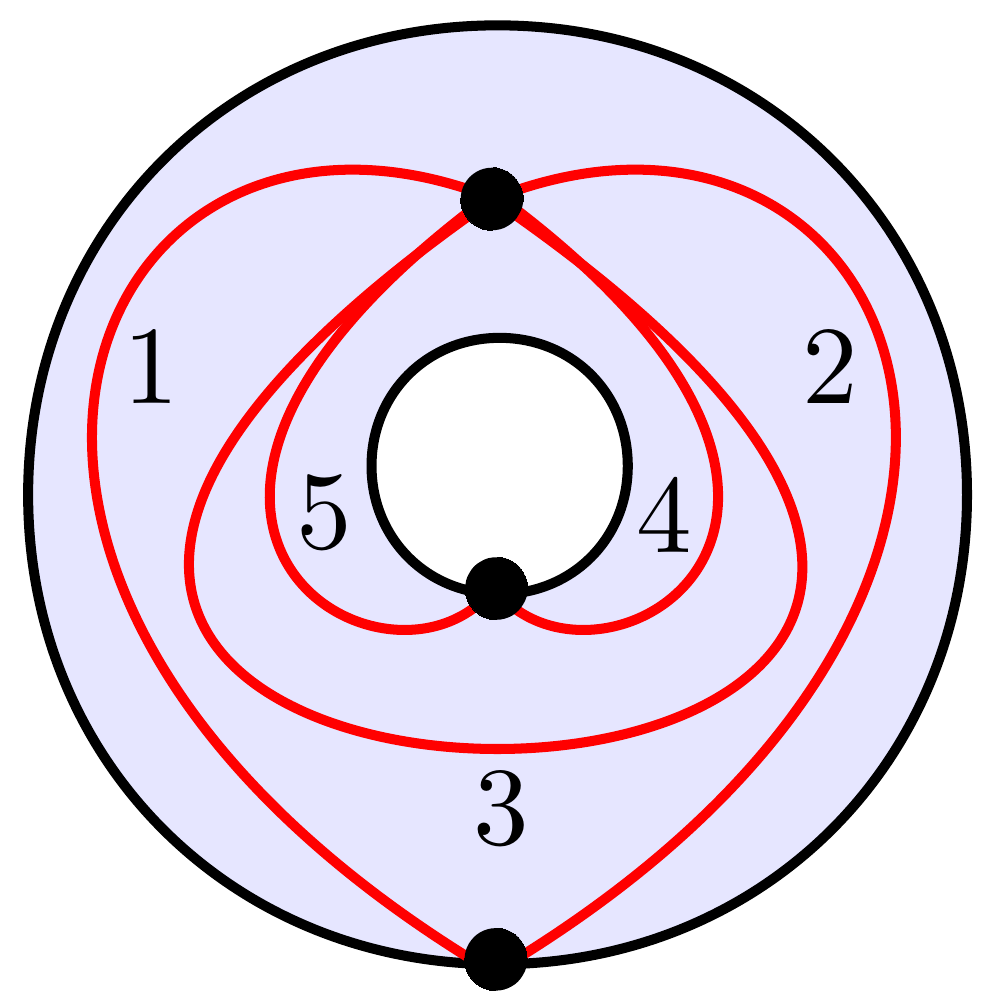}
  \end{minipage}
  \begin{minipage}{0.4\textwidth}
\centerline{
\begin{xy} 0;<1pt,0pt>:<0pt,-1pt>::
(0,0) *+{2}*\cir<8pt>{} ="0",
(0,50) *+{1}*\cir<8pt>{} ="1",
(45,25) *+{3}*\cir<8pt>{} ="2",
(90,50) *+{4}*\cir<8pt>{}="5",
(90,0) *+{5}*\cir<8pt>{}="7",
"1", {\ar"0" <1.5pt>},
"1", {\ar"0" <-1.5pt>},
"0", {\ar"2"},
"2", {\ar"1"},
"2",{\ar"5"},
"5", {\ar"7" <1.5pt>},
"5", {\ar"7" <-1.5pt>},
"7", {\ar"2"},
\end{xy}}
\end{minipage}

\noindent The gauge groups and matter content can be read off directly from the quiver. Each $SU(2)$ corresponds to a two-node $SU(2)$ subquiver, and the bifundamental field corresponds to the node which is attached to each $SU(2)$ in the same way as the third node of the $SU(2),N_f=1$ quiver. The theory is asymptotically free. A finite chamber can be found via the mutation method; for example, we find the chamber $\{\gamma_3,\gamma_1+\gamma_3,\gamma_1+\gamma_2+\gamma_3,\gamma_1+\gamma_2+2\gamma_3+\gamma_4,\gamma_5,\gamma_3+\gamma_4,\gamma_2,\gamma_1+\gamma_3+\gamma_4,\gamma_1,\gamma_4\},$ in decreasing phase order. This follows from the following mutation sequence: $3,1,2,4,5,2,1,3,2,4.$ This chamber includes the nodes themselves along with several bound states; because this quiver contains cycles, there is no chamber without bound states, as there was for Argyres-Douglas. Nonetheless, we have exhibited a finite chamber for this theory.

\item \emph{$SU(2), N_f=4.$}\\
The quiver of this theory is associated to a sphere with four regular punctures, and was given along with the appropriate superpotential in subsection \ref{SQCD}. It is well known that this theory is conformal. Again the mutation method yields a finite chamber: in decreasing phase order, $\{\gamma_3,\gamma_4,\gamma_5,\gamma_6,\gamma_1+\gamma_4+\gamma_6,\gamma_2+\gamma_3+\gamma_5,\gamma_2+\gamma_3,\gamma_1+\gamma_4,\gamma_2+\gamma_5,\gamma_1+\gamma_6,\gamma_1,\gamma_2\}.$ The mutation sequence for this chamber is $3,4,5,6,1,2,3,4,5,6,1,2.$ This finite chamber is particularly interesting because it occurs in the moduli space of a conformal theory. If we tune to the conformal point, by turning off all the masses of the flavor fields, it is expected that the BPS structure becomes highly intricate, respecting some large conformal duality group. In spite of this, we have exhibited a region of moduli space where the BPS spectrum is very simple, and consists of 12 hypermultiplet states.

\end{itemize}

These three cases are neatly representative of the types of theories for which we will find finite chambers. As described above, the existence of finite chambers for Argyres-Douglas theories is already clear, since they all correspond to acyclic Dynkin diagrams. The discussion below will extend this to all complete theories associated to surfaces with boundary; this class includes, in particular, Argyres-Douglas theories, as well as all complete asymptotically free $SU(2)^k$ gauge theories. We will also find finite chambers for the conformal $SU(2)^k$ theories associated to spheres and tori.

\subsection{Quiver Glueing Rule}\label{glue}
Consider two quivers, $A,B$ which separately have finite chambers $\mathcal{G}_A,\mathcal{G}_B$; in each quiver, choose a distinguished node, $a,b$ respectively. We will consider the composite quiver $A\oplus_a^b B$ which is given by drawing one arrow from $a\rightarrow b$. More generally, we might choose several nodes from each quiver, $\{a_i\},\{b_i$\} (where we allow repeats in the chosen nodes), and consider the composite quiver $A\oplus_{\{a_i\}}^{\{b_i\}}B$ formed by drawing arrows between pairs of nodes, $a_i\rightarrow b_i$. Note that all arrows must point from $A$ to $B$. The resulting quiver will contain a finite chamber whose BPS states are precisely the union of the BPS states $\mathcal{G}_A\cup\mathcal{G}_B.$ To specify such a chamber, we simply consider the ordering within each quiver $A,B$ to be given by the known finite chambers $\mathcal{G}_A,\mathcal{G}_B,$ and in addition we require for any nodes $\alpha\in A, \beta\in B$ we have $\mathrm{arg}\mathcal{Z}(\alpha)<\mathrm{arg}\mathcal{Z}(\beta).$


The representation theory makes this fact completely transparent. Consider any representation of the composite quiver. It is given by some representations $\mathcal{A},\mathcal{B}$ respectively of quivers $A,B$ along with a set of maps ${\phi_i:V_{a_i}\rightarrow V_{b_i}}$ corresponding to the arrows $a_i\rightarrow b_i.$ We will denote this rep as $R=(\mathcal{A},\mathcal{B},\{\phi_i\}).$ Let $\mathcal{A},\mathcal{B}$ be nonzero. Now we may consider the subrep $S=(0,\mathcal{B},\{0_i\}).$ This is always a valid subrep, as can be seen by the following commutative diagram:
$$\begin{CD}
\mathcal{A} @>\phi_i>>\mathcal{B}\\
@AA0A @AA\mathrm{id}A\\
0 @>0>>\mathcal{B}
\end{CD}$$
Note that by our choice of chamber, $\mathrm{arg}\mathcal{Z}(S)>\mathrm{arg}\mathcal{Z}(R),$ so that this is automatically a destabilizing subrep. Consequently, any representation that has support on both subquivers $A,B$ will be unstable, leaving only the stable reps of the subquiver $A,B$ separately. This rule can be checked as a simple exercise using the mutation method. Note that we have made no reference to $A,B$ being quivers of complete theories. The glueing rule is completely general and can be applied to any pair of quivers that are known to have finite chambers. 

As a first application of the glueing rule, we study acyclic quivers. Any acyclic quiver can be built up by glueing in one-node quivers, one at a time. Simply pick an ordering of the nodes consistent with the arrows - this is possible because the quiver is acyclic. Then we may glue the nodes to each other one-by-one in the given ordering. Since each one node quiver has only the node itself as a BPS state, we can build up a finite chamber which consists only of the nodes of the quiver. This immediately confirms the claim in subsection \ref{exfinite}, and allows us to conclude that all Argyres-Douglas theories have such chambers. In fact, acyclic finite mutation type quivers were classified by \cite{BR}, and consist precisely of usual $ADE$ and affine $\widehat{A}\widehat{D}\widehat{E}$ Dynkin diagrams. These are the only complete theories containing a minimal chamber in which only the nodes of the quiver are stable BPS states.

\subsection{Triangulation Glueing Rule}
The relation between complete theories and triangulated surfaces allows us to translate the above quiver glueing rule to a glueing rule at the level of the triangulation. First, we define an \emph{augmented quiver} associated to the triangulation of a surface, in which we include nodes corresponding to the boundary edges in the triangulation, and draw arrows as given by the rules of section \ref{triangles}, treating boundary edges and interior diagonals on equal footing. The nodes corresponding to boundary edges will be referred to as \emph{augmented nodes}. Then when we glue together two triangulations along their boundaries, the new augmented quiver of the full surface is given by identifying some pair of augmented nodes in the augmented quivers of the two surfaces.

Notice that if the augmented quiver has a finite chamber, then so does the usual, unaugmented quiver: the usual quiver is a subquiver of the augmented one, and the finiteness of a chamber is preserved by taking subquivers. This can be seen via representation theory. Stability for a representation of a subquiver is equivalent to stability for the same rep considered in the full quiver, since in either case we need to study the same set of destabilizing subreps. So the BPS spectrum of a subquiver is just the restriction of the BPS spectrum of the full quiver to states that have support only on the subquiver of interest.

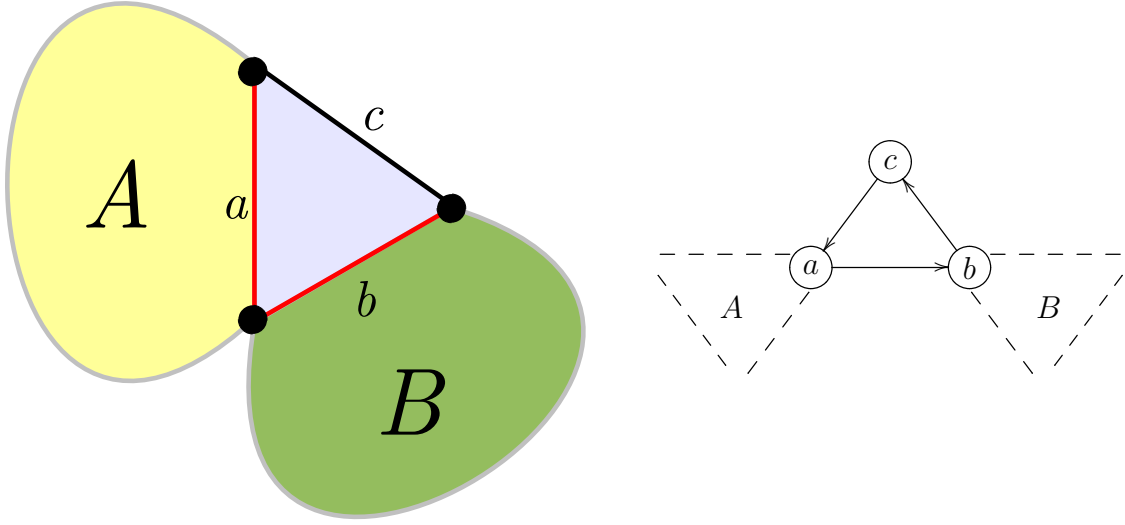
\begin{figure}[here!]
  \centering

  \begin{minipage}{0.45\textwidth}\includegraphics[scale=0.6]{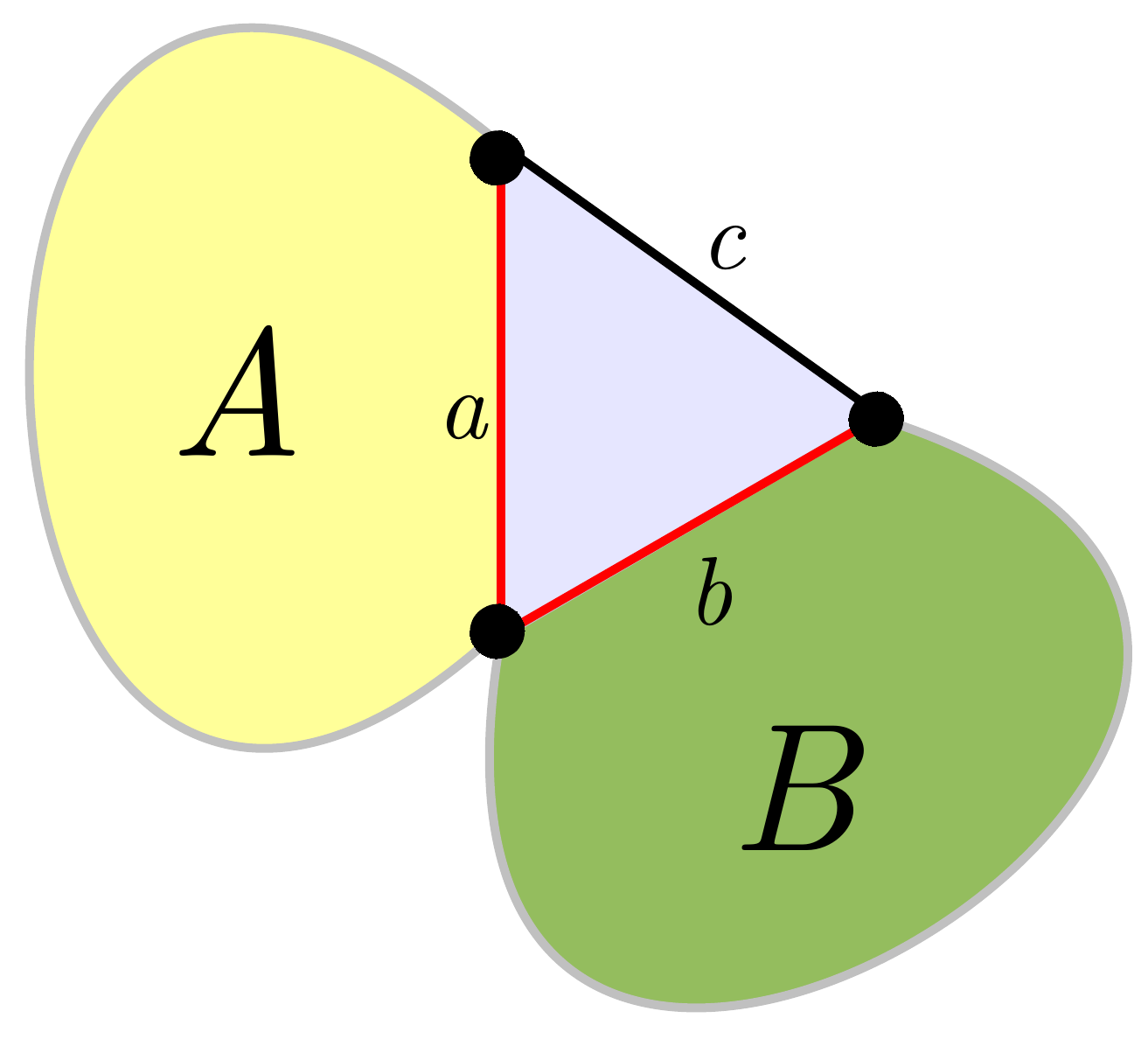}\end{minipage}
\begin{minipage}{0.54\textwidth}
\centerline{
\begin{xy} 0;<1pt,0pt>:<0pt,-1pt>::
(90,-40) *+{c}*\cir<8pt>{} ="0",
(60,0) *+{a}*\cir<8pt>{} ="1",
(120,0) *+{b}*\cir<8pt>{} ="2",
(30,40) *+{}="4",
(0,0) *+{}="3",
(30,15) *+{A},
(150,40) *+{}="5",
(180,0) *+{}="6",
(150,15) *+{B},
"0", {\ar"1"},
"2", {\ar"0"},
"1", {\ar"2"},
"3", {\ar@{--} "1"<5pt>},
"1",{\ar@{--} "4"<5pt>},
"3",{\ar@{--} "4"},
"5",{\ar@{--} "6"},
"6",{\ar@{--} "2"<-5pt>},
"2",{\ar@{--} "5"<-5pt>}
\end{xy}}
\end{minipage}
  \caption{General glueing rule for triangulations. $A$ and $B$ indicate surfaces with boundary, glued along one component of their respective boundaries to a triangle. Red lines indicate interior diagonals, which give nodes of the adjacency quiver. Black lines indicate boundary edges which give augmented nodes in the augmented quiver.}
  \label{fig:aplusb}
\end{figure}

Consider two triangulated surfaces $A,B,$ each with at least one boundary component. We will use the same symbols $A,B$ to denote the associated \emph{augmented} quivers. To achieve the glueing of quivers described above, we consider glueing the two triangulated surfaces along one component of their respective boundary components to two sides of a triangle, as in Figure~\ref{fig:aplusb}. Let us denote by $a,b$ the augmented nodes corresponding to the glued boundary edges of $A,B$ respectively, and let $c$ be the augmented node corresponding to the unglued edge of the triangle. The augmented quiver of the full surface is given in Figure~\ref{fig:aplusb} as well.

\begin{figure}
\centerline{
\begin{xy} 0;<1pt,0pt>:<0pt,-1pt>::
(120,40) *+{c}*\cir<8pt>{}="0",
(60,40) *+{a}*\cir<8pt>{}="1",
(180,40) *+{b}*\cir<8pt>{}="2",
(30,0) *+{}="4",
(0,40) *+{}="3",
(30,25) *+{A},
(210,0) *+{}="5",
(240,40) *+{}="6",
(210,25) *+{B},
"1", {\ar"0"},
"0", {\ar"2"},
"3", {\ar@{--} "1" <-5pt>},
"1",{\ar@{--} "4" <-5pt>},
"3",{\ar@{--} "4" },
"5",{\ar@{--} "6"},
"6",{\ar@{--} "2"<5pt>},
"2",{\ar@{--} "5"<5pt>}
\end{xy}}
\caption{Mutated form of quiver shown in Figure~\ref{fig:aplusb}, obtained by mutating at node $c$}
\label{fig:aplusbmut}
\end{figure}
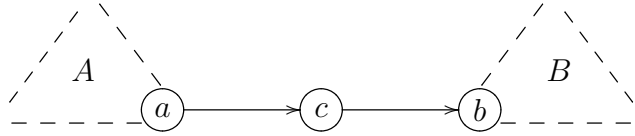

Note that $c$ is an augmented node, so that the unaugmented quiver is already a subquiver of $A\oplus_a^b B.$ Hence if $A$ and $B$ have finite chambers, then so does the resulting unaugmented composite quiver corresponding to the glueing described. However, in order to induct and continue glueing more pieces to this composite quiver, we would like to check that the \emph{augmented} quiver also has a finite chamber. This will be the case if both $A,B$ have finite chambers, and there is a finite chamber of $A$ (or $B$) such that no bound state has coefficient of $a$ (resp. $b$) greater than 1.

To see this, begin by mutating on node $c.$ We find a quiver $A\oplus_a^c\{c\}\oplus_c^b B$ (Figure~\ref{fig:aplusbmut}), which has a finite chamber consisting of $\mathcal{G}_A\cup\{c\}\cup\mathcal{G}_B,$ with $\mathrm{arg}\mathcal{Z}(b_i)>\mathrm{arg}\mathcal{Z}(c)>\mathrm{arg}\mathcal{Z}(a_j)$ for all $a_i\in A,b_j\in B,$ as described above. Now if we do a sequence of wall crossings to let $\mathrm{arg}\mathcal{Z}(c)>\mathrm{arg}\mathcal{Z}(b_j),$ then we will be in a region covered by the quiver form of Figure~\ref{fig:aplusb}. This can be seen by the mutation algorithm: $c$ is now the left-most node, so we mutate at $c$ first, away from the direct sum form in Figure~\ref{fig:aplusbmut}, resulting Figure~\ref{fig:aplusb}. We then see that we are in a region of moduli space covered by the quiver Figure~\ref{fig:aplusb}. As long as this  wall-crossing procedure only goes through pentagon-type crossings, we will only generate finitely many new bound states. Since $c$ only has inner product with $b$ in $B,$ the condition is just that there are no bound states in $\mathcal{G}_B$ with more than one $b$. A similar argument with inverse mutation yields an analogous conclusion for $A$.

To reiterate, we have developed a glueing rule for triangulations, depicted in Figure \ref{fig:aplusb}. The glueing rule provides a finite chamber for the composite triangulated surface, given finite chambers for the two separate triangulated surfaces, subject to an additional mild conditions that there be no bound states of multiple $a$'s or $b$'s.

\subsection{Surfaces with Boundary}
In this section we will explore the quiver glueing rule and its implication for triangulations, to attempt to build up a large class of Riemann surfaces whose quivers contain a finite chamber. In fact, we will find that any surface with boundary has a quiver with finite chamber. Recall from \cite{CV11} that surfaces with boundary correspond to asymptotically free theories along with the conformal Argyres-Douglas theories. Aside from the Argyres-Douglas cases, these theories have negative beta function because they are constructed by taking certain decoupling limits of the conformal theories that correspond to boundaryless Riemann surfaces.

A surface in this context is characterized completely by its genus $g$, number of punctures $n$, and number of boundary components $b$, along with some number of marked points $k_i\ge 1$ for every boundary component, $i=1\dots b.$ The $k_i$ are identified with the orders of poles as given in section \ref{triangulation}.  In order to build up new surfaces, we will glue triangulated pieces $B$ to some existing surface $A$ with finite chamber, as in Figure \ref{fig:aplusb}, all while making sure to preserve the finite chamber. Suppose we have some surface $(g,n,b,\{k_i\}_{i=1}^b)$ whose quiver, $A$, has a finite chamber. There are four types of operations we will need to consider:

\begin{itemize}
\item \emph{Add a marked point on the boundary}\\
An unpunctured triangle glued to boundary component $i$ of the surface $A$ will increase the number of marked points on $i$ by one ($k_i\rightarrow k_i+1$) and leave the other parameters of the surface unchanged.

\noindent
  \begin{minipage}{0.5\textwidth}
  \includegraphics[scale=0.6]{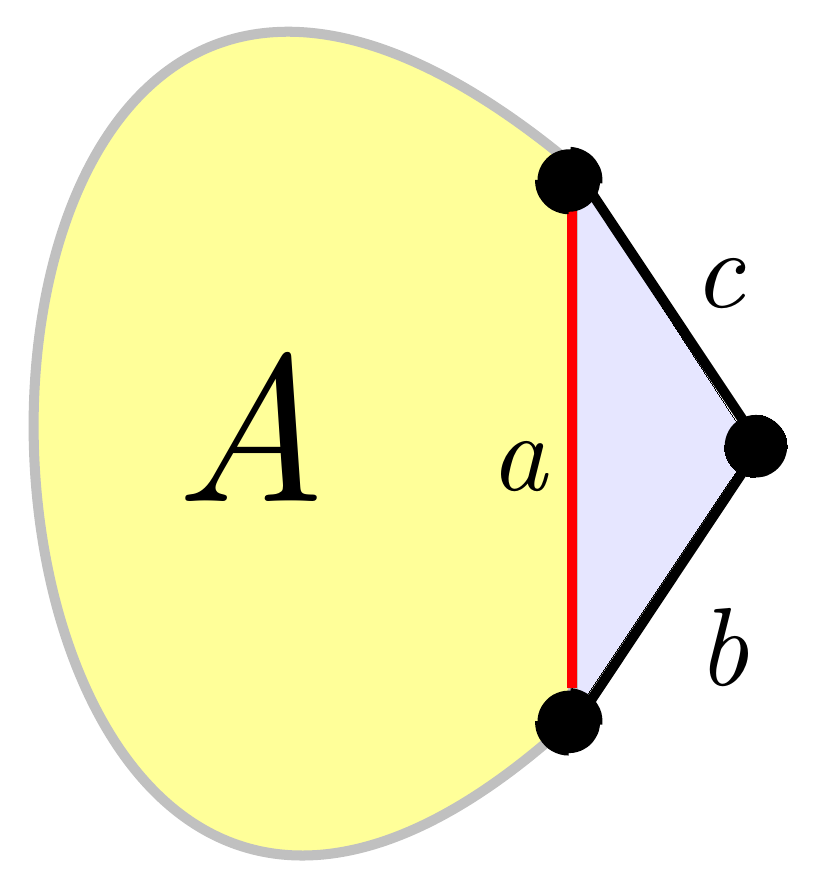}
  \end{minipage}
  \begin{minipage}{0.4\textwidth}
  \centerline{\begin{xy} 0;<1pt,0pt>:<0pt,-1pt>::
(90,-40) *+{c}*\cir<8pt>{} ="0",
(60,0) *+{a}*\cir<8pt>{} ="1",
(120,0) *+{b}*\cir<8pt>{} ="2",
(30,40) *+{}="4",
(0,0) *+{}="3",
(30,15) *+{A},
"3", {\ar@{--} "1"<5pt>},
"1",{\ar@{--} "4"<5pt>},
"3",{\ar@{--} "4"},
"1", {\ar"2"},
"0", {\ar"1"},
"2", {\ar"0"}
\end{xy}}
\end{minipage}

\noindent
On the augmented quivers, this adds an oriented three-node cycle with one node identified with an existing node on the quiver $A.$  This is just the general triangulation glueing described above, in which the surface $B$ is empty and the quiver $B$ is only the node $b$ itself. So this glueing preserves the finite chamber.

%



\item \emph{Add a puncture}\\
To add a puncture, we take $B$ to be a once-punctured monogon. This takes $n\rightarrow n+1,$ leaving everything else unchanged.

\noindent
\begin{minipage}{0.5\textwidth}\includegraphics[scale=0.6]{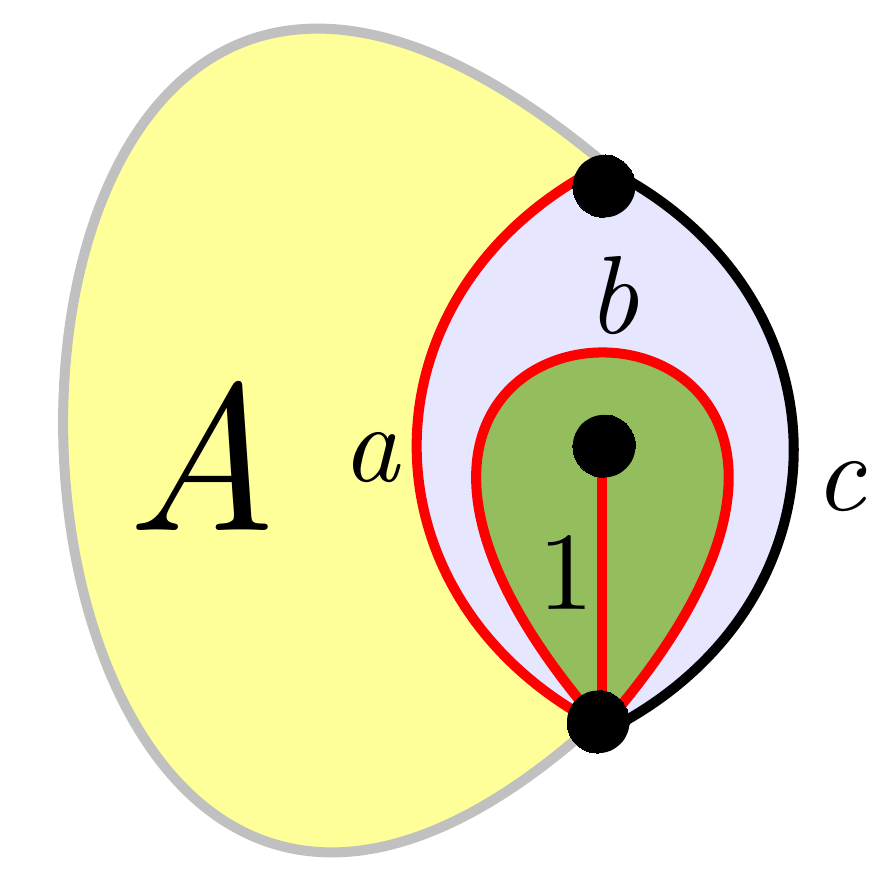}\end{minipage}
\begin{minipage}{0.4\textwidth}  \centerline{
\begin{xy} 0;<1pt,0pt>:<0pt,-1pt>::
(90,-40) *+{c}*\cir<8pt>{}="0",
(60,0) *+{a}*\cir<8pt>{}="1",
(120,0) *+{b}*\cir<8pt>{}="2",
(90,-13) *+{1}*\cir<8pt>{}="5",
(30,40) *+{}="4",
(0,0) *+{}="3",
(30,15) *+{A},
"3", {\ar@{--} "1"<5pt>},
"1",{\ar@{--} "4"<5pt>},
"3",{\ar@{--} "4"},
"1", {\ar"2"},
"0", {\ar"1"},
"2", {\ar"0"},
"1", {\ar"5"},
"5", {\ar"0"}
\end{xy}}
\end{minipage}

\noindent
The quiver $B$ is just two copies of the node $b$. Here we have encountered a self-folded triangle in the triangulation, so we must refer to the extended rules given in the appendix \ref{selffold}. The quiver has a finite chamber by the general glueing rule.



\item \emph{Add a boundary component}\\
For this we let $B$ be the annulus with one marked point on each boundary component. This glueing adds one boundary with one marked point, and leaves everything else fixed. That is, $b\rightarrow b+1$ and $k_{b+1}=1.$

\noindent
  \begin{minipage}{0.5\textwidth}\includegraphics[scale=0.45]{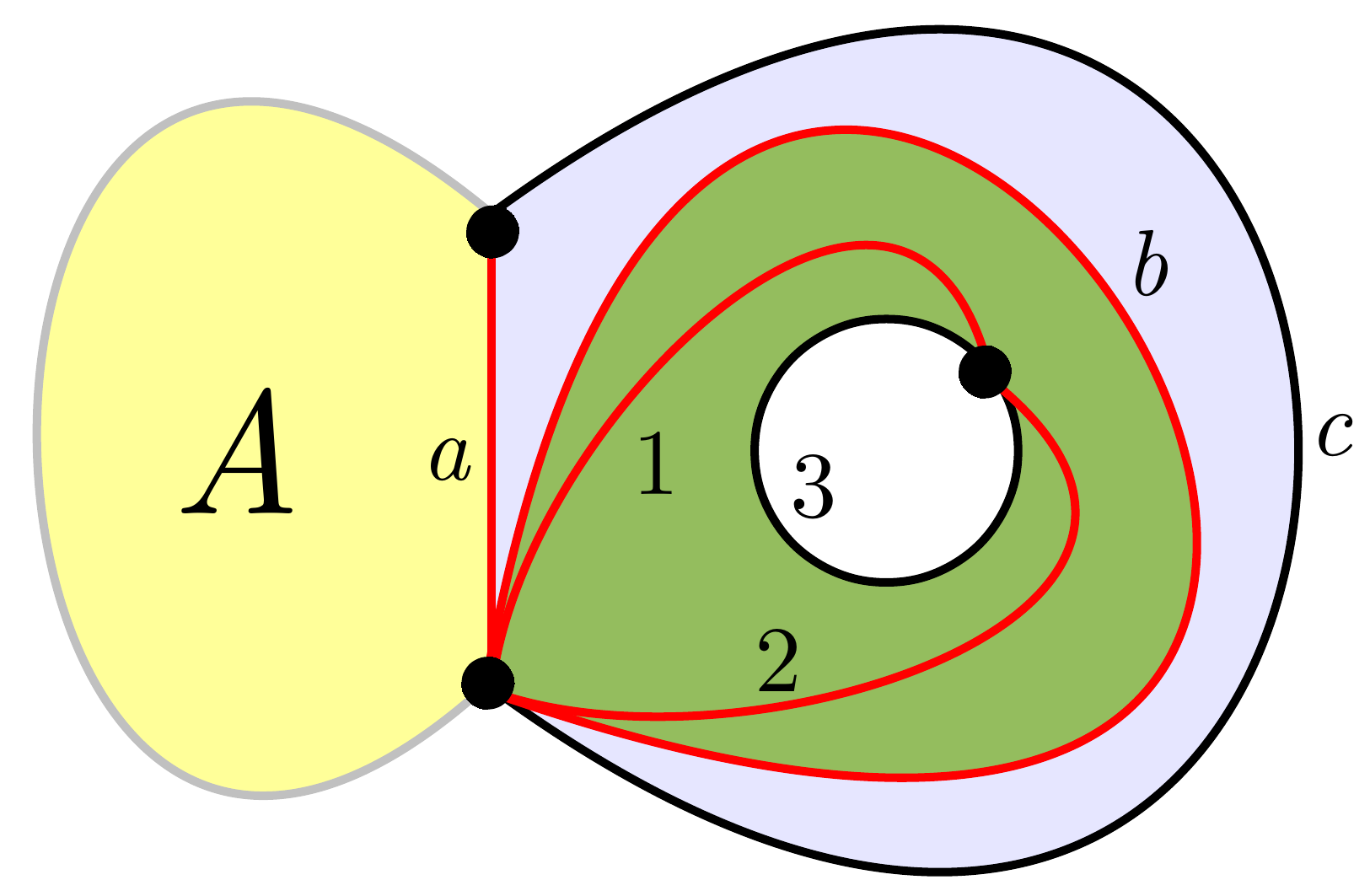}\end{minipage}
\begin{minipage}{0.4\textwidth}  \centerline{
\begin{xy} 0;<1pt,0pt>:<0pt,-1pt>::
(60,-50) *+{c}*\cir<8pt>{} ="0",
(60,0) *+{a}*\cir<8pt>{} ="1",
(105,-25) *+{b}*\cir<8pt>{} ="2",
(150,0) *+{1}*\cir<8pt>{}="5",
(195,-25) *+{3}*\cir<8pt>{}="6",
(150,-50) *+{2}*\cir<8pt>{}="7",
(15,25) *+{}="4",
(15,-25) *+{}="3",
(30,0) *+{A},
"3", {\ar@{--} "1"<5pt>},
"1",{\ar@{--} "4"<5pt>},
"3",{\ar@{--} "4"},
"1", {\ar"2"},
"0", {\ar"1"},
"2", {\ar"0"},
"2",{\ar"5"},
"5",{\ar"7" <1.5pt>},
"5",{\ar"7" <-1.5pt>},
"7",{\ar"2"},
"7",{\ar"6"},
"6",{\ar"5"}
\end{xy}}
\end{minipage}

\noindent
$B$ as a quiver is the $SU(2)\;N_f=2$ quiver.
By hand, we can check that $B$ has a finite chamber in which there is no bound state with multiple $b$'s, using the mutation method. For example, we find a chamber with states in decreasing phase order $\{b,\gamma_3,\gamma_1+b+\gamma_3,\gamma_2,\gamma_1+b,\gamma_1+\gamma_3,\gamma_2\}.$ Thus, the full augmented quiver also has a finite chamber.

%


\item \emph{Increase genus}\\
We may increase the genus of the surface by taking $B$ to be a torus with boundary with one marked point. This gives $g\rightarrow g+1$ with all other parameters fixed.

\noindent
  \begin{minipage}{0.4\textwidth}\includegraphics[scale=0.5]{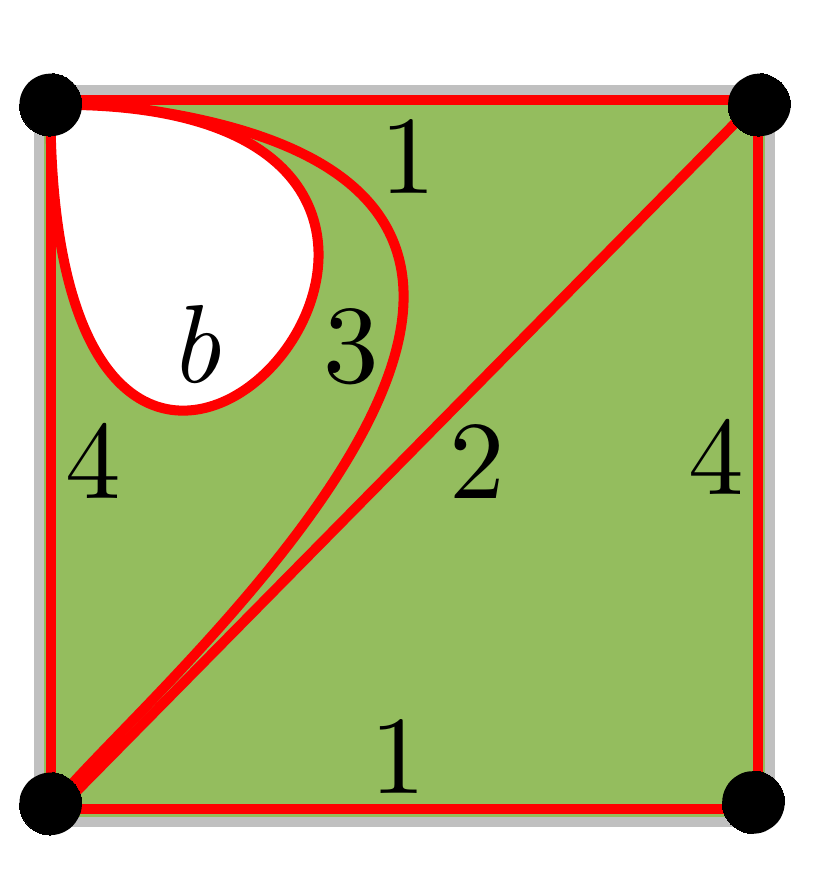}\end{minipage}
\begin{minipage}{0.5\textwidth}\centerline{
\begin{xy} 0;<1pt,0pt>:<0pt,-1pt>::
(60,0) *+{c}*\cir<8pt>{} ="0",
(60,50) *+{a}*\cir<8pt>{} ="1",
(105,25) *+{b}*\cir<8pt>{} ="2",
(150,50) *+{3}*\cir<8pt>{}="5",
(180,25) *+{1}*\cir<8pt>{}="6",
(150,0) *+{4}*\cir<8pt>{}="7",
(225,25) *+{2}*\cir<8pt>{}="8",
(15,25) *+{}="4",
(15,75) *+{}="3",
(30,50) *+{A},
"3", {\ar@{--} "1"<-5pt>},
"1",{\ar@{--} "4"<-5pt>},
"3",{\ar@{--} "4"},
"1", {\ar"2"},
"0", {\ar"1"},
"2", {\ar"0"},
"5",{\ar"2"},
"7",{\ar"5"},
"2",{\ar"7"},
"6",{\ar"7"},
"6",{\ar"5"},
"7",{\ar"8"},
"8",{\ar"6" <1.5pt>},
"8",{\ar"6" <-1.5pt>},
"5",{\ar"8"}
\end{xy}}
\end{minipage}

\noindent
Note that we have only drawn $B$, the torus with boundary, which must be glued into the surface $A$ as in Figure~\ref{fig:aplusb}. Again we can check by hand that $B$ contains a finite chamber with no bound states of multiple $b$'s.  For example, the mutation method gives a finite chamber with states in decreasing phase order $\{b,\gamma_3,\gamma_1,\gamma_2+\gamma_3,\gamma_1+\gamma_4+b,\gamma_1+\gamma_4,\gamma_2,\gamma_4+b,\gamma_4\}.$ So the resulting augmented quiver has a finite chamber.
\end{itemize}

Finally, we need to check that we have sufficient base cases in order to build up all possible surfaces with boundary. Again we will be parameterizing surfaces as $(g,n,b,\{k_i\}).$ The following are the base cases we need: once-punctured monogon $(0,1,1,\{1\}),$ unpunctured triangle $(0,0,1,\{3\})$, annulus with one marked point on each boundary $(0,0,2,\{1,1\})$, torus with one boundary component and one marked point $(1,0,1,\{1\}).$ It is straightforward to see any surface not generated by increasing the four parameters $(g,n,b,\{k_i\})$ starting from one of these base cases is either a surface without boundary or a surface that cannot be triangulated. For example, if we try to reduce $n$ in punctured monogon $(0,1,1,\{1\})$, we see that the unpunctured monogon, $(0,0,1,\{1\})$ cannot be triangulated. Notice that the base cases are precisely the pieces that we used in the glueings above, so we have already checked that the corresponding augmented quivers all contain the desired finite chambers. So we conclude that all surfaces with boundary (and thus all asymptotically free complete theories) have at least one chamber in their parameter space with finitely many states. 

Using the glueing rule and the wall-crossing formulae given at the beginning of this section, computing explicit spectra for these theories is now a completely algorithmic process. For any surface with boundary, we take a decomposition into the pieces used above: punctured monogon, unpunctured annulus, and torus with boundary. The pieces should all be glued together using unpunctured triangles as in Figure \ref{fig:aplusb}. The choice of decomposition will specify the mutation form of the quiver we must study, along with a point of parameter space, fixed by the ordering of central charges compatible with the glueing rule. Now we simply take the union of the finite spectra associated to each of these pieces; this gives the resulting spectrum of the total surface, according to the glueing rule. Finally, we can use wall-crossing formulae to move to other points in parameter space.



\subsection{Conformal Theories}
For surfaces without boundary (that is, conformal theories), there seems to be an essential complication in trying to decompose these quivers using the techniques above. Very generally, quivers for boundaryless surfaces have large cyclic structures that prevent such a decomposition. In particular, no node for a boundaryless surface can be a sink or source; consequently, the quiver glueing rule is of little use.

Nonetheless, some progress has been made in searching for finite chambers using the mutation method. We have extracted a finite chamber for genus $g=0,1$ with arbitrary punctures, which we give below. First, we recall some reasoning introduced in \cite{Gaiotto}, which allows us to deduce a Lagrangian description for these theories. For any of these rank 2 Gaiotto-type theories, we can understand the gauge groups and matter contained in the theory as follows. Take a pair-of-pants decomposition of the boundaryless Riemann surface $\mathcal{C}$. Each pair of pants corresponds to a half-hypermultiplet charged under 3 $SU(2)$'s, where each $SU(2)$ is represented by one of the boundary components of the pair of pants. Each glueing of a pair of pants identifies the corresponding $SU(2)$'s and gauges that $SU(2)$ symmetry. Given a boundaryless surface, one can use this recipe to deduce the gauge group and matter content of the corresponding theory.

\subsubsection{Sphere with $n\ge 4$ Punctures}
The sphere with $n < 3$ punctures cannot be triangulated; for $n=3$ punctures, it corresponds to three nodes with no arrows, which yields no interesting structure.
The sphere with $n\ge 4$ punctures has a Lagrangian description as an $SU(2)^{n-3}$ theory with bifundamentals charged under the $i$th and $i+1$th $SU(2)$s for $i=1,\dots,n-4$, and 2 additional fundamentals each for the first and last $SU(2)$.

\begin{figure}
\centering
\includegraphics[scale=0.65]{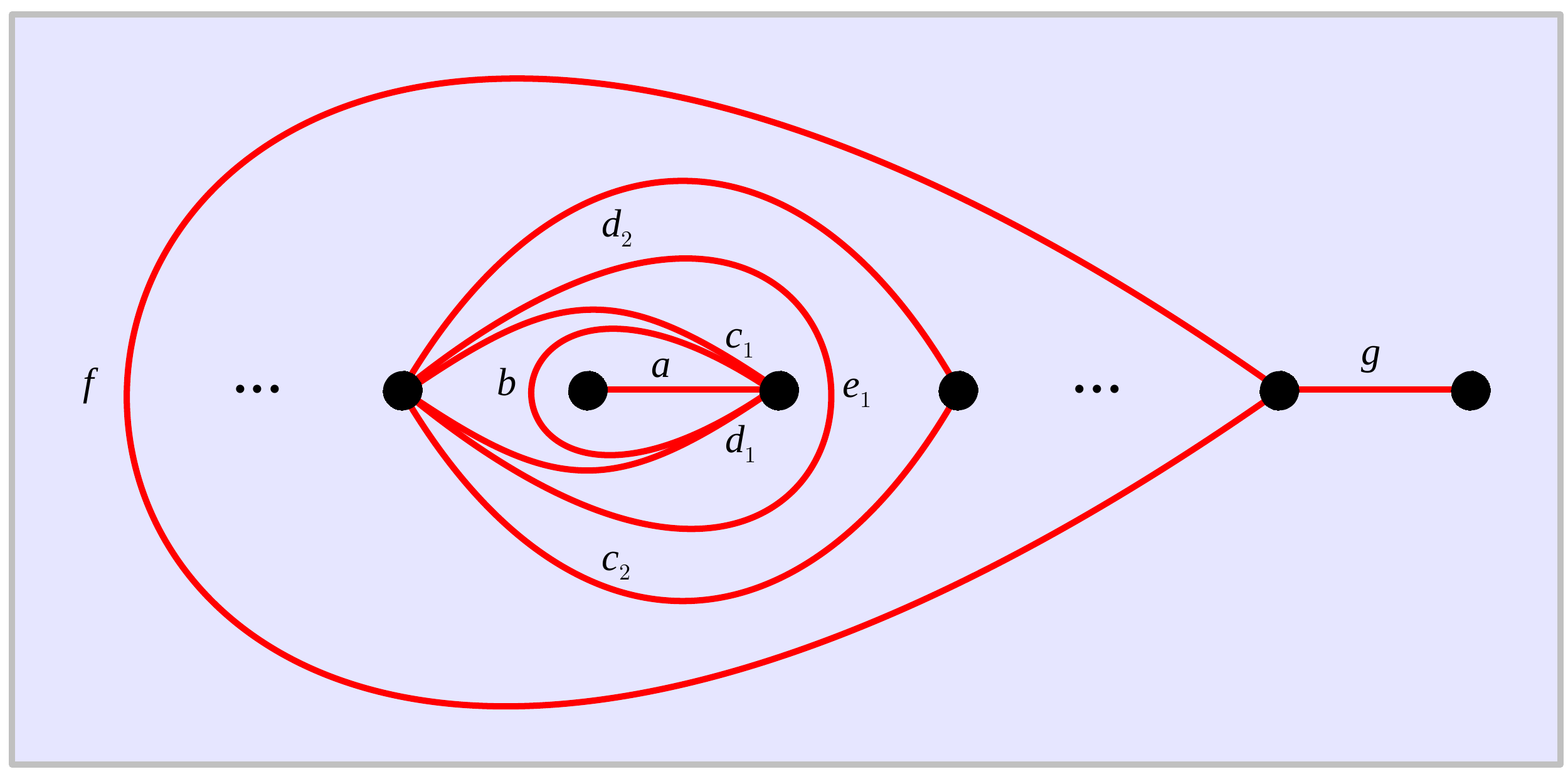}
\centerline{
\begin{xy} 0;<1pt,0pt>:<0pt,-1pt>::
(-45,12) *+{a}*\cir<8pt>{}="13",
(-45,38) *+{b}*\cir<8pt>{}="12",
(0,0) *+{d_1}*\cir<8pt>{} ="0",
(0,50) *+{c_1}*\cir<8pt>{} ="1",
(45,25) *+{e_1}*\cir<8pt>{} ="2",
(90,50) *+{c_2}*\cir<8pt>{}="5",
(135,25) *+{e_2}*\cir<8pt>{}="6",
(165,25) *+{\dots\dots},
(90,0) *+{d_2}*\cir<8pt>{}="7",
(240,0) *+{d_{n-3}}*\cir<12pt>{}="8",
(240,50) *+{c_{n-3}}*\cir<12pt>{}="9",
(285,12) *+{f}*\cir<8pt>{}="10",
(285,38) *+{g}*\cir<8pt>{}="14",
(195,25) *+{e_{n-4}}*\cir<12pt>{}="11",
"1", {\ar"0" <1.5pt>},
"1", {\ar"0" <-1.5pt>},
"0", {\ar"2"},
"2", {\ar"1"},
"2",{\ar"5"},
"5", {\ar"7" <1.5pt>},
"5", {\ar"7" <-1.5pt>},
"7",{\ar"2"},
"7",{\ar"6"},
"6",{\ar"5"},
"12",{\ar"1"},
"13",{\ar"1"},
"0",{\ar"12"},
"0",{\ar"13"},
"10",{\ar"9"},
"14",{\ar"9"},
"11",{\ar"9"},
"8",{\ar"10"},
"8",{\ar"14"},
"8",{\ar"11"},
"9", {\ar"8" <1.5pt>},
"9", {\ar"8" <-1.5pt>},
\end{xy}}
\caption{Triangulation and quiver for the sphere with $n\ge 4$ punctures. The triangulation is drawn on a plane with the point at infinity omitted. Note there are self-folded triangles formed by the interior of $a$ and the exterior of $f$ (see appendix \ref{selffold}). In both the triangulation and the quiver, the dots indicate repetition of the 3-node structure, $c_i d_i e_i$. The sphere with $n$ punctures has $n-4$ such pieces, and $3n-6$ nodes.}
\label{fig:sphere}
\end{figure}
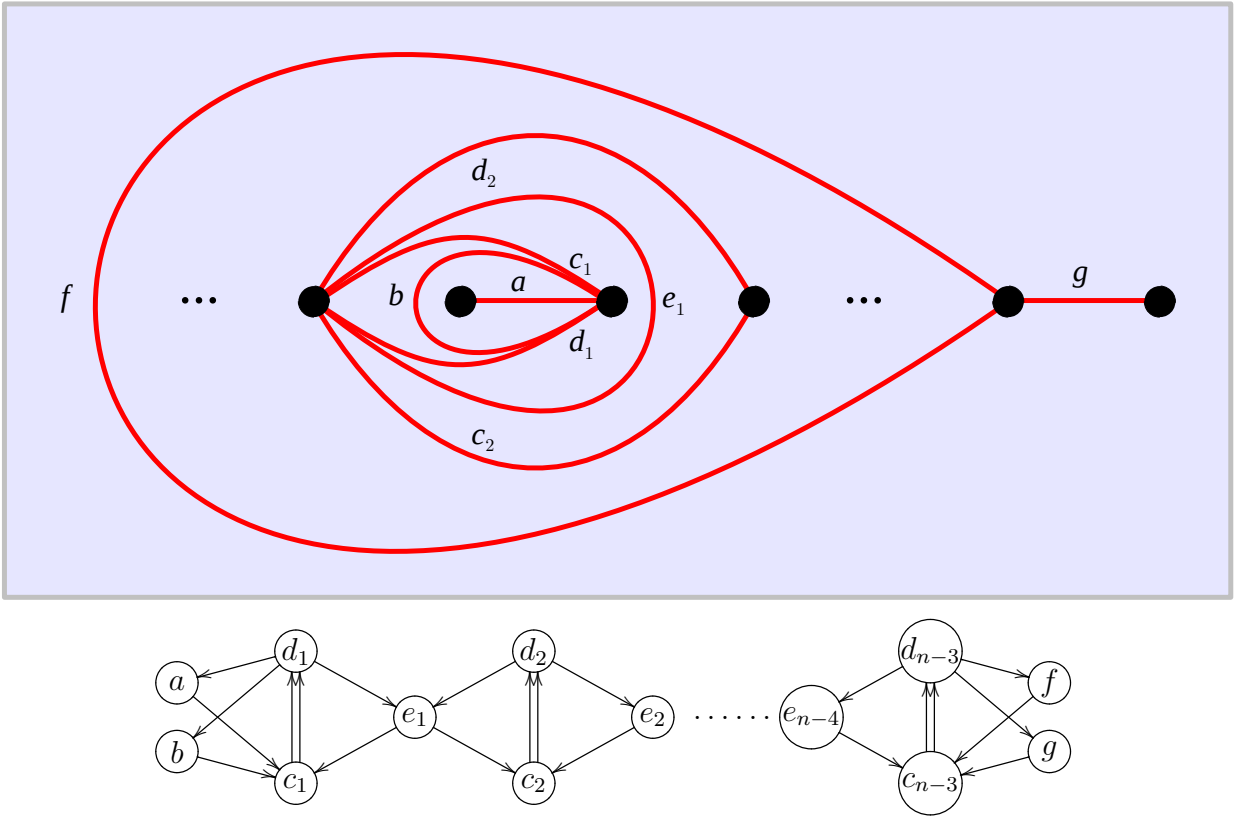

A triangulation and quiver of a sphere with $n\ge 4$ punctures is given in Figure~\ref{fig:sphere}.
In fact, the Lagrangian description can be read off directly from this quiver, forgetting the surface and triangulation. Each two-node structure $c_i d_i$ is precisely a pure $SU(2)$ subquiver, and thus indicates an independent $SU(2)$. The nodes $a,b,f,g$ appear just the flavor nodes in subsection \ref{asympex}, and correspond to flavors charged under the first and last $SU(2).$ Finally the nodes $e_i$ appear as flavor nodes for two adjacent $SU(2)$s, and thus correspond to bifundamental flavors. So, we have reconstructed the description of the gauge group and matter given above. This type of reasoning was discussed further in \cite{CV11}.

A finite chamber for $n\ge 4$ is given by the following sequence of states, in decreasing phase order:
$$\displaystyle\begin{array}{l}
a,b,a+b+c_1,d_1,a+c_1,b+c_1,\\
d_1+e_1,c_1,d_2,e_1+d_2,c_1+d_1+e_1+c_2,d_1+e_1+c_2,e_1,e_1+c_2,\\
\centerline{\vdots}\\
d_k+e_k,c_k,d_{k+1},e_k+d_{k+1},c_k+d_k+e_k+c_{k+1},d_k+e_k+c_{k+1},e_k,e_k+c_{k+1}, \\
\centerline{\vdots}\\
d_{n-4}+e_{n-4},c_{n-4},d_{n-3},e_{n-4}+d_{n-3},c_{n-4}+d_{n-4}+e_{n-4}+c_{n-3},d_{n-4}+e_{n-4}+c_{n-3},\\
\quad\quad e_{n-4},e_{n-4}+c_{n-3},\\
f+d_{n-3},g+d_{n-3},f+g+d_{n-3},c_{n-3},f,g
\end{array}$$
which is a chamber with $8n-20$ states.
This can be verified by applying the mutation method with the following mutations, in order:
 $$\begin{array}{l}a,b,c_1,d_1,a,b,\\e_1,c_1,d_2,d_1,c_2,c_1,d_2,e_1,\\
e_2,d_2,d_3,d_1,c_3,d_2,d_3,e_2,\\
\quad\quad\quad\quad\quad\quad \vdots \\
e_k,d_k,d_{k+1},d_1,c_{k+1},d_k,d_{k+1},e_k,\\ \quad\quad\quad\quad\quad\quad \vdots \\
e_{n-4},d_{n-4},d_{n-3},d_1,c_{n-3},d_{n-4},d_{n-3},e_{n-4},
\\f,g,d_{1},d_{n-3},f,g\end{array}$$

\subsubsection{Torus with $n\ge 2$ Punctures}
The torus with one puncture is the $\mathcal{N}=2^*$ theory, which has no finite chamber; this theory is explored further in \cite{ACCERV}.
The torus with $n\ge 2$ punctures has a Lagrangian description as an $SU(2)^n$ gauge theory with a bifundamental between the $i$th and $i+1$th $SU(2)$ for $i=1,\dots, n-1$ and a bifundamental between the first and last $SU(2)$.

\begin{figure}
\centering
\includegraphics[scale=0.75]{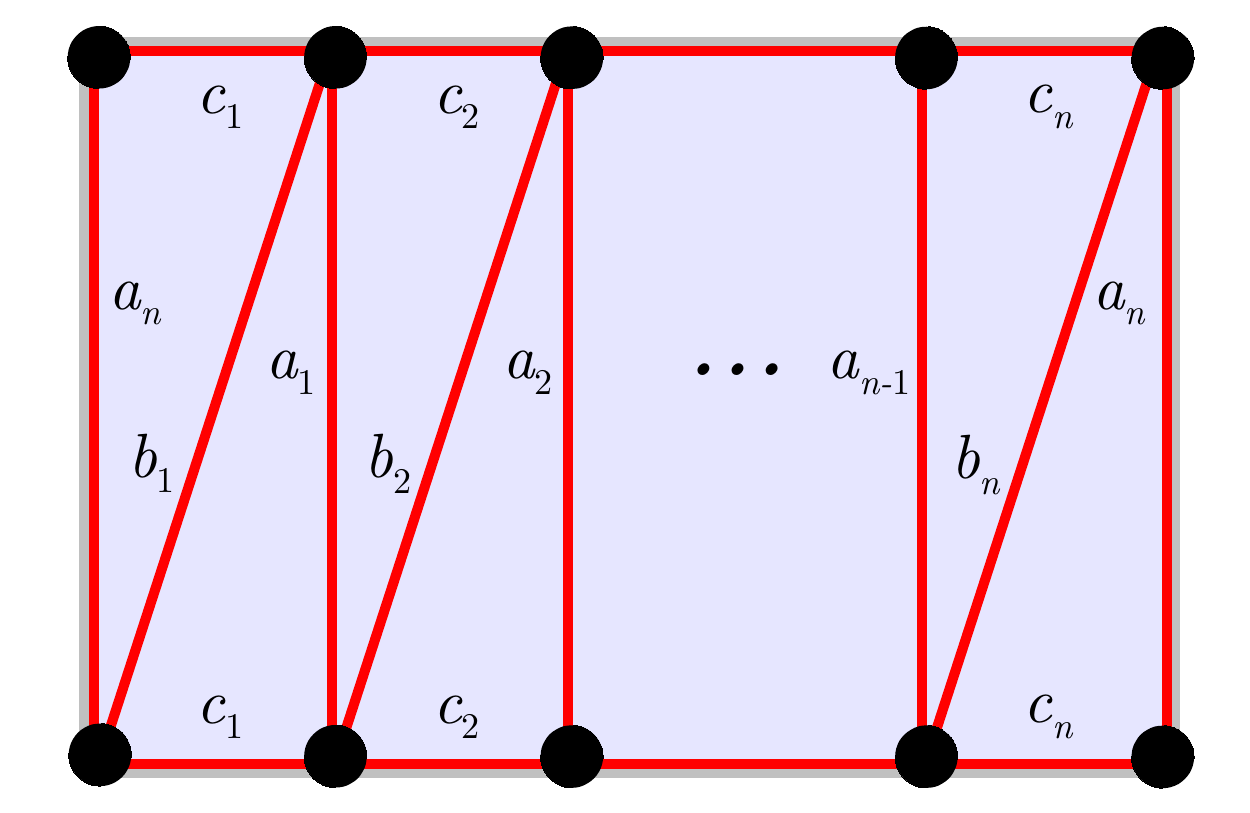}
\centerline{
\begin{xy} 0;<1pt,0pt>:<0pt,-1pt>::
(-45,25) *+{a_n}*\cir<8pt>{}="12",
(0,0) *+{c_1}*\cir<8pt>{} ="0",
(0,50) *+{b_1}*\cir<8pt>{} ="1",
(45,25) *+{a_1}*\cir<8pt>{} ="2",
(90,50) *+{b_2}*\cir<8pt>{}="5",
(135,25) *+{a_2}*\cir<8pt>{}="6",
(165,25) *+{\dots\dots},
(90,0) *+{c_2}*\cir<8pt>{}="7",
(240,0) *+{c_n}*\cir<8pt>{}="8",
(240,50) *+{b_n}*\cir<8pt>{}="9",
(285,25) *+{a_n}*\cir<8pt>{}="10",
(195,25) *+{a_{n-1}}*\cir<12pt>{}="11",
"1", {\ar"0" <1.5pt>},
"1", {\ar"0" <-1.5pt>},
"0", {\ar"2"},
"2", {\ar"1"},
"2",{\ar"5"},
"5", {\ar"7" <1.5pt>},
"5", {\ar"7" <-1.5pt>},
"7",{\ar"2"},
"7",{\ar"6"},
"6",{\ar"5"},
"12",{\ar"1"},
"0",{\ar"12"},
"10",{\ar"9"},
"11",{\ar"9"},
"8",{\ar"10"},
"8",{\ar"11"},
"9", {\ar"8" <1.5pt>},
"9", {\ar"8" <-1.5pt>},
\end{xy}}
\caption{Triangulation and quiver for the torus with $n\ge 2$ punctures. The triangulation is drawn on a rectangle with opposite sides identified.  In both the triangulation and the quiver, the dots indicate repetition of the 3-node structure $a_i b_i c_i$. The torus with $n$ punctures has $n$ sets of double arrows, and $3n$ nodes. Note that the two nodes labelled $a_n$ should be identified, producing a quiver with cyclic symmetry.}
\label{fig:torus}
\end{figure}
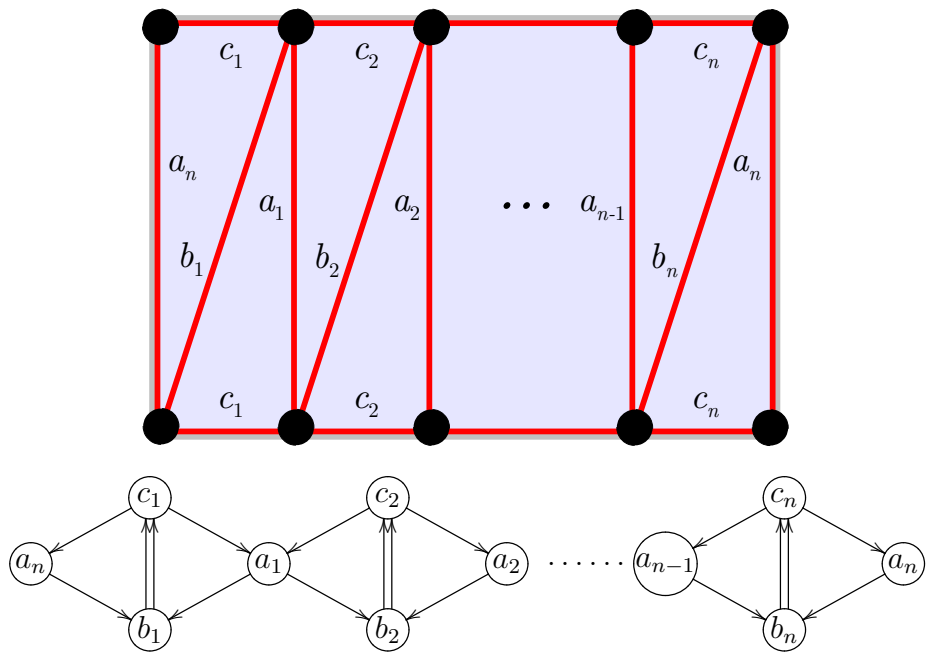

A triangulation and quiver for the torus with $n$ punctures is given in Figure~\ref{fig:torus}.
Again from the triangulation the gauge group and matter content can be directly read off. We have $n$ $SU(2)$ subquivers, giving gauge group $SU(2)^n,$ with bifundamental matter arranged cyclically between every adjacent pair of $SU(2)$s.

A finite chamber for this theory is given by the following sequence of states, in decreasing phase order:
$$\begin{array}{l} a_1,a_1+b_1,a_1+b_1+c_1,2a_1+b_1+c_1+b_2,c_2,a_1+b_2,c_1,a_1+b_1+b_2, \\ a_2+c_2,b_2,a_2,2a_2+b_2+c_2+b_3,c_3,a_2+b_2+c_2+b_3,a_2+c_2+b_3,a_2+b_3, \\
\centerline{\vdots} \\
a_k+c_k,b_k,a_k,2a_k+b_k+c_k+b_{k+1},c_{k+1},a_k+b_k+c_k+b_{k+1},a_k+c_k+b_{k+1},a_k+b_{k+1}, \\
\centerline{\vdots}\\
a_{n-1}+c_{n-1},b_{n-1},a_{n-1},2a_{n-1}+b_{n-1}+c_{n-1}+b_{n},c_{n},a_{n-1}+b_{n-1}+c_{n-1}+b_{n},\\
\quad\quad a_{n-1}+c_{n-1}+b_{n},a_{n-1}+b_{n}, \\
a_n+c_n+c_1,b_n,a_n+c_1,2a_n+b_n+c_n+c_1,b_1,a_n+b_n+c_n,a_n+c_n,a_n\end{array}$$
which is a chamber with $8n$ states.

This can be verified by applying the mutation method with the following mutations, in order:
$$\begin{array}{l} a_1, b_1,c_1,b_2,c_2,c_1,b_1,a_1,a_2,b_2,c_2,b_3,c_3,c_2,b_2,a_2,\dots, a_k,b_k,c_k,b_{k+1},c_{k+1},c_k,b_k,a_k,\dots,\\ a_n, b_n,c_n,b_1,c_1,c_n,b_n,a_n.\end{array}$$

\section{Exceptional Complete Theories}\label{exceptional}
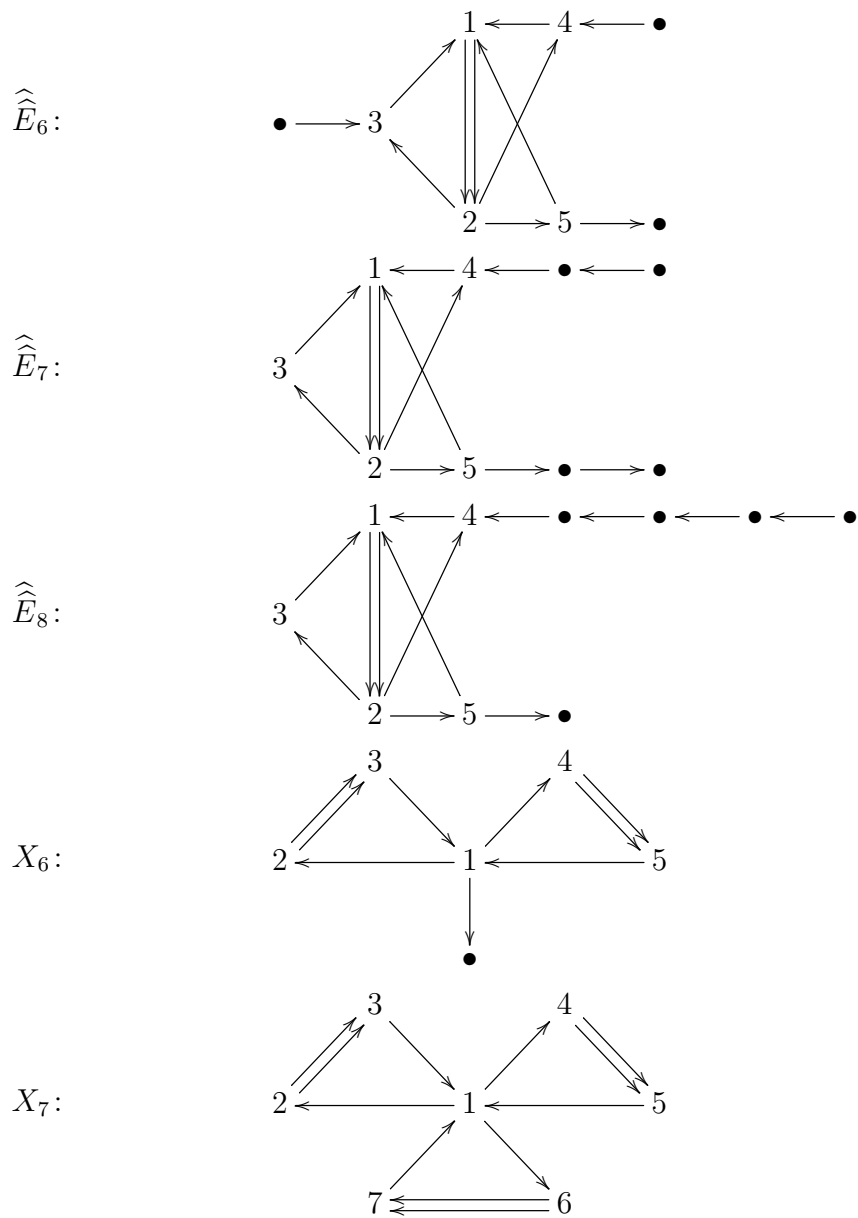
\begin{figure}

 \begin{align*}
 &\widehat{\widehat{E}}_6\colon &&\begin{gathered}
\xymatrix{ & & {1}\ar@<-0.35ex>[dd]\ar@<0.35ex>[dd] &{4}\ar[l]& \bullet\ar[l]\\
\bullet \ar[r] &{3}\ar[ur] & &\\
&& {2} \ar[lu]\ar[ruu]\ar[r] & {5}\ar[uul] \ar[r] &\bullet}
                            \end{gathered}\\
&\widehat{\widehat{E}}_7\colon &&\begin{gathered}
\xymatrix{ & {1}\ar@<-0.35ex>[dd]\ar@<0.35ex>[dd] &{4}\ar[l]& \bullet\ar[l]&\bullet\ar[l]\\
{3}\ar[ur] & & &\\
& {2} \ar[lu]\ar[ruu]\ar[r] & {5}\ar[uul] \ar[r] &\bullet\ar[r]&\bullet}
                            \end{gathered}\\
&\widehat{\widehat{E}}_8\colon &&\begin{gathered}
\xymatrix{ & {1}\ar@<-0.35ex>[dd]\ar@<0.35ex>[dd] &{4}\ar[l]& \bullet\ar[l]&\bullet\ar[l]& \bullet\ar[l] &\bullet\ar[l]\\
{3}\ar[ur] & & & & &\\
& {2} \ar[lu]\ar[ruu]\ar[r] & {5}\ar[uul]\ar[r] &\bullet&&&}
                          \end{gathered}\\
&X_6\colon && \begin{gathered}
\xymatrix{& {3}\ar[rd] && {4}\ar@<0.4ex>[dr]\ar@<-0.4ex>[dr]  &\\
{2} \ar@<0.4ex>[ur]\ar@<-0.4ex>[ur] && {1} \ar[ll]\ar[ur]\ar[d] && {5}\ar[ll]\\
&& \bullet &&}\end{gathered}\\
&X_7\colon &&
\begin{gathered} \xymatrix{& {3}\ar[rd] & & {4}\ar@<0.4ex>[dr]\ar@<-0.4ex>[dr] &\\
{2}\ar@<0.4ex>[ur]\ar@<-0.4ex>[ur]&&{1}\ar[ll]\ar[ur]\ar[rd] && {5}\ar[ll]\\
& {7}\ar[ru] && {6} \ar@<0.4ex>[ll]\ar@<-0.4ex>[ll] &}\end{gathered}
 \end{align*}
\caption{\label{except}The three elliptic $E$--type Dynkin diagrams oriented as to give finite mutation quivers, and the two Derksen--Owen quivers.}
\end{figure}

Thus far in our analysis in this paper we have studied complete gauge theories that are canonically related to Riemann surfaces.  These Riemann surface examples constitute all but finitely many of the complete theories with BPS quivers.  More generally, the full classification of complete theories consists of \cite{FST08, CV11}:
\begin{itemize}
\item All quivers associated to triangulated surfaces, as described in subsection \ref{triangles}.
\item 9 quivers corresponding to $E_n, \widehat{E}_n,\widehat{\widehat{E}}_n$ type Dynkin diagrams, for $n=6,7,8$. $E_n$ and $\widehat{E}_n$ correspond to the usual finite and affine Dynkin diagrams; $\widehat{\widehat{E}}_n$ is given in Figure~\ref{except}.
\item Derksen-Owen quivers, $X_6,X_7$, given in Figure~\ref{except}\cite{DO}.
\end{itemize}
Having thoroughly investigated the BPS quivers and spectra for complete theories associated to Riemann surfaces, we now take our investigation to its logical conclusion and investigate the BPS spectra of the 11 exceptional cases.  By construction, the examples of quivers described here have no interpretation in terms of triangulated surfaces.  Thus a priori we have no independent method for fixing the superpotential, and we simply proceed with an ad hoc case by case investigation.\footnote{After completing the manuscript, we were informed that these potentials (excluding $X_7$) were independently obtained in \cite{Lad} from slightly different considerations.} 

\subsection{$E_n,\widehat{E}_n,\widehat{\widehat{E}}_n$}
The $E_n$ quivers correspond to physical theories that are generalizations of the Argyres-Douglas superconformal theories, and were studied with the affine $\widehat{E}_n$ quivers in \cite{CNV}. These quivers are acyclic, and thus have no superpotential. As described in section \ref{glue}, acyclic quivers always contain a chamber in which the only stable states are those given by the nodes themselves. Thus these theories have finite chambers, where the BPS spectra consists of only the nodes themselves.

\begin{figure}
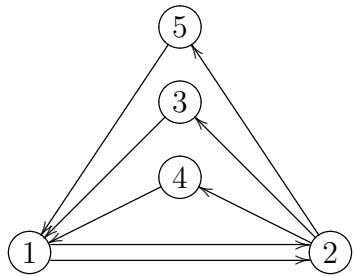

\centerline{\xy
(-20,0)*+{1}*\cir<8pt>{}="a" ; (20,0)*+{2}*\cir<8pt>{}="b" ;(0,20)*+{3}*\cir<8pt>{}="c";(0,30)*+{5}*\cir<8pt>{}="e"; (0,10)*+{4}*\cir<8pt>{}="d";
\ar @{->} "a"; "b" <3pt>
\ar @{->} "a"; "b" <-3pt>
\ar @{->} "b"; "c"
\ar @{->} "b"; "d"
\ar @{->} "b"; "e"
\ar @{->} "c"; "a"
\ar @{->} "d"; "a"
\ar @{->} "e"; "a"
 \endxy}
 \caption{Quiver of $SU(2),\,N_f=3.$ The superpotential is given by $\mathcal{W}=X_{12}X_{23}X_{31}+Y_{12}X_{24}X_{41}+(X_{12}+Y_{12})X_{25}X_{51}$. Notice that this quiver is embedded as a subquiver of the $\widehat{\widehat{E}}_n$ quivers, as shown in Fig. \ref{except}. A decoupling argument indicates that this gives the correct superpotential for studying the $\widehat{\widehat{E}}_n$ quivers.}
 \label{nf3}
\end{figure}

The $\widehat{\widehat{E}}_n$ quivers were also explored in \cite{CV11}. They are given by glueing linear acyclic quivers to the quiver of $SU(2),\, N_f=3,$ (see Figure \ref{nf3}). The only cycles available in these quivers are those of the $SU(2),\, N_f=3$ quiver; thus we can decouple the acyclic linear pieces as described in subsection \ref{super}. The linear subquivers do not participate in the superpotential, since they are not involved in any cycles of the full quiver; therefore this decoupling does not change the superpotential at all.  The superpotential for these quivers is  simply the one given by $SU(2),\,N_f=3,$ shown in Figure \ref{nf3}. Since the quivers involved in the glueing (i.e. $A_n$ linear quivers and $SU(2),N_f=3$) have finite chambers \footnote{We have not described an explicit finite chamber for the $SU(2),\,N_f=3$ quiver. However, since it corresponds to a Riemann surface with boundary, namely the disc with two marked points on the boundary and two punctures, we know that a finite chamber exists.} we conclude that the $\widehat{\widehat{E}}_n$ quivers also have finite chambers.

\subsection{$X_6,X_7$}
The corresponding theories to the Derksen-Owen quivers were also studied in \cite{CV11}. The $X_7$ theory is an $SU(2)^3$ gauge theory with a massive hypermultiplet trifundamental. The $X_6$ theory is a certain decoupling limit of the $X_7$.

\begin{figure}
\centerline{
\begin{xy} 0;<1pt,0pt>:<0pt,-1pt>::
(0,0) *+{2}*\cir<8pt>{} ="0",
(0,50) *+{1}*\cir<8pt>{} ="1",
(45,25) *+{3}*\cir<8pt>{} ="2",
(90,50) *+{4}*\cir<8pt>{}="5",
(90,0) *+{5}*\cir<8pt>{}="7",
"1", {\ar"0" <1.5pt>},
"1", {\ar"0" <-1.5pt>},
"0", {\ar"2"},
"2", {\ar"1"},
"2",{\ar"5"},
"5", {\ar"7" <1.5pt>},
"5", {\ar"7" <-1.5pt>},
"7", {\ar"2"},
\end{xy}}
\caption{Quiver of the annulus with one marked point on each boundary and one puncture, $(0,1,2,\{1,1\}).$ The superpotential is given by $\mathcal{W}=X_{12}X_{23}X_{31}+X_{34}X_{45}X_{53}+Y_{12}X_{23}X_{34}Y_{45}X_{53}X_{31}.$ Note that this quiver is embedded as a subquiver in $X_6,X_7.$}
\label{bifund}
\end{figure}
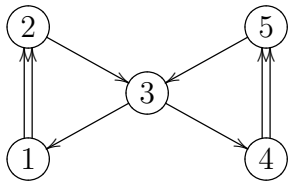

The $X_6$ theory can be decoupled to the quiver corresponding to a punctured annulus, with one marked point on each boundary $(0,1,2,\{1,1\})$ without losing any cycles. Thus its superpotential is simply given by the triangulation construction for that theory, as shown in Figure \ref{bifund}. Since $X_6$ can be obtained from a quiver glueing of the punctured annulus quiver to a one-node quiver, this theory also has a finite chamber.

Finally, we consider $X_7.$ No node of this quiver can be decoupled without removing an oriented cycle, so the approaches used for the other exceptional quivers will not apply. However, the mutation class consists of only two quivers\cite{DO}; thus it is easy to check by hand that a propsed superpotential provides a quadratic mass term for all two-cycles generated under mutation. Furthermore, decoupling node $7$ should yield the quiver $X_6$, with the superpotential given there. From this we are able to guess the superpotential, $\mathcal{W}=X_{12}X_{23}X_{31}+X_{14}X_{45}X_{51}+X_{16}X_{67}X_{71}+Y_{12}X_{23}X_{34}Y_{45}X_{51}+Y_{45}X_{53}X_{36}Y_{67}X_{73}X_{34}+Y_{67}X_{73}X_{31}Y_{12}X_{23},$ which has the desired properties. In principle there are infinitely many higher order terms that could be added to this potential and preserve these properties; this is simply the minimal guess.  Exhaustive computational searches via the mutation method have failed to yield a finite chamber for this quiver. Although we have no proof of this statement, it appears that this quiver does not admit any finite chamber.

\section{Conclusions}
\label{conclu}
In this paper, we have explicitly constructed quivers and corresponding superpotentials for the class of $\mathcal{N}=2$ complete BPS quiver theories.  In hindsight we can see that there are two separate reasons why complete theories have simple and frequently determinable BPS data. 

First, the defining feature of these complete theories is that the dimension of their parameter space coincides with the rank of their charge lattice.  This means that locally in the physical parameter space we may think of the central charges as coordinates.  This has the effect of trivializing any intricate special geometry \cite{STROMINGER} that may have been associated with the Seiberg-Witten solution.  In terms of quiver representation theory this simplification means that in order to find a sensible physical spectrum we may freely choose \emph{any} value of the central charges and compute the corresponding spectrum.  Since the central charges form coordinate functions on parameter space, we are then ensured that the resulting spectrum we have found is indeed the BPS spectrum somewhere.  By contrast in the case of a non-complete theory, a general choice of central charge simply has no a priori relation to any meaningful physical spectrum.  In order to determine which regions of central space correspond to honest physical regimes of parameters one must in general solve a Schottky type problem.

A second simplification which occurs for complete theories is their close relationship with Riemann surfaces and triangulations.  There is a natural mathematical relationship between quivers and triangulations of surfaces and here we have seen a physical interpretation of this fact in terms of BPS state counting.

From this discussion it is clear that a natural extension of our work would be to find a constructive framework for generating quivers for higher rank Gaiotto theories. In \cite{ACCERV} the subject of BPS quiver theories was addressed more broadly, and indeed some progress was made in this direction. In particular, a conjecture was given for the quiver associated with the E6 superconformal theory, which is a building block for the rank 3 Gaiotto case. A more general story one would hope to find would include the appropriate generalization of the triangulation of the Riemann surface $\mathcal{C}$, and a similar map from so called generalized triangulation to the set of quivers that respects the operation of mutation in the necessary way.

The other interesting outcome of this work was the discovery of finite chambers in a large subset of these complete theories. The knowledge of even a single finite chamber in a theory allows us to explore a large number of chambers in its parameter space by applying known wall crossing formulas. (While generally intractable, wall crossing for hypermultiplets in complete BPS quiver theories is completely understood and in fact constructive.) While we indeed found such finite chambers in a large subset of these theories, we would like to complete this classification. This would mean, for each of the remaining complete theories, either demonstrating the existence of a finite chamber or showing that no finite chambers can exist.

\section*{Acknowledgements}

We thank Daniel Labardini-Fragoso for helpful discussions. MA, SC, CC, SE, AR
and CV thank the 2011 Simons workshop in Mathematics and Physics and the Simons
Center for Geometry and Physics for hospitality during the completion of this
work. The work of MA is supported by DFG fellowship AL 1407/1-1. The work of CV
is supported by NSF grant PHY-0244821

\appendix
\section{Self-Folded Triangles}\label{selffold}
In our discussion above we have left out a minor technicality involving \emph{self-folded} triangles. A self-folded triangle is one in which two sides become identified, resulting in the degenerate structure seen below.
\begin{equation}\label{selfoldedT}
\begin{gathered}{\xy {(20,0)*+{\bullet}; (20,20)*+{\bullet} **\crv{}\POS?(0.8)*^+!L{int}};
{(20,0)*+{}; (20,0)*+{} **\crv{(5,20)&(20,40)&(35,20)}\POS?*_+!D{ext}}
\endxy}\end{gathered}
\end{equation}
We will call the edge labeled \emph{ext} exterior, and the edge labeled \emph{int} interior.
The framework of triangulations above requires allowance of self-folded triangles. In particular, some triangulations obtained from special lagrangian flows will require self-folded triangles, and similarly, some flips will force self-folded triangles to occur.

To properly include these structures, we must slightly augment the rules for obtaining a quiver $Q$ and superpotential $\mathcal{W}$ from a triangulation $\mathcal{T}$. First, it is useful to note that self-folded triangles, while necessary for the formalism, are a bit of an extraneous complication. It is a theorem from \cite{FST} that every surface admits a triangulation without self-folded triangles. Thus, having carefully understood the map from triangulations and quivers, which maps flips to mutations, the rules for self-folded triangles can be derived from the rules given in the body of the paper. We would simply apply flips of the triangulation to remove all self-folded triangles, use the given rules to obtain $Q$ and $\mathcal{W}$, and then invert the flips with the appropriate inverse mutations on the quiver. For completeness, we give the relevant rules here.

To obtain the quiver $Q$, we apply the usual rules as given in section \ref{triangles} to all diagonals, except for interior edges of self-folded triangles. For the interior edge of each self-folded triangle, we draw a node corresponding to it, and draw arrows that duplicate the arrows of the node corresponding to the exterior edge of the same self-folded triangle. For clarity, let us define a function $e$ on diagonals $\delta$: if $\delta$ is an interior edge, $e(\delta)$ is the exterior edge of the self-folded triangle whose interior edge is $\delta$; otherwise, $e(\delta)$ is simply $\delta$. Similarly, we define $i(\delta)$ to give the associated interior edge if $\delta$ is an exterior one. Thus the full rules are:
\begin{itemize}
\item  For each diagonal $\delta$ in the triangulation, draw exactly one node of the quiver.
\item For each pair of diagonals $\delta_{1}, \delta_{2}$ find all triangles for which $e(\delta_1),e(\delta_2)$ are both edges.  Then for each such triangle draw one arrow from $\delta_{1}$ to $\delta_{2}$ if $e(\delta_{1})$ immediately precedes $e(\delta_{2})$ going counter-clockwise around the triangle.
\end{itemize}

Similarly, we should also extend the superpotential to include self-folded triangles. We use $\alpha, \beta,\gamma\dots$ to denote both the diagonals and their respective nodes in the quiver, and $B_{\alpha\beta}$ to denote both an arrow from $\alpha$ to $\beta$ and the associated bifundamental matter field. The full rules are as follows:
\begin{itemize}

\item For each internal, non-self-folded triangle $\alpha\beta\gamma$, we add the associated three cycle $B_{\alpha\beta}B_{\beta\gamma}B_{\gamma\alpha}.$
\item For each internal, non-self-folded triangle $\alpha\beta\gamma$ adjacent to exactly two self-folded triangles enclosed by $\alpha,\beta$ respectively, we add an additional three cycle $B_{i(\alpha)i(\beta)}B_{i(\beta)\gamma}B_{\gamma i(\alpha)}.$
\item For each internal, non-self-folded triangle $\alpha\beta\gamma$ adjacent to exactly three self-folded triangles, we add three additional terms $B_{i(\alpha)i(\beta)}B_{i(\beta)\gamma}B_{\gamma i(\alpha)}+B_{i(\alpha)\beta}B_{\beta i(\gamma)}B_{i(\gamma) i(\alpha)}+B_{\alpha i(\beta)}B_{i(\beta)i(\gamma)}B_{i(\gamma) \alpha}.$
\item For each internal, regular puncture adjacent to exactly one internal diagonal $\alpha$, we must have a self-folded triangle. The diagonal $e(\alpha)$ occurs in at most one non-self-folded triangle. If that triangle is internal, $e(\alpha)\beta\gamma,$ we add the three cycle $B_{\alpha\beta}B_{\beta\gamma}B_{\gamma\alpha}$.
\item For each internal, regular puncture adjacent to more than one internal diagonal, we remove all the exterior edges of self-folded triangles incident on the puncture. Now let $n$ be the number of remaining diagonals incident on the puncture. The quiver must have an $n$ cycle $\alpha_1\dots\alpha_n$; we add the term $B_{\alpha_1\alpha_2}\dots B_{\alpha_{n-1}\alpha_n}B_{\alpha_n\alpha_1}$.
\end{itemize}

\bibliography{quivers.bib}{}
\bibliographystyle{utphys}
\end{document}